\documentclass[10pt, aps,pra,twocolumn, superscriptaddress,longbibliography,footinbib]{revtex4-2}
\usepackage[dvipsnames]{xcolor}
\usepackage[colorlinks=true, linkcolor=purple, urlcolor=purple, citecolor=purple]{hyperref}
\usepackage[english,main=english]{babel}
\usepackage[T1]{fontenc}
\usepackage[utf8]{inputenc} 
\usepackage{ragged2e}
\usepackage{graphicx}
\usepackage{dcolumn}
\usepackage{booktabs}
\usepackage{dsfont, amsmath,amsfonts, bm}
\usepackage{MnSymbol}
\usepackage{algorithm,algorithmicx,algpseudocode}

\newcolumntype{C}[1]{>{\centering\arraybackslash}p{#1}}
\algrenewcommand\algorithmicrequire{\textbf{Input:}}
\algrenewcommand\algorithmicensure{\textbf{Output:}}
\newcommand{\const}[1]{\textcolor{darkgray}{#1}}
\newcommand{\varp}[1]{\textcolor{black}{#1}}
\newcommand{\breakablealgcaption}[2]{\par\noindent\hrulefill\par\refstepcounter{algorithm}\noindent {Algorithm~\thealgorithm. \textbf{#1}}\label{#2}\par\noindent\hrulefill\par}

\begin{document}

\title{Autonomous oscillations in quantum electromechanics: tensor network treatment}
\author{Mahasweta Pandit}
\email{m.pandit@um.es}
\affiliation{Departamento de Física - CIOyN, Universidad de Murcia, Murcia E-30071, Spain}
\author{Sheikh Parvez Mandal}
\email{sheikhparvez.mandal@um.es}
\affiliation{Departamento de Física - CIOyN, Universidad de Murcia, Murcia E-30071, Spain}
\author{Mark T. Mitchison}
\email{mark.mitchison@kcl.ac.uk}
\affiliation{School of Physics, Trinity College Dublin, College Green, Dublin 2, D02 K8N4, Ireland}
\affiliation{Department of Physics, King’s College London, Strand, London, WC2R 2LS, United Kingdom}
\author{Javier Prior}
\email{javier.prior@um.es}
\affiliation{Departamento de Física - CIOyN, Universidad de Murcia, Murcia E-30071, Spain}

\begin{abstract}
    Transport-induced self-sustained oscillations in electromechanical systems convert a static electrochemical bias into robust, autonomous oscillatory motion in the absence of any external periodic drive. However, an exact description of such self-oscillations remains challenging in nanoscale electromechanical devices featuring a simultaneously large bosonic Hilbert space, strong interactions, and structured fermionic leads. We formulate a tensor-network framework that combines a binary representation of the vibrational mode with mesoscopic reservoir embeddings that enable controlled access to the self-oscillatory steady states and relevant transport observables without explicit real-time propagation. We demonstrate the emergence of mechanical self-oscillations across a broad set of operating conditions, in which strong electromechanical backaction, nonadiabatic oscillator dynamics, and energy-dependent electronic tunneling processes compete. Furthermore, we observe that for both slow and fast vibrating mechanical modes, suppressed vibrational occupation fluctuations in the self-oscillation window along the electromechanical coupling strength sweep are preceded by a peak in the occupation fluctuations. Collectively, we explore how both intrinsic system properties and environmental parameters govern such autonomous oscillations over a broad range of operating conditions. The broad applicability of our framework will facilitate further extension of the method to more complex and experimentally relevant scenarios to study the thermodynamics of quantum autonomous devices with potential applications in quantum sensing, timekeeping, information processing, and energy conversion.
\end{abstract}

\maketitle
\section{Introduction} \label{ch::introduction}
In nonlinear dissipative systems, self-sustained oscillations are stable periodic motions that arise without external periodic driving when a constant supply of nonequilibrium free energy autonomously supports a limit cycle through a balance of gain and nonlinear dissipation \cite{ginoux2012van, Jenkins2013PhysRep}. Such dynamical states are ubiquitous in nature and underpin autonomous signal generation \cite{guzman2023useful}, timekeeping \cite{Erker2017PRX}, and energy conversion \cite{sevitz2026autonomous, Culhane2022PRE}, among other important phenomena. This mechanism underlies models such as the van der Pol oscillator \cite{van1926lxxxviii} and lasers \cite{haken1964nonlinear}, and more broadly informs descriptions of chemical oscillators and biological clocks, including circadian rhythms \cite{BenArosh2021PRResearch, EpsteinShowalter1996JPhysChem, BellPedersen2005NatRevGenet, Patke2020NatRevMolCellBiol}. In mesoscopic electromechanical and optomechanical systems, analogous behavior arises when backaction from tunneling electrons or cavity photons produces negative effective damping, with saturation by nonlinear damping, yielding autonomous mechanical motion \cite{blanter2004single, a2005quantum, Aspelmeyer2014RMP, Urgell2020NatPhys, Wen2020NatPhys, Culhane2024NJP}. 

Nanoelectromechanical systems (NEMS), where a vibrational mode is coupled to electronic, photonic, or spin degrees of freedom, provide a natural platform for realizing and studying autonomous limit cycles \cite{Aspelmeyer2014RMP, PootVanDerZant2012PhysRep, OConnell2010Nature, Teufel2011Nature, ArrangoizArriola2019Nature, Rossi2018Nature, Kolkowitz2012Science}. Their mechanical mode can simultaneously act as a dynamical subsystem, an internal energy reservoir, and a quantum sensor, while enabling electrical or optical readout of the vibrational state \cite{PootVanDerZant2012PhysRep, Aspelmeyer2014RMP}. In NEMS, charge transport through a resonator-coupled quantum dot generates backaction forces that can cool, amplify, or destabilize the mechanical mode \cite{blanter2004single,a2005quantum, Piovano2011PRB} and are well established in platforms ranging from molecular junctions \cite{Park2000Nature, Koch05, Koch96, mitra2004phonon, avriller2011unified, Schinabeck2020HQMEFCS} to suspended carbon nanotubes (CNT)~\cite{Urgell2020NatPhys, Wen2020NatPhys, Willick2020PRR, Vigneau2022PRR, Aresbistability24}.\,Transport-induced limit cycles have also been explored in voltage-sustained electron shuttles \cite{Fedorets2004PRL, Novotny2003PRL, Gorelik1998PRL, Utami2006PRB}, multimode electromechanical instabilities \cite{Jonsson2008}, single-electron-transistor resonator theories \cite{Blencowe2005NJP, blanter2004single, Pistolesi2004PRB}, Kondo-assisted backaction \cite{Song2014}, and strong-coupling polaronic regimes \cite{Skorobagatko2013}, highlighting delayed electronic feedback and nonlinear transport \cite{Koenig2012} as mechanisms for negative damping and limit-cycle stabilization. They have also been directly observed in CNT devices for long-lived and bistable oscillatory states near transport-regime boundaries \cite{Urgell2020NatPhys, a2005quantum, Willick2020PRR, Aresbistability24, Aresbistability24}.
    
From the perspective of quantum thermodynamics, these devices are minimal realizations of \textit{autonomous machines} in which a constant electrochemical bias is converted into periodic mechanical motion entirely without time-dependent control. This connection makes NEMS self-oscillators a unique platform for investigating precision-dissipation tradeoffs at the intersection of quantum transport, open quantum systems, and the thermodynamics of timekeeping \cite{Wachtler2019NJP, Erker2017PRX, Culhane2024NJP, Karwat2018PRA, sevitz2026autonomous, 2026roadmap, mahadeviya26}. A key parameter of the underlying physics is the dimensionless ratio $\gamma = \Gamma/\omega_0$ between the electronic tunneling rate $\Gamma$ and the mechanical frequency $\omega_{0}$. In the slow-vibration limit $\gamma \gg 1$, effective semiclassical descriptions with backaction-induced drift and diffusion are often useful, whereas in the fast-vibration limit $\gamma \ll 1$, dynamical memory and quantum fluctuations become central, adiabatic elimination fails \cite{Blencowe2005NJP, Piovano2011PRB, sevitz2025quantum}, and non-Markovian reservoir correlations~\cite{deVegaAlonso2017RMP, Tamascelli2018, Strathearn2018NatComm} can decisively influence whether a limit cycle forms at all. 

In this paper, we bridge these separate parameter regimes in \textit{a single, unbiased computational framework} to determine which features of the dynamics and the self-oscillatory steady state are universal and which are regime-specific. This provides a unified characterization of the self‑oscillation windows, in which the parameter bounds manifest as continuous intervals that would otherwise need to be analyzed as distinct, separate regimes.  We consider a standard model that captures the most important ingredients of an NEMS device, namely the Anderson--Holstein (AH) impurity model. In this model, a localized electronic level is linearly coupled to a quantized vibrational mode, connected to voltage-biased fermionic leads, and weakly damped by a bosonic bath \cite{Anderson1961, Holstein1959, HewsonMeyer2002, hubener2007vibrational, de2020manipulation}. Despite its conceptual simplicity, this model has a broad experimental relevance. 

However, controlled access to the nonequilibrium steady state (NESS) of the AH model over broad parameter regimes remains computationally demanding. Strong electron--phonon coupling and nonequilibrium vibrational excitation can generate long transients before the NESS is reached \cite{mitra2004phonon, Schmidt2008, Albrecht2013}. Sequential-tunneling and Markovian master-equation approaches give useful physical insight into weak dot--lead coupling regimes, but rely on perturbative tunneling expansions and often on secular or diagonal vibronic approximations \cite{Flindt2010PRB, FlindtNovotny2005NoiseSpectrum, avriller2011unified, Thomas2013PRB}. Semiclassical Fokker--Planck or Langevin methods are appropriate when the oscillator is slow and highly occupied, but do not directly provide the full quantum oscillator state in the coherent crossover regime \cite{blanter2004single, a2005quantum,  BennettClerk2010Scattering, Culhane2022PRE, jovchev2013}. Nonequilibrium Green's function methods offer a general transport framework, but practical treatments of electron--phonon coupling usually require diagrammatic or self-consistent approximations whose accuracy must be assessed in strongly coupled, highly excited regimes \cite{HaugJauho2008, KamenevBook2011, HartleThoss2009PRL, Thoss2011, ChenReichman2016NCA}. Hierarchical quantum master-equation methods provide controlled benchmarks for parts of this problem, including nonadiabatic vibrational dynamics and current fluctuations \cite{Schinabeck2016PRB, Schinabeck2018HQMEVibBath, Schinabeck2020HQMEFCS}, but their cost grows rapidly with the vibronic cutoff and with the number of bath-correlation terms required for structured or low-temperature reservoirs.

We develop a numerically controlled, steady-state tensor-network framework that tackles the three central difficulties of simulating an impurity model simultaneously: (i) the need for large phonon occupation cutoffs in the limit-cycle regime, (ii) the structured (non-Markovian) nature of the fermionic reservoirs, and (iii) strong electron-phonon coupling beyond perturbation theory. Our approach combines an exact binary (pseudo-site) encoding of the truncated bosonic Hilbert space \cite{JeckelmannWhite1998PRB} with mesoscopic reservoir embedding \cite{Imamoglu1994, dalton2001theory, subotnik2009nonequilibrium, Dzhioev2011, ajisaka2012nonequilibrium, Arrigoni2013PRL, Dorda2014PRB, Schwarz2016PRB, elenewski2017master, Brenes2020, lacerda2024entropy,alford2025subtleties}, which provides us remarkable computational efficiency while allowing a large number of phonon levels without loss of accuracy and exactly represents structured fermionic environments through a finite chain of damped auxiliary modes, respectively. We use a variational NESS solver \cite{Mascarenhas2015, CuiCiracBanuls2015PRL, Weimer2021RMP} in Liouville space \cite{Verstraete2004PRL, ZwolakVidal2004PRL, Prosen2008NJP} with tensor-network compression \cite{White1992DMRG, White1993DMRG, Schollwoeck2005RMP, Schollwoeck2011, Stolpp2021CPC, Zhao2023PRR} that directly targets the steady state without long real-time evolution. This unified framework grants simultaneous access to phase-space, number-statistics, energetic, and transport diagnostics of the mechanical mode for both slow and fast phonon oscillations and operates in both weak and strong coupling regimes. It offers a complementary perspective to \cite{Schinabeck2018HQMEVibBath} by retaining direct access to the nonequilibrium \textit{oscillator state} and its phase-space structure, particularly enabling us to distinguish transport-induced vibrational excitation and fluctuations from coherent self-oscillation dynamics. Employing our approach across a wide parameter landscape, we establish under what conditions the device exhibits a limit cycle. We show that the clearest self-oscillatory states emerge in an intermediate window of electromechanical coupling and low-to-moderate temperature. The finite lead bandwidth (memory) can either assist or suppress limit-cycle formation depending on the mechanical frequency, thereby reshaping the self-oscillation window. 

The article is organized as follows. In Sec.~\ref{sec:model}, we formally introduce the AH model for NEMS and point out key parameter regimes that have been explored in this work. In Sec.~\ref{sec::methods}, we present our numerically controlled methodology, including the binary encoding of the phonon mode, the mesoscopic-lead embedding of structured reservoirs, the superfermionic representation, and the tensor network implementation. In Sec.~\ref{sec::results}, we begin by defining the relevant observables and diagnostics used to characterize the limit cycle behavior in the NESS. In this section, we then analyze the steady-state behavior across several parameter regimes and present the interplay between the self-sustained oscillations with adiabaticity, lead structure, thermal broadening, and bosonic dissipation. We also present benchmarks with established bounds and methodologies and discuss the computational complexity of our simulations. Finally, in Sec.~\ref{sec::conclusion}, we summarize our main results and discuss their implications for identifying and controlling self-sustained oscillations in autonomous nanoelectromechanical systems.

\section{Modeling systems with electromechanical coupling at the nanoscale} \label{sec:model}
\begin{figure*}[htbp]
    \centering
    \includegraphics[width=\textwidth]{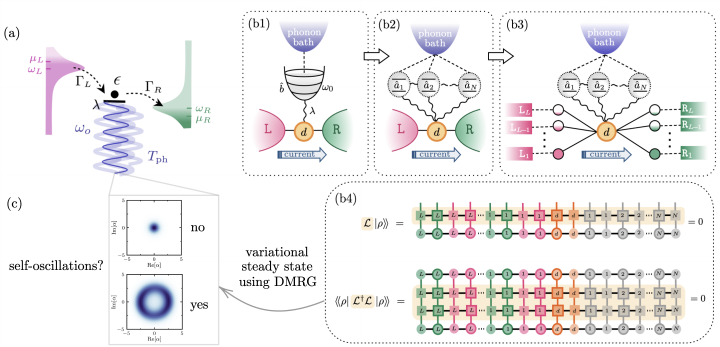}
    \parbox{\textwidth}{\caption{\justifying (a) Schematic of a NEMS consisting of a quantum dot, with onsite energy $\epsilon$, coupled to two fermionic reservoirs that are driven out of equilibrium by an electrochemical bias $\mu_L \ge \mu_R$. The imbalance induces electron tunneling through the dot, transferring electrical energy between the reservoirs. The dot additionally couples to a vibrational mode with frequency $\omega_{0}$ (denoted as a solid violet wavy line). The vibrational mode is weakly damped by a bosonic vibrational bath of temperature $T_{\rm ph}$ (depicted here as light wavy lines in the background). Self-sustained oscillations in the phonon mode can be induced by the backaction of electron transport through the dot. (b1) The quantum dot $d$ is coupled to a vibrational mode $\hat b$ of frequency $\omega_0$, two fermionic environments ($\mathtt{L}$ and $\mathtt{R}$), and a weakly coupled phonon bath (AH model). (b2) Binary encoding of the truncated phonon Hilbert space with a cutoff $M$ is represented by $N=\log_2 M$ hard-core-boson pseudosites $\hat a_j$, as in Eq.~\eqref{eq:binary_map}.
    (b3) Mesoscopic-lead embedding of the structured fermionic reservoirs into damped auxiliary modes governed by Lindblad dissipators, Eq.~\eqref{eq:ME_aug}. (b4) Variational steady-state condition depicted in tensor network form after Liouville-space vectorization of the density matrix $\rho$. The NESS satisfies $\mathcal{L}|\rho\rangle\!\rangle=0$ and is obtained by variationally minimizing the positive semidefinite form $\langle\!\langle\rho|\mathcal{L}^\dagger\mathcal{L}|\rho\rangle\!\rangle$. (c) Signature of self-sustained oscillations captured by the Husimi $Q$-distribution: a unimodal bell-shaped profile signifies the absence of oscillatory dynamics, whereas a ring-shaped structure is indicative of the presence of self-oscillations. \label{fig::schematic}}}
\end{figure*}
\subsection{Setup}
The total Hamiltonian $\hat{H}$ in Eq.\,\eqref{eq::Htot} for the nano-electromechanical device model comprises a system part $\hat{H}_{S}$ that describes a quantum dot hosting a spinless electronic level coupled to a local phonon (vibrational) mode [see Fig.~\ref{fig::schematic}(a)]. The macroscopic bath Hamiltonian $\hat{H}_{B}$ accounts for two non-interacting electronic leads, representing source (left) and drain (right) reservoirs, respectively, connected to the dot, as well as a phonon bath weakly damping the phonon mode. The interaction term $\hat{H}_{I}$ captures the coupling of the system to these three baths, encompassing both fermionic and bosonic interactions. In total,
\begin{equation}
     \hat{H}  = \hat{H}_{S}  + \hat{H}_{B}  + \hat{H}_{I}, \label{eq::Htot}
\end{equation}
where each component reads as
\begin{eqnarray} 
    \hat{H}_{S} &=& \epsilon \, \hat{n}_{d}
    + \omega_0\, \left(\hat{b}^{\dagger }\hat{b} + \frac{1}{2}\right)
    + \lambda\omega_0\,
    \left(\hat{n}_{d}-n_g\right)
    (\hat{b}^{\dagger } + \hat{b}), \label{eq::Hs}\\
    \hat{H}_{B} &=& \sum _{k, \alpha = L,R} {\Omega}_{k\alpha} \ \hat{c}_{k\alpha}^{\dagger }\hat{c}_{k\alpha}
    + \sum _{j} {\Omega}_{bj} \ \hat{b}_{j}^{\dagger }\hat{b}_{j}, \label{eq::HB}\\
    \hat{H}_{I} &=& \sum _{k, \alpha = L,R} g_{k\alpha}\hat{d}^{\dagger }\hat{c}_{k\alpha}
    + \sum _{j} h_{j} \hat{b}^{\dagger }\hat{b}_{j}
    +\text{H.c.} \label{eq::HI}
\end{eqnarray}
We consider that the dot hosts a single spinless resonant level with occupation $\hat n_d=\hat d^\dagger\hat d$, where the allowed charge states are $n_d=0,1$. This corresponds to an AH model in the Coulomb-blockade regime, where the charging energy is the largest energy scale and double occupation is excluded. The parameter $\epsilon$ denotes the effective gate-tunable addition energy of the dot, and the dimensionless charge induced by the gate voltage $V_g$ is
\begin{equation}
    n_g =\frac{C_g V_g}{e}
\end{equation}
where $C_g$ is the gate capacitance, and $e>0$ is the magnitude of the electron charge. In this work, we set $\hbar=k=e=1$.

The vibrational mode has frequency $\omega_0$ with creation (annihilation) operator $\hat{b}^{\dagger}$ ($\hat{b}$), $[ \hat{b}, \,\hat{b}^{\dagger} ] = 1$  and displacement
\begin{equation}
    \hat x =  \frac{x_0}{2}(\hat b^\dagger+\hat b), \label{eq:position_op}
\end{equation}
with $x_0=\sqrt{2/(m\omega_0)}$ being the mechanical zero-point fluctuation for an oscillator of mass $m$. The electromechanical interaction is linear in displacement and couples to the effective dot charge $\hat n_d-n_g$ with dimensionless coupling $\lambda$. In this paper, we take the symmetric choice $n_g=1/2$, for which the empty and occupied dot states exert equal and opposite forces on the vibrational mode.

The fermionic operator $\hat{c}_{k\alpha}$ annihilates an electron in lead $\alpha\in\{L, R\}$ with frequency $\Omega_{k\alpha}$, and $\hat{b}_j$ annihilates bath bosons with frequency $\Omega_{bj}$ and tunneling rate $h_{j}$. The electron tunneling amplitudes $g_{k\alpha}$ hybridize the dot with the fermionic leads with spectral densities featuring a finite Lorentzian bandwidth,
\begin{equation}
    J_\alpha(\omega) = 2\pi \sum_{k} |g_{k\alpha}|^{2} \delta(\omega - \Omega_{k\alpha}) = \frac{\Gamma_\alpha\, \delta_\alpha^2}{(\omega - \omega_\alpha)^2+\delta_\alpha^2}, \label{eq:spectral_densities}
\end{equation}
where $\omega_\alpha$ is the peak position, $\delta_{\alpha}$ is the bandwidth and $\Gamma_{\alpha}$ the height. A bias voltage $V_{s}$ is applied between the baths, opening up a bias energy window $e V_{s} = \mu_{L} - \mu_{R}$, with $\mu_L$ ($\mu_R$) being the position of the Fermi surface of the left (right) lead. Given an electrochemical potential level $\mu$ corresponding to a transition that involves the charge state of the quantum dot within this window, electrons can tunnel from one bath to the other via the quantum dot. Since the dot charge is coupled to the vibrational displacement, the oscillator equilibrium position changes, and the tunneling operator is therefore dressed by a displacement operator. This can be seen through the Franck--Condon matrix elements for transitions between phonon number states
$|m\rangle\to |n\rangle$,
\begin{equation}
    X_{nm} = \langle n |e^{-\lambda(\hat b^\dagger-\hat b)}|m\rangle, \qquad F_{nm}=|X_{nm}|^2 .
    \label{eq:FC_factor}
\end{equation}
Such displacement-dressed tunneling is a standard mechanism in molecular and nanoelectromechanical transport \cite{wingreen1989inelastic, Koch05, Galperin2007JPCM} and its relevance in the formation of limit cycle dynamics in our NEMS model is discussed in Sec.~\ref{subsec:heating_vs_limitcycle}.

For the symmetric choice $n_g=1/2$, the Lang--Firsov reorganization energy is an occupation-independent constant and does not shift the relative dot level \cite{Holstein1959, LangFirsov1963, Mahan2000}. The vibronic transition energy is then $\Delta_{nm}=\epsilon+(n-m)\omega_0 $. For general $n_g$, the Lang--Firsov transformation generates the reorganization term $-\lambda^2\omega_0(\hat n_d-n_g)^2$. In the dot, $\hat n_d^2=\hat n_d$, so this is equivalent to a constant term and a gate-dependent renormalization of the dot level, $\epsilon_{\rm eff}=\epsilon-\lambda^2\omega_0(1-2n_g)$.

Real mechanical resonators have a finite quality factor. For example, in suspended NEMS, the vibrational mode can lose energy to the substrate, clamping, electromagnetic, and other environmental degrees of freedom \cite{Koch96, PootVanDerZant2012PhysRep, lau2016redox}. To model a finite \textit{mechanical quality factor}, 
\begin{equation}
    Q_m = \frac{\omega_0}{\Gamma_{\rm ph}}, \label{eq:quality_factor}
\end{equation}
we therefore couple the phonon mode to a weak thermal bosonic bath~\cite{Siddiqui2007} at temperature $T_{\rm ph}$ (details in Section \ref{sec::methods}), with damping rate $\Gamma_{\rm ph}=2\pi \sum_j |h_j|^2 \delta(\omega_0-\Omega_{bj})$.
We take $\Gamma_{\rm ph}\ll\omega_0$, so that the mechanical mode has a high quality factor. The bosonic bath sets a thermal reference for the vibrational mode, and in Sec.~\ref{subsec::params_interplay} we study the effect of mechanical dissipation on limit cycle formation in this model.

\subsection{Parameter regimes}
\begin{figure}[!tbp]
     \includegraphics[clip, width=\columnwidth]{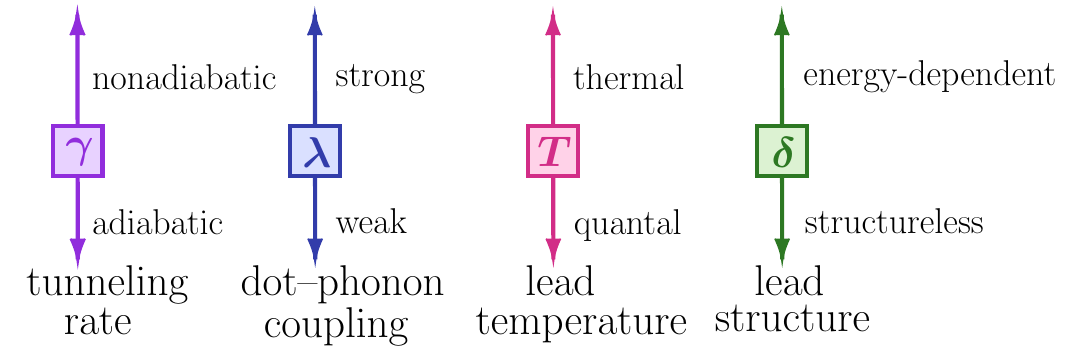}
     \parbox{\columnwidth}{\caption{\justifying Major parameter regimes explored in this work. The leftmost diagram featuring the adiabaticity parameter, $\gamma=\Gamma/ \omega_0$, depicts how increasing $\Gamma$ with respect to $\omega_0$ takes the model to the nonadiabatic regime. In the other diagrams, $\lambda$ denotes the dimensionless electromechanical coupling strength, the fermionic lead temperature is denoted by $T$, and $\delta$ is the bandwidth of the fermionic lead. The methodology presented in this work allows us to uncover the physics at the interface of the denoted regimes in a unified framework.\label{fig::schematic_regime}}}
\end{figure}

NEMS exhibit rich physics in a number of qualitatively different regimes, depending on parameters such as the vibration frequency, the dot--phonon or dot--leads coupling strengths, the vibrational relaxation rate, etc., where the interplay between the single-electron transport through the dot and the mechanical motion of the vibrational mode plays a crucial role \cite{Koch05, WegewijsNowack2005NJP, Koch96, RyndykHartungCuniberti2006PRB}. In Fig.~\ref{fig::schematic_regime}, we point out some parameters such as the electron tunneling rate $\Gamma$, mechanical oscillation frequency $\omega_0$, dot--phonon coupling strength $\lambda$, fermionic lead temperature $T$, and the lead bandwidth $\delta$.  We use the \textit{adiabaticity parameter},
\begin{equation}
    \gamma =  \frac{\Gamma}{\omega_{0}}, \label{eq:gamma_definition}
\end{equation}
and distinguish three dynamical regimes as $\gamma$ is varied \cite{Piovano2011PRB}. 

In the \textit{fast-vibration}, or slow-transport, regime $\gamma\ll 1$, the mechanical oscillator or the phonon mode completes many periods during the typical electronic tunneling time. The electronic dynamics, therefore, averages over the oscillator phase and is primarily sensitive to vibronic transition energies rather than to the instantaneous oscillator displacement. In this limit, the oscillator density matrix becomes approximately diagonal in the energy eigenbasis \cite{Piovano2011PRB}.  This regime underlies rate-equation descriptions of Franck--Condon blockade \cite{Leturcq2009NatPhys}, super-Poissonian current noise \cite{PhysRevB.74.195305}, and nonequilibrium vibrational distributions \cite{mitra2004phonon, avriller2011unified}. It is also the regime considered in a recent autonomous particle-exchange model \cite{sevitz2025quantum} of the conversion of slow sequential tunneling into mechanical self-oscillation.

The opposite limit, $\gamma\gg 1$, corresponds to \textit{slow vibrations}, or the quasi-adiabatic regime. Here, the electronic relaxation is fast compared with the mechanical motion, so the dot occupation can approximately adjust to the instantaneous oscillator coordinate. In this regime, the electronic degrees of freedom can often be adiabatically eliminated, leading to effective semiclassical descriptions in terms of current-induced forces, damping, and diffusion acting on the oscillator \cite{armour2004classical, blanter2004single, a2005quantum, Bode2012Review}. This quasi-adiabatic limit is also used in quantum-thermodynamic descriptions of electromechanical work extraction and autonomous clock operation, where the mechanical mode acts as a slowly evolving work-storage or timing degree of freedom \cite{Culhane2022PRE, Culhane2024NJP}.

In the intermediate regime, $\gamma \sim 1$, the electronic response is neither slow enough for a purely energy-resolved sequential-tunneling picture nor fast enough for a fully adiabatic semiclassical elimination. In this crossover regime, coherences between vibrational states can become relevant. Piovano \textit{et al.} showed that off-diagonal density-matrix elements can modify the redistribution of vibrational occupations and lead to an effective vibrational temperature below that inferred from a purely diagonal rate-equation treatment \cite{Piovano2011PRB}. Related intermediate-regime physics, where tunneling backaction and mechanical motion must be treated on comparable time scales, has also been discussed in coherent nanoelectromechanical and shuttle-type transport settings \cite{a2005quantum, Novotny2003PRL, Wen2020NatPhys, Aresbistability24}. 

Experiments with NEMS devices are generally located in high-bias, high-quality-factor regimes.  In the recent experiment of Wen \emph{et al.}~\cite{Wen2020NatPhys} on a strongly coupled nanotube single-electron transistor, the mechanical quality factor is \(Q_m\simeq 1.8\times10^4\), and the device shows bias-driven, self-sustained oscillations.  Our minimal spinless AH model does not include all the nonlinear components of that experiment, but it captures the same central requirement that the current must supply strong backaction to stabilize the self-oscillations. A related CNT experiment by Tabanera-Bravo \emph{et al.}~\cite{Aresbistability24} reports bistable and long-sustained self-oscillations with $f_m\simeq270\,{\rm MHz}$, $Q_m\sim 2000$, and tunneling rates much larger than the mechanical frequency, placing the device deep in the slow-vibration or quasi-adiabatic side of Eq.~\eqref{eq:gamma_definition}. We will explore this regime in Sec.~\ref{sec::results}. In the experiment of Urgell \emph{et al.}~\cite{Urgell2020NatPhys}, the resonator has $f_0\simeq 90.6\,{\rm MHz}$ and $Q_m=6.8\times10^6$, corresponding to $\Gamma_{\rm ph}/\omega_0 \sim 10^{-7}$.  The experiment observes both cooling and self-oscillation under a constant electric bias, with phase-space distributions described as a mixture of a ring-like state and a central thermal peak near the switching regime. This is qualitatively consistent with the distinction made in this paper between vibrational excitation, high vibrational occupation fluctuations, and finite-$T_Q$ self-oscillatory NESS. We show the contribution of the mechanical quality factor in facilitating self-sustained oscillations in Fig.~\ref{plot::environment}(c).

While the present model is not a device-specific simulation of a particular experiment, it identifies the minimal conditions under which transport-induced backaction produces a limit cycle in the parameter regimes accessed by these devices. All of these regimes often require different approaches and distinct simulation strategies \cite{Culhane2022PRE, Piovano2011PRB, sevitz2025quantum}, whereas our methodology provides a unified framework capable of representing a majority of these parameter regimes with controlled accuracy using a powerful tensor network methodology. In Sec.~\ref{sec::results}, we perform a systematic analysis of these dynamical regimes as functions of the pertinent system and environmental parameters such as electromechanical coupling $\lambda$, mechanical frequency $\omega_0$, lead bandwidth $\delta$, lead temperature $T$, and phonon damping $\Gamma_{\rm ph}$, to identify the criteria for stable self-sustained oscillations in a NEMS model.

\section{A tensor-network framework for NEMS transport with emergent limit cycles} \label{sec::methods}

A central challenge in describing our transport-driven NEMS model \,[Eqs.~\eqref{eq::Htot}-\eqref{eq::HI}] is to obtain the NESS in parameter regimes where the following difficult features coexist:
\begin{itemize}
    \item[(i)] \emph{highly excited} vibrational states, which naturally arise in the limit-cycle regime \cite{armour2004classical, Blencowe2005NJP,a2005quantum, Fedorets2004PRL, Pistolesi2004PRB, PootVanDerZant2012PhysRep},
    \item[(ii)] \emph{structured} fermionic environments with finite reservoir correlation times and memory \cite{Schwarz2016PRB, sevitz2025quantum, luo2013non, ribeiro2015non},
    \item[(iii)] \emph{strong local dot-phonon coupling} between the quantum dot and the vibrational mode \cite{sapmaz2006tunneling,Vigneau2022PRR}. 
\end{itemize}
These ingredients are precisely those that make electromechanical feedback capable of producing transport instabilities and coherent self-oscillations in NEMS \cite{Wen2020NatPhys, Urgell2020NatPhys, blanter2004single, a2005quantum}. They also make the problem difficult for approaches that rely on small phonon cutoffs, weak coupling, wide-band reservoirs, or a clear separation of electronic and mechanical time scales.

We address this problem using a unified tensor-network construction built from three controlled ingredients:
\begin{itemize}
    \item[(i)] a \emph{binary encoding} of the truncated phonon Hilbert space, allowing large vibrational cutoffs to be represented with logarithmically many pseudosites \cite{JeckelmannWhite1998PRB},
    \item[(ii)] a \emph{mesoscopic-lead embedding}, which maps structured fermionic reservoirs onto damped auxiliary modes. This yields an augmented Markovian open-system dynamics \cite{Schwarz2016PRB, Brenes2020},
    \item[(iii)] a \emph{variational Liouville-space algorithm} that directly targets the NESS within this augmented Hilbert space \cite{JeckelmannWhite1998PRB,  Verstraete2004PRL, CuiCiracBanuls2015PRL}.
\end{itemize}
The resulting framework [see Fig.~\ref{fig::schematic}(b)] enables direct and computationally efficient access to transport observables, phonon-number statistics, and the vibrational nonequilibrium steady state (NESS). In particular, it enables the identification and analysis of dynamical phenomena such as limit cycles, whose characterization critically relies on detailed information about the state of the phonon mode. Below, we briefly summarize these three central components of our methodology. While each of these elements is essential in its own right, their combination renders the NESS accessible beyond the level of transport cumulants and conventional perturbative approaches.
    

\subsection{Binary encoding of the vibrational mode}
\label{subsec::binary}

A single phonon mode spans an infinite-dimensional Hilbert space. Except in special, exactly solvable limits, numerical calculations require truncating this space to a finite Fock basis. Under nonequilibrium transport conditions, this cutoff may need to be large, since the vibrational mode can be driven far from equilibrium and develop broad, non-thermal phonon-number distributions. A converged phonon-resolved calculation, therefore, requires an occupation cutoff large enough to contain the support of the steady-state distribution, rather than only the thermal occupation scale.

As discussed in Sec.~\ref{subsec::L^+L}, we use the density matrix renormalization group (DMRG) \cite{White1992DMRG} method for variationally targeting the NESS directly. However, applying DMRG can be challenging for bosonic systems with a large number of states per site. As the computational cost of DMRG scales linearly with the number of lattice sites, with other parameters held fixed, DMRG typically performs better when the individual sites have the fewest possible states, i.e., the number of states required per block to achieve a desired accuracy. Therefore, directly treating the oscillator as one bosonic site of local dimension \(M\) is increasingly costly as it grows. Given the truncated Fock space
\begin{equation}
    \{|m\rangle\}_{m=0}^{M-1},
    \qquad
    M=2^N,
    \label{eq:boson_basis}
\end{equation}
the binary pseudosite encoding method introduced by Jeckelmann and White \cite{JeckelmannWhite1998PRB} allows us to exactly transform a boson site into several 2-dimensional pseudosites. Given that $k$ is the number of the most important states for forming the ground state in each block, the DMRG targets $2k$ number of states in each of the $M$ blocks instead of treating $kM$ at once, making it significantly more efficient and accurate. 

The integer \(m\) is represented by \(N\) hard-core-boson occupations \(r_j\in\{0,1\}\),
\begin{equation}
    |m\rangle
    \longleftrightarrow
    |r_1 r_2 \dots r_N\rangle,
    \qquad
    m=\sum_{j=1}^{N}2^{j-1}r_j .
    \label{eq:binary_map}
\end{equation}
Each pseudosite is a two-level system with hard-core-boson operators \(\hat a_j,\hat a_j^\dagger\) satisfying
\begin{equation}
    \hat a_j^2=(\hat a_j^\dagger)^2=0,
    \qquad
    \hat a_j\hat a_j^\dagger+\hat a_j^\dagger\hat a_j=1 ,
\end{equation}
while operators on different pseudosites commute. Following \cite{JeckelmannWhite1998PRB}, one may write the truncated creation operator as
\begin{equation}
    \hat b^\dagger
    =
    \hat B^\dagger
    \sqrt{\hat n_{\rm ph}+1},
    \label{eq:bdag_factorized}
\end{equation}
where \(\hat B^\dagger|m\rangle=|m+1\rangle\) for \(m<M-1\), while
\(\hat B^\dagger|M-1\rangle=0\). We have
\begin{equation}
    \sqrt{\hat n_{\rm ph}+1}
    =
    \sum_{m=0}^{M-1}
    \sqrt{m+1}
    \prod_{j=1}^{N} P_j\!\left[r_j(m)\right],
    \label{eq:sqrt_number_pseudo}
\end{equation}
where \(r_j(m)\) is the \(j\)th binary digit of \(m\) [Eq.~\eqref{eq:binary_map}], $P_j(1)=\hat a_j^\dagger\hat a_j$, $P_j(0)=\hat a_j\hat a_j^\dagger$, and
\begin{equation}
    \hat B^\dagger
    =
    \sum_{\ell=1}^{N}
    \hat a_\ell^\dagger
    \prod_{j=1}^{\ell-1}\hat a_j .
    \label{eq:binary_increment}
\end{equation} In this basis, the truncated phonon number operator is diagonal,
\begin{equation}
    \hat b^\dagger \hat b
    \equiv
    \hat n_{\rm ph}
    =
    \sum_{j=1}^{N}2^{j-1}\hat n_j,
    \qquad
    \hat n_j=\hat a_j^\dagger\hat a_j .
    \label{eq:number_pseudo}
\end{equation}
Due to this transformation, the ladder operators ($\hat b$, $\hat b^\dagger$) become non-local over the pseudosites. We discuss how to efficiently handle these nonlocal operators in Sec.~\ref{ape:numerical_strategies}. Eqs.~\eqref{eq:bdag_factorized}-\eqref{eq:binary_increment} give an \textit{exact} representation of \(\hat b^\dagger\) and $\hat b$ (by Hermitian conjugation) within the truncated Fock space. The transformation scheme discussed above maps the AH model shown in Fig.~\ref{fig::schematic}(b1) to Fig.~\ref{fig::schematic}(b2).
\subsection{Mesoscopic-leads embedding of the macroscopic reservoirs}
\label{subsec::mlead}

The fermionic leads represent macroscopic reservoirs with temperatures $T_\alpha$, chemical potentials $\mu_\alpha$, and structured spectral density functions $J_\alpha(\omega)$, where $\alpha\in\{L,R\}$. We replace each fermionic lead by a finite set of damped auxiliary fermionic modes $\{\hat a_{k\alpha}\}_{k=1}^{L_\alpha}$, namely the \textit{mesoscopic leads} \cite{Imamoglu1994, dalton2001theory, subotnik2009nonequilibrium,Dzhioev2011,  ajisaka2012nonequilibrium, Arrigoni2013PRL, Dorda2014PRB, Schwarz2016PRB, elenewski2017master, Brenes2020, lacerda2024entropy,alford2025subtleties}. This mesoscopic-lead or pseudomode embedding is chosen so that the augmented system reproduces the target reservoir correlation functions in a controlled manner, mapping the original dot-reservoir model further to Fig.~\ref{fig::schematic}(b3). The exact dynamics of the original system may be non-Markovian, but the enlarged system consisting of the dot, phonon mode, and auxiliary lead modes evolves under a Markovian master equation.

For each lead $\alpha$, we introduce $L_\alpha$ auxiliary fermionic modes $\hat c_{l\alpha}$ with energies $\varepsilon_{l\alpha}$ and dot-auxiliary couplings $\kappa_{l\alpha}$. The augmented total Hamiltonian can be written as, 
\begin{multline}
    \hat{H}_{\rm aug} = \hat{H}_S + \sum_{\alpha=L,R} \sum_{l=1}^{L_\alpha} \varepsilon_{l\alpha} \hat c_{l\alpha}^\dagger \hat c_{l\alpha}\\ + \sum_{\alpha=L,R} \sum_{l=1}^{L_\alpha} \left( \kappa_{l\alpha}\hat d^\dagger \hat c_{l\alpha} + {\rm H.c.} \right),\label{eq:Haug_meso}
\end{multline}
where $H_S$ is the system Hamiltonian defined in Sec.~\ref{sec:model}. Each auxiliary mode is thermalized by an independent Markovian residual bath that drives it toward the Fermi occupation
\begin{equation}
    f_\alpha(\varepsilon_{l\alpha}) = \frac{1}{e^{(\varepsilon_{l\alpha}-\mu_\alpha)/T_\alpha}+1}.
    \label{eq:fermi_aux}
\end{equation}
In the mesoscopic-lead embedding, the dynamics of the augmented density matrix $\rho$ is described by the Markovian master equation
\begin{equation}
\partial_t\rho = \mathcal{L}[\rho] = -i[\hat{H}_{\rm aug},\rho] + \sum_{\alpha=L,R} \sum_{l=1}^{L_\alpha} \mathcal{D}_{l\alpha}[\rho] +
\mathcal{D}_{\rm ph}[\rho], \label{eq:ME_aug}
\end{equation}
where the fermionic dissipator acting on the auxiliary mode $k_\alpha$ has the form,
\begin{align}
\mathcal{D}_{l\alpha}[\rho] &= \gamma_{l\alpha} \left[ 1-f_\alpha(\varepsilon_{l\alpha}) \right] \left( \hat c_{l\alpha}\rho \hat c_{l\alpha}^\dagger - \frac{1}{2} \{ \hat c_{l\alpha}^\dagger \hat c_{l\alpha},\rho \} \right) \nonumber\\ &\quad + \gamma_{l\alpha}
f_\alpha(\varepsilon_{l\alpha}) \left( \hat c_{l\alpha}^\dagger\rho \hat c_{l\alpha} - \frac{1}{2}\{ \hat c_{l\alpha}\hat c_{l\alpha}^\dagger,\rho \} \right). \label{eq:Dlead_meso}
\end{align}
Here $\gamma_{l\alpha}$ is the damping rate of the auxiliary mode into its residual Markovian bath. The coherent dot-auxiliary coupling in $H_{\rm aug}$ retains the finite reservoir correlation time of the original structured bath. However, the residual baths enforce the desired occupations of the auxiliary modes \cite{Brenes2020, brenes2023particle}. The reduced state of the dot--phonon subsystem is obtained by tracing out the mesoscopic lead modes, $\rho_S(t) = {\rm Tr}_{\rm leads}[\rho(t)]$.

The mesoscopic embedding produces the effective hybridization function
\begin{equation}
J_{\alpha}^{\rm eff}(\omega) = \sum_{l=1}^{L_\alpha} \frac{ |\kappa_{l\alpha}|^2\,\gamma_{l\alpha} }{ (\omega -\varepsilon_{l\alpha})^2 + (\gamma_{l\alpha}/2)^2}. \label{eq:Jeff}
\end{equation}
Thus, for uncoupled auxiliary modes, $J_{\alpha}^{\rm eff}(\omega)$ is a sum of Lorentzians peaked at $\varepsilon_{l\alpha}$, with widths set by $\gamma_{l\alpha}$. Each Lorentzian is generated by one damped auxiliary mode in Eq.~\eqref{eq:Dlead_meso}. More structured reservoirs can be represented by adding more auxiliary modes, by optimizing the positions, widths, and couplings of each Lorentzian component, or by allowing more general pseudomode geometries with coupled auxiliary modes \cite{alford2025subtleties}. In such cases, the quality of the embedding must be checked by the convergence of $J_\alpha^{\rm eff}(\omega)$ and of the finite-temperature correlation functions over the transport-relevant energy window. For a Lorentzian spectral density of the fermionic leads in the form of Eq.~\eqref{eq:spectral_densities}, each fermionic lead can be represented by a single ($L_\alpha=1$) damped auxiliary mode at energy $\varepsilon_{\alpha}=\omega_\alpha$. Matching Eq.~\eqref{eq:Jeff} to Eq.~\eqref{eq:spectral_densities} gives
\begin{equation}
    \gamma_\alpha=2\delta_\alpha, \qquad
    \kappa_\alpha = \sqrt{\frac{\Gamma_\alpha\delta_\alpha}{2}} .
    \label{eq:lorentzian_matching}
\end{equation}

Lastly, to model a finite quality factor of the mechanical oscillator, we include a weak thermal Lindblad channel for the phonon mode,
\begin{align}
\mathcal{D}_{\rm ph}[\rho] &= \Gamma_{\rm ph} (\bar n_{\rm ph}+1)
\left( \hat b\rho \hat b^\dagger - \frac{1}{2} \{ \hat b^\dagger \hat b,\rho \} \right) \nonumber\\
&\quad + \Gamma_{\rm ph} \bar n_{\rm ph} \left( \hat b^\dagger\rho\hat b - \frac{1}{2} \{ \hat b \hat b^\dagger,\rho \}\right), \label{eq:Dph}
\end{align}
with
\begin{equation}
    \bar n_{\rm ph} = \frac{1}{e^{\omega_0/T_{\rm ph}}-1}.
\end{equation}
This is the standard weak-coupling thermalization model for a harmonic mode. It provides a minimal description of intrinsic mechanical damping in regimes with or without net electromechanical backaction.

\subsection{Superfermion representation} \label{subsec::superf}

\begin{figure}[!tbp]
\centering
\includegraphics[width=0.95\columnwidth, trim=0.cm 0.0cm 0.0cm 0.cm]{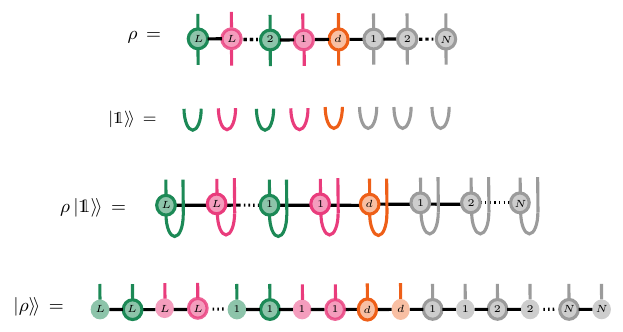} \vspace{.3cm}

\parbox{\linewidth}{\caption{\justifying  Liouville-space vectorization of the density matrix $\rho$ is initially written as an MPO over the mesoscopic lead sites, dot, and phonon pseudosites (top row). Acting on the left vacuum $|\mathds{1}\rangle\!\rangle$ (second row) gives the vectorized state $|\rho\rangle\!\rangle=\rho|\mathds{1}\rangle\!\rangle$ (third row), represented as an MPS in the doubled space. In the bottom row, pink and green sites denote left and right mesoscopic lead sites, orange denotes the dot, and gray denotes phonon pseudosites; sites with(-out) boundaries indicate the physical (tilde) copies. \label{fig::tn_vec}}}
\end{figure}

To solve the Lindblad equation $\partial_t\rho = \mathcal{L}[\rho]$ within a tensor-network framework, we work in the doubled Liouville space $\mathcal{H}_{\rm phys}\otimes\mathcal{H}_{\rm anc}$ and represent $\rho$ as a vector $|\rho\rangle\!\rangle$~\cite{Verstraete2004PRL, ZwolakVidal2004PRL, Prosen2008NJP, CuiCiracBanuls2015PRL}. We use the superfermion representation ~\cite{Dzhioev2011, Brenes2020}, which introduces a tilde copy of each physical fermionic mode and uses conjugation rules to rewrite right multiplication as an operation on the auxiliary copy. This maps the master equation to a linear evolution equation in a doubled space. A key property of this formalism is that the Lindblad dissipator for mode $j$ involves only operators $\hat f_j$ and $\hat{\tilde f}_j$ of the same mode. Thus, the dissipator is local in operator space independently of the one-dimensional chain ordering. When this superfermion structure is embedded into a 1D chain, we additionally choose an interleaved ordering of physical and tilde modes (details in Appendix~\ref{subsec:geometry_aware_ordering}), placing each physical mode immediately adjacent to its tilde partner \cite{Brenes2020}. The dissipators then become on-site or nearest-neighbor terms in the chain, avoiding the long Jordan--Wigner strings that would appear if all physical modes were placed before all auxiliary modes. The augmented fermionic chain of our model contains $M_f = 2L+1$ spinless physical modes, comprised of the quantum dot and $L$ mesoscopic lead modes for each of the left and right reservoirs. Without loss of generality and to simplify the notation, we consider a single auxiliary lead for each bath, so that $L=1$ and $M_f=3$. We use the unified notation
\begin{equation}
    \{\hat f_j\}_{j=1}^{3} \equiv \{\hat c_L,\hat d,\hat c_R\}.
\end{equation}

We then double the fermionic Hilbert space by introducing one tilde mode $\hat{\tilde f}_j$ for each physical mode $\hat f_j$. The tilde modes satisfy canonical anticommutation relations, and physical and tilde fermions anticommute with each other. Following Ref.~\cite{Brenes2020}, we adopt the interleaved ordering
\begin{equation}
|\mathbf{n}|\mathbf{m}\rangle_f = \bigl(\hat{f}_1^\dagger\bigr)^{n_1}
  \bigl(\hat{\tilde{f}}_1^\dagger\bigr)^{m_1} \bigl(\hat{f}_2^\dagger\bigr)^{n_2}
  \bigl(\hat{\tilde{f}}_2^\dagger\bigr)^{m_2}
  \bigl(\hat{f}_3^\dagger\bigr)^{n_3}
  \bigl(\hat{\tilde{f}}_3^\dagger\bigr)^{m_3}
  |\mathrm{vac}\rangle_f ,
\label{eq:interleaved_basis}
\end{equation}
with $n_j,m_j\in\{0,1\}$. As shown in Fig.~\ref{fig::tn_vec}, the physical $j$th mode and its tilde partner occupy adjacent sites in the 1D chain. The fermionic trace state, or \textit{left vacuum}, is
\begin{equation}
|\mathds{1}\rangle\!\rangle_f = \sum_{\mathbf{n}\in\{0,1\}^{M_f}}
|\mathbf{n}|\mathbf{n}\rangle_f , \label{eq:left_vac_f}
\end{equation}
an equal-weight superposition of states in which the tilde copy mirrors the physical copy. Together with the bosonic trace state
\begin{equation}
|\mathds{1}\rangle\!\rangle_b = \sum_{\mathbf{r}\in\{0,1\}^{N}} |\mathbf{r}|\mathbf{r}\rangle_b ,
\end{equation}
where the sum runs over the $2^N$ binary-encoded phonon basis states, the total trace state is
\begin{equation}
|\mathds{1}\rangle\!\rangle = |\mathds{1}\rangle\!\rangle_f
\otimes |\mathds{1}\rangle\!\rangle_b.
\end{equation}
The vectorized density matrix is then
\begin{equation}
|\rho(t)\rangle\!\rangle = \rho(t)|\mathds{1}\rangle\!\rangle ,
\end{equation}
and expectation values are evaluated as
\begin{equation}
\langle\hat O(t)\rangle = \frac{ \langle\!\langle\mathds{1}|\hat O
|\rho(t)\rangle\!\rangle}{\langle\!\langle\mathds{1}|\rho(t)\rangle\!\rangle}.
\label{eq:vectorization}
\end{equation}

Here $\hat O$ acts on the physical copy unless stated otherwise. Acting on the left vacuum state with a physical creation or annihilation operator can be equivalently represented for an operation on the tilde copy. For the interleaved ordering of Eq.~\eqref{eq:interleaved_basis}, one obtains the \textit{conjugation rules}~\cite{Dzhioev2011,Brenes2020}
\begin{equation}
\hat{f}_j^\dagger|\mathds{1}\rangle\!\rangle = -\hat{\tilde{f}}_j|\mathds{1}\rangle\!\rangle, \qquad \hat{f}_j|\mathds{1}\rangle\!\rangle = \hat{\tilde{f}}_j^\dagger|\mathds{1}\rangle\!\rangle .
\label{eq:conjugation_f}
\end{equation}
The corresponding conjugate relations are
\begin{equation}
\langle\!\langle\mathds{1}|\hat f_j = -\langle\!\langle\mathds{1}|\hat{\tilde f}_j^\dagger,
\qquad \langle\!\langle\mathds{1}|\hat f_j^\dagger = \langle\!\langle\mathds{1}|\hat{\tilde f}_j .
\end{equation}
The original Dzhioev--Kosov formulation~\cite{Dzhioev2011} uses a different (but algebraically equivalent) phase convention compared to the one we follow~\cite{Brenes2020}. The conjugation rules for the hardcore bosonic pseudosites, denoted here by $\hat a_j$, are
\begin{equation}
\hat a_j^\dagger|\mathds{1}\rangle\!\rangle = \hat{\tilde a}_j|\mathds{1}\rangle\!\rangle,
\qquad \hat a_j|\mathds{1}\rangle\!\rangle = \hat{\tilde a}_j^\dagger|\mathds{1}\rangle\!\rangle .
\label{eq:conjugation_b}
\end{equation}
Acting $\partial_t\rho=\mathcal{L}[\rho]$ on
$|\mathds{1}\rangle\!\rangle$ and applying
Eqs.~\eqref{eq:conjugation_f}-\eqref{eq:conjugation_b} converts the master equation into
\begin{equation}
\partial_t|\rho(t)\rangle\!\rangle = \hat{\mathcal L} |\rho(t)\rangle\!\rangle ,
\label{eq:schrodinger_liouville}
\end{equation}
where $\hat{\mathcal L}$ is the Liouvillian superoperator in the doubled space.

\subsection{Variational solution for the NESS} \label{subsec::L^+L}

Rather than relying on long-time propagation of the master equation in Eq.~\eqref{eq:ME_aug}, which can become inefficient when the Liouvillian gap is small and relaxation is slow, we target the nonequilibrium steady state $\rho_{\rm ss}$ directly as a vector in Liouville space, defined by
\begin{equation}
\mathcal{L}|\rho_{\rm ss}\rangle\!\rangle=0.    \label{eq:ness_condition}
\end{equation}
We cast this as a variational problem \cite{Mascarenhas2015, CuiCiracBanuls2015PRL}
\begin{equation}
\min_{|\rho\rangle\!\rangle}\ \|\mathcal{L}|\rho\rangle\!\rangle\|^2,
\label{eq:LL_objective}
\end{equation}
and the NESS $|\rho_{\rm ss}\rangle\!\rangle$ can be obtained by minimizing the positive semidefinite functional
\begin{equation}
    E_{\mathcal A}[\rho] = \langle\!\langle \rho|\mathcal{A}|\rho\rangle\!\rangle =    \|\mathcal{L}|\rho\rangle\!\rangle\|^2 ,\qquad  \mathcal{A}=\mathcal{L}^{\dagger}\mathcal{L}.
    \label{eq:variational_energy}
\end{equation}
for a normalized trial state $|\rho\rangle\!\rangle$ [see Fig.~\ref{fig::schematic}(b4)]. 

Since $\mathcal{A}$ is Hermitian and positive semidefinite, its ground-state manifold coincides with the kernel of $\mathcal{L}$, and in the generic case of a unique steady state, this ground state is nondegenerate, and minimizing Eq.~\eqref{eq:LL_objective} yields $|\rho_{\rm ss}\rangle\!\rangle$ up to normalization. We solve Eq.~\eqref{eq:LL_objective} using standard DMRG sweeps \cite{Mascarenhas2015, CuiCiracBanuls2015PRL, ZwolakVidal2004PRL, Schollwoeck2011}. Convergence to the desired NESS is certified by a small value for the energy functional $ E_{\mathcal A}[\rho]$ together with the stability of independent target observables under DMRG sweeps, as we describe in the next section, Sec.~\ref{subsec:tn_implementation}. 

\subsection{Tensor-network implementation}
\label{subsec:tn_implementation}

The variational NESS condition in Eq.~\eqref{eq:variational_energy} is implemented as a Liouville-space MPS optimization.  The numerical procedure requires choices of site ordering, initialization, and convergence diagnostics. The full algorithmic details are provided in the Appendix \ref{ape:numerical_strategies}. Here, we summarize the main implementation choices used in the calculations.

The MPS ordering is chosen to keep the dominant couplings short-ranged.\,Since the efficiency of an MPS representation depends sensitively on the entanglement growth across the chosen one-dimensional ordering \cite{Schollwoeck2005RMP, Schollwoeck2011, Wolf2014PRB, HeMillis2017PRB}, each physical fermionic mode is placed adjacent to its tilde partner. In the superfermion representation, the reservoir dissipators then act locally on physical--tilde pairs \cite{Dzhioev2011, Brenes2020}, avoiding unnecessary long-range Jordan--Wigner strings from the thermalization terms. The binary pseudosites representing the oscillator are grouped into a compact physical--tilde block that confines the MPO representation of $\hat{b}$ and $\hat b^{\dagger}$ to a contiguous region and keeps the dot--phonon coupling short-ranged in the MPS ordering. The dot pair is placed between the oscillator block and the auxiliary lead modes, so that both dot--phonon and dot--lead couplings are represented by short MPO strings. This follows the standard principle used in tensor-network treatments of impurity problems, where site orderings are chosen to reduce the bond dimensions required to represent the dominant correlations~\cite{Wolf2014PRB, KohnSantoro2021PRB, rams2020breaking}.

To improve stability in parameter regimes with slow relaxation, we use a continuation strategy in parameter space. Direct minimization of Eq.~\eqref{eq:variational_energy} can be sensitive to the initial MPS when the phonon distribution is broad or when the Liouvillian relaxation gap is small.  We therefore use a continuation strategy whereby we first converge on a nearby parameter point and then consecutively use the resulting MPS as the initial state for the next point, while simultaneously ramping the phonon-bath parameters $\Gamma_{\rm ph}$ and $T_{\rm ph}$ through intermediate values before reaching the target parameter values.  Related continuation strategies are commonly used to stabilize variational searches for nonequilibrium steady states \cite{CuiCiracBanuls2015PRL, Mascarenhas2015}. At each step, two-site DMRG sweeps \cite{White1992DMRG, White1993DMRG} are used to minimize the variational energy. The Krylov subspace dimension and the MPS bond dimension are increased when the discarded weight or the stability of observables indicates that the current approximation is insufficient. At the final target point, any auxiliary density-matrix perturbation or subspace-expansion term introduced during the optimization is gradually removed, following the logic of stabilized
single-site and subspace-expansion methods \cite{White2005SingleSiteDMRG, Hubig2015SubspaceExpansion}. Convergence is assessed using both numerical and physical diagnostics. We monitor the Liouvillian residual
\begin{equation}
    r = \frac{\|\mathcal{L}|\rho\rangle\!\rangle\|}{\| |\rho\rangle\!\rangle\|},
    \label{eq:residual}
\end{equation} 
but do not use it as the sole criterion, since finite bond dimension and MPO compression can leave observables insufficiently converged even when the residual is small. We therefore also verify current balance, $I_L\simeq J$, where $I_L$ is the particle current injected by the Markovian relaxation of the left mesoscopic lead mode, while $J$ is the coherent tunneling current across the left lead-mode--dot interface. We also check the stability of $\langle n_{\rm ph}\rangle$, ${\rm Var}(n_{\rm ph})$, $F_{\rm ph}$, and $g^{(2)}(0)$ under multiple additional DMRG sweeps, moderate increases of bond dimensions, larger oscillator Hilbert-space cutoff $M$, and tighter MPO compression.  If the vibrational steady state distribution has appreciable weight near the largest retained Fock state, the calculation is repeated with larger $M$. This truncation check is particularly important near the onset of self-oscillation, where broad oscillator distributions are most susceptible to truncation artifacts. The convergence properties of relevant transport, steady state, and fluctuation observables with respect to the vibrational Hilbert-space cutoff and the maximum available bond dimensions of the steady state are studied in detail in Sec.~\ref{sec::convergence}.

\section{Capturing limit cycle NESS in the autonomous NEMS model} \label{sec::results}
The AH model, being a fundamental model for describing NEMS transport, exhibits a range of nonequilibrium vibrational phenomena, including incoherent heating and transport-induced self-sustained oscillations. Identifying the emergence of such oscillations remains particularly difficult, especially when quantitatively comparing distinct dynamical and transport regimes. In this section, we characterize the steady-state vibrational NESS using complementary diagnostics that identify these transport phenomena. Specifically, we combine torotropy $T_Q$, recently introduced in Ref.~\cite{sevitz2026autonomous} and the Husimi-$Q$ function (see Fig. \ref{plot::Husimis}) ~\cite{schleich2015quantum}, which directly probes annular phase-space structure, with the zero-delay second-order coherence $g^{(2)}(0)$ and the phonon-number Fano factor $F_{\rm ph}$, which probe number fluctuations. By scanning the dimensionless electromechanical coupling $\lambda$, mechanical frequency $\omega_{0}$, electronic relaxation rate $\Gamma$, lead spectral width $\delta$, lead temperature $T$, and phonon damping strength $\Gamma_{\rm ph}$, we identify the regimes where transport backaction gives rise to a self-oscillatory NESS. Table~\ref{tab::results} summarizes the main conclusions before the detailed parameter scans.

\subsection{Diagnosing phase-space structure and vibrational fluctuations} \label{subsec::phase-space struct}
\begin{table*}[!tbp]
\centering
\renewcommand{\arraystretch}{1.16}
\setlength{\tabcolsep}{5pt}
\caption{\textbf{Summary of key findings.}}
\label{tab::results}
\begin{tabular}{lll}
\hline
\parbox[t]{0.22\textwidth}{\textbf{Observation}} &\parbox[t]{0.38\textwidth}{\textbf{Evidence}} & \parbox[t]{0.34\textwidth}{\textbf{Inference}} \\
\hline
 \vspace{-.2cm} \\
\parbox[t]{0.22\textwidth}{\raggedright Adiabaticity controls self-sustained oscillations}
&\parbox[t]{0.38\textwidth}{\raggedright Finite \(T_Q\) appears only in an intermediate region of the \((\lambda,\omega_0)\) plane; increasing \(\Gamma\) shifts this region toward larger \(\omega_0\) [Fig.~\ref{plot::TQ_lambda_Omega0}].}
&\parbox[t]{0.34\textwidth}{\raggedright Self-oscillation requires both sufficient electromechanical backaction and matched electronic and mechanical timescales.}
\\ \hfill \vspace{-.2cm} \\
\parbox[t]{0.22\textwidth}{\raggedright High mechanical occupation fluctuations precedes self-sustained oscillations}
&\parbox[t]{0.38\textwidth}{ \raggedright While increasing $\lambda$, the phonon-number Fano factor peaks just before finite $T_Q$ emerges for both slow and fast vibrational regimes  [Fig.~\ref{plot::transport_vs_lambda}(a, c)].}
&\parbox[t]{0.34\textwidth}{\raggedright Transport can broaden the phonon distribution without yet stabilizing a self-oscillating steady state.}
\\ \hfill \vspace{-.2cm} \\
\parbox[t]{0.22\textwidth}{\raggedright Heating does not imply self-sustained oscillations}
&\parbox[t]{0.38\textwidth}{\raggedright While increasing $\omega_0$, large $\langle n_{\rm ph}\rangle$ can occur with negligible $T_Q$ whereas $T_Q$ can increase while $J$ and $\zeta$ decrease, and  [Fig.~\ref{plot::transport_vs_lambda}(b)].}
&\parbox[t]{0.34\textwidth}{\raggedright Energy pumping into the vibrational mode is necessary but not sufficient for self-oscillations.}
\\ \hfill \vspace{-.2cm} \\
\parbox[t]{0.22\textwidth}{\raggedright Environment parameters reshape the self-oscillation window}
&\parbox[t]{0.38\textwidth}{\raggedright $\Gamma$, $\delta$, $T$, and $\Gamma_{\rm ph}$ affect $T_Q$ non-trivially. [Fig.~\ref{plot::environment}].}
&\parbox[t]{0.34\textwidth}{\raggedright Lead memory, thermal fluctuations, and bosonic dissipation act as control parameters.}
\\ \hfill \vspace{-.3cm} \\

\hline
\end{tabular}
\end{table*}
\begin{table*}[!tbp]
\centering
\caption{\textbf{Parameters explored in the figures (unless stated otherwise)}}
\label{tab::param}
\renewcommand{\arraystretch}{1.25}
\setlength{\tabcolsep}{2pt}
\scriptsize

\resizebox{\textwidth}{!}{

\begin{tabular}{@{}cccccccccccc@{}}
\toprule 
Fig. &  $\omega_{0}$ & $\lambda$ & $\Gamma$ & $T$ & $\delta$ &
$\omega_L$ & $\omega_R$ & $\mu_{L}$ & $\mu_{R}$ &
$T_{\mathrm{ph}}$ & $\Gamma_{\mathrm{ph}}$ \\
\midrule
\ref{plot::Husimis} &
\const{1} & \varp{$[0.01, 0.5, 1]$} & \const{1} & \const{1} & \const{1} & \const{1} & \const{$-1$} & \const{5} & \const{$-5$} & \varp{0.1} & \varp{0.01} \\
\ref{plot::TQ_lambda_Omega0} &
\varp{$[0.5-1.4]$} & \varp{$[0.5-1.4]$} &
\varp{$[0.5, 1, 1.5, 2]$} & \const{1} & \const{1} & \const{1} & \const{$-1$} & \const{5} & \const{$-5$} & \const{0.01} & \const{0.001} \\
\ref{plot::transport_vs_lambda}(a) &
 \const{1} & \varp{$[0.01-0.8]$} & \const{1} & \const{1} & \const{1} & \const{1} & \const{$-1$} & \const{5} & \const{$-5$} & \const{0.01} & \const{0.001} \\
\ref{plot::transport_vs_lambda}(b) &
 \varp{$[0.3-1.53]$} & \const{1} & \varp{1.5} & \const{1} & \const{1} & \const{1} & \const{$-1$} & \const{5} & \const{$-5$} & \const{0.01} & \const{0.001} \\
\ref{plot::transport_vs_lambda}(c) &
 \varp{$[0.66, 1, 1.5]$} & \varp{$[0.01-1.15]$} & \const{1} & \const{1} & \const{1} & \const{1} & \const{$-1$} & \const{5} & \const{$-5$} & \const{0.01} & \const{0.001} \\
\ref{plot::environment}(a) &
\varp{$[0.5-1.4]$} & \varp{$[0.5-1.4]$} & \const{1} & \const{1} & \varp{$[0.5, 1]$} & \const{1} & \const{$-1$} & \const{5} & \const{$-5$} & \const{0.01} & \const{0.001} \\ 
\ref{plot::environment}(b) &
\const{1} & \varp{$[0.01-1.5]$} & \const{1} & \varp{$[1, 5, 7, 10]$} & \const{1} & \const{1} & \const{$-1$} & \const{5} & \const{$-5$} & \const{0.01} & \const{0.001} \\
\ref{plot::environment}(c) &
\const{1} & \varp{$[0.01-1.4]$} & \const{1} & \const{1} & \const{1} & \const{1} & \const{$-1$} & \const{5} & \const{$-5$} & \const{0.01} & \varp{$[0.02-0.1]$} \\
\ref{plot::environment}(d) &
\const{1} & \const{1} & \const{1} & \const{1} & \const{1} & \const{1} & \const{$-1$} & \const{5} & \const{$-5$} & \const{0.01} & \varp{$[0.001-0.1]$} \\
\ref{plot::oisin_benchmark}(a) &
 \const{1} & \varp{0.5} & \const{1} & \const{1} & \varp{$[1-40]$} & \varp{0} & \varp{0} & \varp{10} & \varp{$-1000$} & \const{0.01} & \const{0.001} \\
\ref{plot::oisin_benchmark}(b) &
\const{1} & \varp{0.5} & \varp{5} & \varp{10} & \const{1} & \const{1} & \const{$-1$} & \varp{$[2.5, 50]$} & \varp{$[-2.5, -50]$} & \const{0.01} & \varp{$10^{-5}$} \\
\ref{plot::convergence_check} &
\const{1} & \varp{0.5} & \const{1} & \const{1} & \const{1} & \const{1} & \const{$-1$} & \const{5} & \const{$-5$} & \const{0.01} & \varp{$10^{-6}$} \\
\bottomrule
\end{tabular}}
\end{table*}

To characterize self-sustained oscillations, we employ the quantity ``torotropy'', \(T_Q\) , a phase-space-based measure of the emergence of a limit-cycle in the Husimi \(Q\)-function of the reduced vibrational steady state~\cite{sevitz2026autonomous}. \(T_Q\) is defined based on a rationale that is formally analogous to the one underlying the concept of ergotropy \cite{allahverdyan2004}, which quantifies the work extractable by cyclic unitary operations. Ergotropy can serve as an order parameter for both the phonon lasing transition and the emergence of a limit cycle. However, it can be misleading, for instance, in the quantum regime, by suggesting the presence of a limit cycle, since a non-zero ergotropy, i.e., a finite amount of extractable work, may originate from mechanisms other than limit-cycle dynamics. Torotropy quantifies the extent to which a phase-space distribution develops an annular, torus-like structure. In this work, the relevant phase-space distribution is the Husimi \(Q\)-function~\cite{schleich2015quantum} of the reduced phonon steady state,
\begin{equation}
    Q(\alpha) = \frac{1}{\pi} \langle \alpha | \rho_{\rm ph} |\alpha\rangle ,
    \qquad \int d^2\alpha\, Q(\alpha)=1 ,
    \label{eq:Husimi}
\end{equation}
where \(|\alpha\rangle\) is a coherent state and \(\rho_{\rm ph}={\rm Tr}_{d,{\rm leads}}[\rho_{\rm ss}]\) is the reduced steady state of the phonon mode. The barycenter of the distribution is
\begin{equation}
    \alpha_c = \int d^2\alpha\, \alpha Q(\alpha). 
    \label{eq:Q_barycenter}
\end{equation}
For each angle \(\varphi\in[0,2\pi)\), we consider the half-line
\begin{equation}
    a_\varphi = \left\{ \alpha_c+r e^{i\varphi}\,:\, r\geq 0 \right\},
    \label{eq:torotropy_ray}
\end{equation}
emanating from \(\alpha_c\).  The normalized restriction of \(Q\) to this ray is
\begin{equation}
    Q_\varphi(r) = \frac{ Q\!\left(\alpha_c+r e^{i\varphi}\right)}{Z_\varphi}, \qquad
    Z_\varphi = \int_0^\infty dr\, Q\!\left(\alpha_c+r e^{i\varphi}\right),
    \label{eq:Q_radial_profile}
\end{equation}
so that \(\int_0^\infty dr\,Q_\varphi(r)=1\).  We denote by \(Q_\varphi^\downarrow(r)\) the monotonically decreasing rearrangement of \(Q_\varphi(r)\), i.e., the radial profile obtained by ordering the same probability weight from largest to smallest as \(r\) increases. A centrally peaked radial profile along the ray therefore satisfies
\(Q_\varphi=Q_\varphi^\downarrow\). 

\begin{figure}[!tbp]
\centering
\includegraphics[width=\columnwidth,
  trim=0.5cm 0.0cm 0.0cm 0.5cm]
  {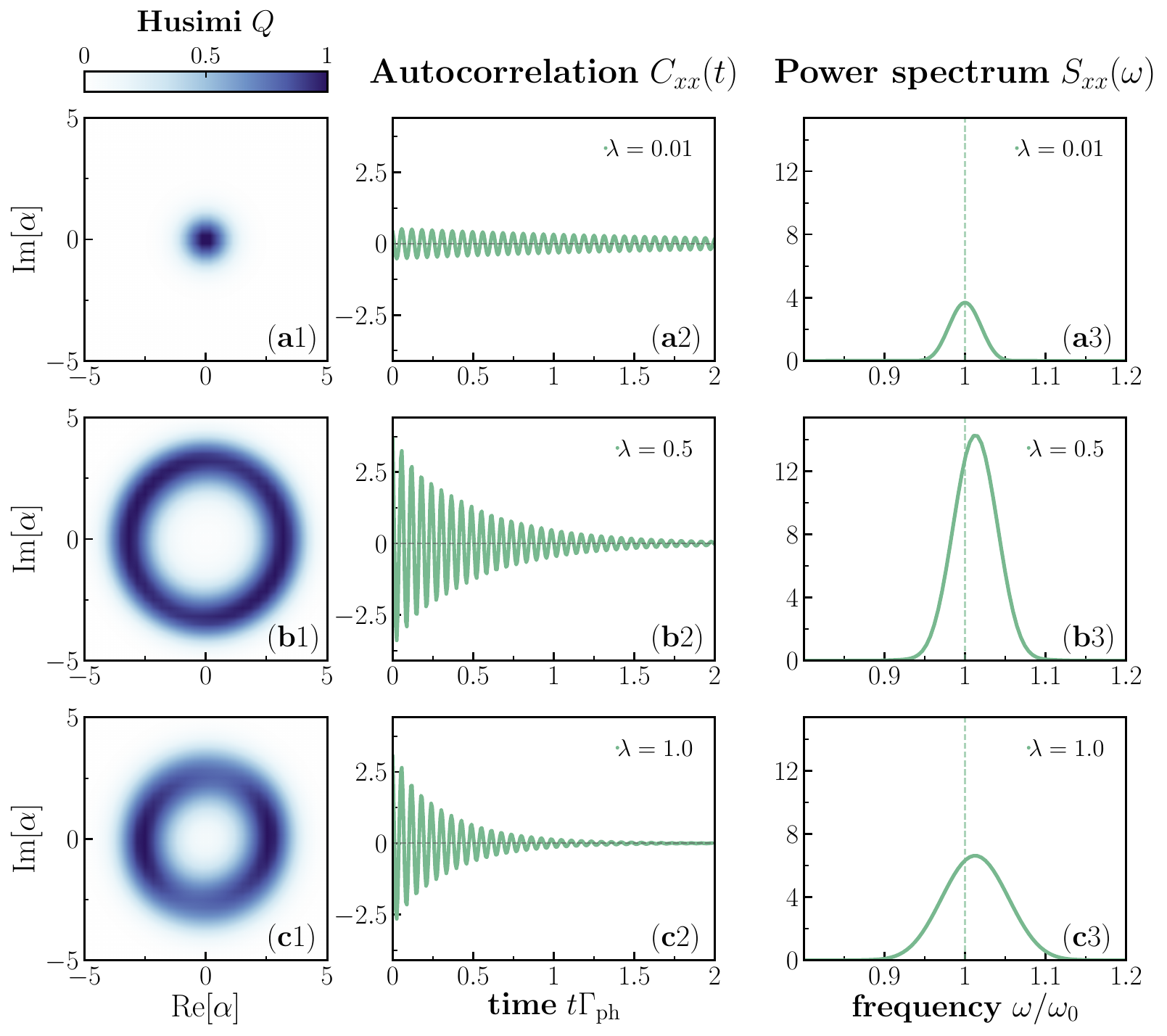}\newline\\
\parbox{\linewidth}{\caption{\justifying Signatures of self-sustained oscillations in the Husimi $Q$-distribution, position autocorrelation, and mechanical power spectrum for different dot--phonon couplings. \textit{Left column:} normalized Husimi $Q$-distribution in the $(x,p)$ phase plane with torotropy values $T_Q=0 \text{ (a1), } 1.4 \text{ (b1), } 0.7 \text{ (c1)} $. \textit{Middle column:} Position autocorrelation as a function of dimensionless time $t\Gamma_{\rm ph}$. \textit{Right column:} PSD and the dashed vertical line marks $\omega = \omega_0$. Note that in (b3, c3), the PSD peaks are slightly shifted from the vertical dashed line.  All other parameters are given in Table~\ref{tab::param}.
\label{plot::Husimis}}}
\end{figure}

The torotropy is then defined as
\begin{equation}
    T_Q = \min_{\varphi\in[0,2\pi)} \frac{1}{S_{Q_\varphi}}
    \int_0^\infty dr\, r\,  \left[Q_\varphi(r)-Q_\varphi^\downarrow(r)\right],
    \label{eq:torotropy}
\end{equation}
where
\begin{equation}
    S_{Q_\varphi} = -\int_0^\infty dr\, Q_\varphi(r)\ln Q_\varphi(r)
    \label{eq:torotropy_entropy}
\end{equation}
is the entropy of the normalized radial profile.  This entropy factor reduces the sensitivity of \(T_Q\) to weak non-monotonic features appearing only at large radius.  A finite \(T_Q\) therefore indicates that the Husimi distribution is not simply peaked around its center, but develops a radial bump along all directions selected by the minimization, consistent with annular phase-space distribution [see Fig.~\ref{plot::Husimis}].

The physical role of \(Q_\varphi^\downarrow\) is to construct a center-peaked reference profile on the same ray.  If the Husimi function is a coherent-state-like or thermal-state-like blob centered at \(\alpha_c\), then each radial cut decreases away from the center and \(Q_\varphi(r)=Q_\varphi^\downarrow(r)\), giving \(T_Q=0\). Heating can broaden such a blob, but as long as the most likely phase-space point remains at the center, the radial profile is still monotone, and torotropy remains zero.  By contrast, a limit-cycle-like state has suppressed weight near the center and enhanced weight at a finite radius. Rearranging the same probability weight into a monotone profile then moves weight inward, and the weighted difference in Eq.~\eqref{eq:torotropy} becomes positive. Due to the minimization over $\varphi$, a finite value requires the finite-radius bump to survive along all radial directions included in the minimization, rather than arising from a single lobe or a localized fluctuation.  

It is also important that $T_Q$ is defined here from the positive Husimi \(Q\)-function, not from the Wigner function.\,A ring-like feature in a Wigner distribution is not by itself sufficient evidence for finite torotropy. Wigner quasiprobabilities can show rings, nodes, or sign-changing interference features that do not correspond to a finite-radius bump in the smoothed Husimi distribution~\cite{sevitz2026autonomous}. We illustrate this distinction explicitly in the benchmark comparison of Fig.~\ref{plot::oisin_benchmark}(b), where a ring-like Wigner feature in \cite{Culhane2022PRE} does not necessarily imply nonzero $T_Q$ and self-sustained oscillations. In the results below, we therefore use $T_Q$ as the quantitative criterion for a self-oscillatory steady state.

We combine this geometric phase-space diagnostic with phonon-number statistics. Given the phonon occupation operator $\hat n_{\rm ph} =  \hat b^\dagger \hat b$ we study the phonon-number Fano factor given by,
\begin{equation}
    F_{\rm ph} = \frac{{\rm Var}(n_{\rm ph})}{\langle n_{\rm ph}\rangle} = \frac{\langle \hat n_{\rm ph}^2\rangle - \langle \hat n_{\rm ph}\rangle^2}{\langle \hat n_{\rm ph}\rangle}. \label{eq:fano}
\end{equation}
It distinguishes sub-Poissonian $(F_{\rm ph}<1)$, Poissonian $(F_{\rm ph}=1)$, and super-Poissonian $(F_{\rm ph}>1)$ phonon-number fluctuations. We also compute the zero-delay second-order coherence of the phonon mode,
\begin{equation}
    g^{(2)}(0) = \frac{\langle \hat b^\dagger \hat b^\dagger \hat b \hat b \rangle}{\langle \hat b^\dagger \hat b\rangle^2} =
    \frac{\langle\hat n_{\rm ph}(\hat n_{\rm ph}-1)\rangle}{\langle \hat n_{\rm ph}\rangle^2}. \label{eq:g2}
\end{equation}
Here \(g^{(2)}(0)>1\) indicates bunching, \(g^{(2)}(0)<1\) indicates antibunching, and \(g^{(2)}(0)\simeq 2\) is characteristic of a thermal state. The Fano factor and \(g^{(2)}(0)\) are related by
\begin{equation}
    g^{(2)}(0) = 1+ \frac{F_{\rm ph}-1}{\langle n_{\rm ph}\rangle}.
    \label{eq:g2_fano_relation}
\end{equation}
Thus, $F_{\rm ph}$ and $g^{(2)}(0)$ probe phonon-number statistics, while $T_Q$ directly diagnoses the limit cycle from the oscillator state.
\begin{figure*}[!tbp]
\hspace{-.4cm}
\includegraphics[
            width=.249\linewidth,
            trim=0.0cm 0.0cm 2.50cm 0.0cm,
            clip
        ]{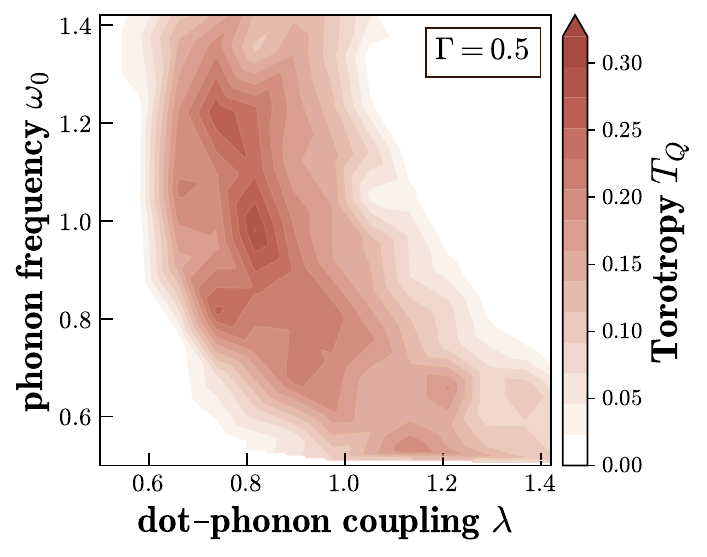} \hspace{-.2cm}
\includegraphics[
            width=.226\linewidth,
            trim=1.0cm 0.0cm 2.50cm 0.0cm,
            clip
        ]{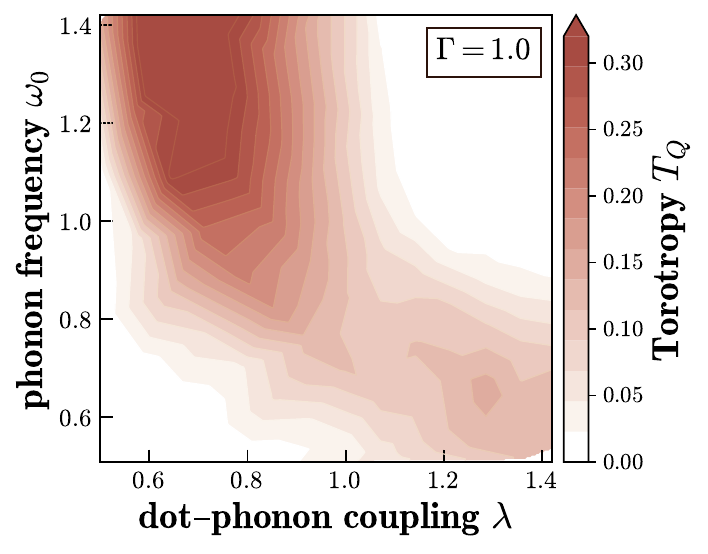}\hspace{-.1cm}
\includegraphics[
            width=.226\linewidth,
            trim=1.0cm 0.0cm 2.50cm 0.0cm,
            clip
        ]{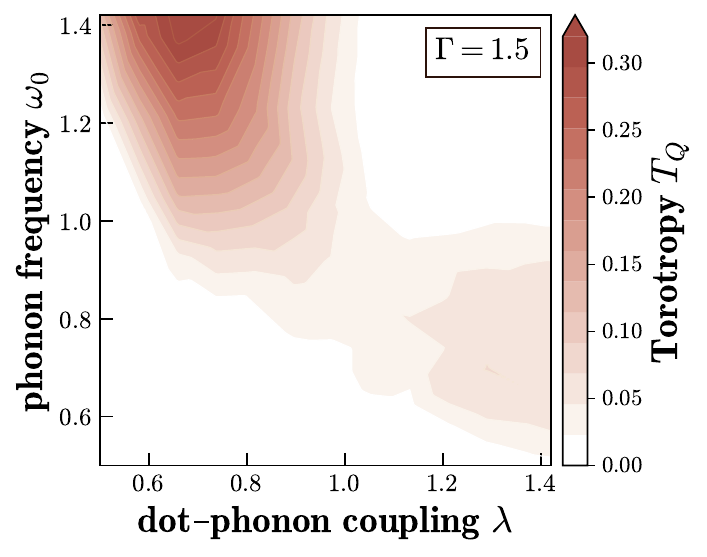}\hspace{-.1cm}
\includegraphics[
            width=.226\linewidth,
            trim=1.0cm 0.0cm 2.50cm 0.0cm,
            clip
        ]{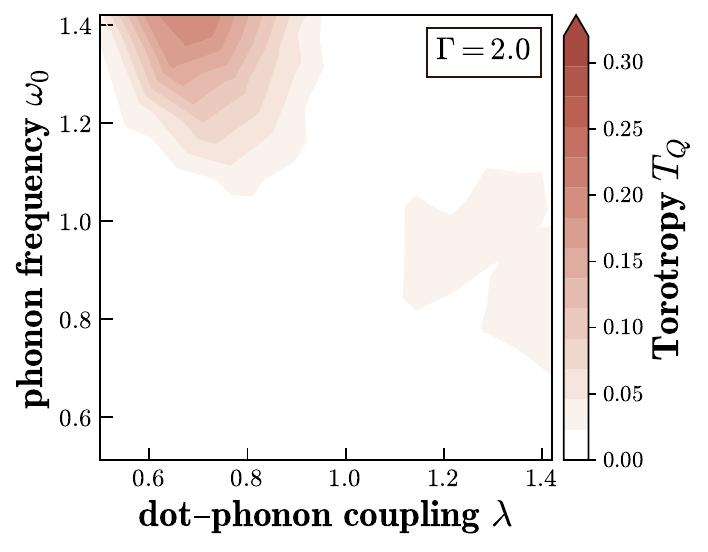}
\includegraphics[
        trim=9.5cm 1.4cm 0.3cm 0.0cm,
        clip,
        width=.067\linewidth
    ]{7contourplotGo5.pdf}
\newline \vspace{0.01cm}
\parbox{\linewidth}{\caption{\justifying Transport-induced self-oscillations across dynamical regimes. Torotropy $T_Q$ is measured in the $(\lambda,\omega_0)$ plane for different electronic relaxation rates $\Gamma$. The color scale indicates the magnitude of $T_Q$, highlighting the regions where the steady state develops limit-cycle behavior. Increasing electronic relaxation progressively shifts the optimal regime towards higher vibrational frequencies. The vibrational cutoff is $M=128$. See Table.~\ref{tab::param} for the parameters used in this figure. \label{plot::TQ_lambda_Omega0}}}
\end{figure*}

The Husimi $Q$-distribution, the position autocorrelation
\begin{equation}
  C_{xx}(t) = \langle x(t)\,x(0)\rangle - \langle x\rangle^2,
\end{equation}
and the corresponding mechanical power spectral density (PSD) 
\begin{equation}
    S_{xx}(\omega) = \int_{-\infty}^{\infty} dt\, e^{i\omega t} C_{xx}(t)
\end{equation}
together provide complementary characterizations of the vibrational steady state. The Husimi $Q$-function captures the static phase-space structure. $C_{xx}(t)$ probes how long the oscillator retains phase memory of its initial quadrature. The power spectral density gives the frequency-domain counterpart, with a narrow spectral peak indicating long coherence and spectral broadening reflecting phase diffusion.

At very weak coupling $\lambda = 0.01$ [Figs.~\ref{plot::Husimis}(a1--a3)], the Husimi $Q$-function is a symmetric blob centred at the phase-space origin, indicating no finite-amplitude limit cycle. The autocorrelation shows low-amplitude oscillations at frequency $\omega_0$, and the PSD displays a correspondingly narrow peak at $\omega_0$. This is an expected response of an underdamped harmonic oscillator, that is, the position amplitude undergoes decay depending on the bosonic damping rate $\Gamma_{\rm ph}$, with negligible transport-induced backaction at weak coupling.

At intermediate coupling $\lambda = 0.5$ [Figs.~\ref{plot::Husimis}(b1--b3)], the Husimi $Q$-function develops a clear annular structure, and the autocorrelation amplitude is markedly enhanced, with an envelope that decays slowly on the $\Gamma_{\rm ph}$ timescale. The PSD shows a tall, narrow peak near $\omega_0$, confirming that the oscillation has a well-defined frequency and a long coherence time. The slight shift of the PSD peak from the bare mechanical frequency $\omega_0$ can be interpreted as a coupling-dependent frequency pulling of the dressed electromechanical oscillator. All three diagnostics are thus mutually consistent with transport-induced self-oscillations.

At stronger coupling $\lambda = 1.0$ [Figs.~\ref{plot::Husimis}(c1--c3)], the Husimi-$Q$ ring persists but is less sharp, the autocorrelation envelope decays appreciably faster, and the PSD peak is lower and broader. The increased spectral linewidth directly reflects enhanced phase diffusion, since the oscillator sustains a large phase-space amplitude while losing temporal phase memory more rapidly than at $\lambda = 0.5$. This behavior is consistent with stronger fluctuations or more intermittent electronic backaction in the strong-coupling regime, which we discuss further in Sec.~\ref{subsec:heating_vs_limitcycle}. The spectral linewidth of $S_{xx}(\omega)$ is in principle accessible via displacement spectroscopy~\cite{PootVanDerZant2012PhysRep, Aspelmeyer2014RMP}, or via current-noise spectroscopy~\cite{liu2024fine, Steele2009Science, Lassagne2009Science, Wen2020NatPhys, Urgell2020NatPhys}, since the mechanical motion modulates the electron tunneling current through the electromechanical coupling. These results can also be connected directly to studying the precision of this device as an autonomous oscillator~\cite{Erker2017PRX, Culhane2024NJP}.

\begin{figure*}[!tbp]
    \centering
    \includegraphics[trim=0.25cm 0.0cm 0.0cm 0.3cm, width=0.76\linewidth]{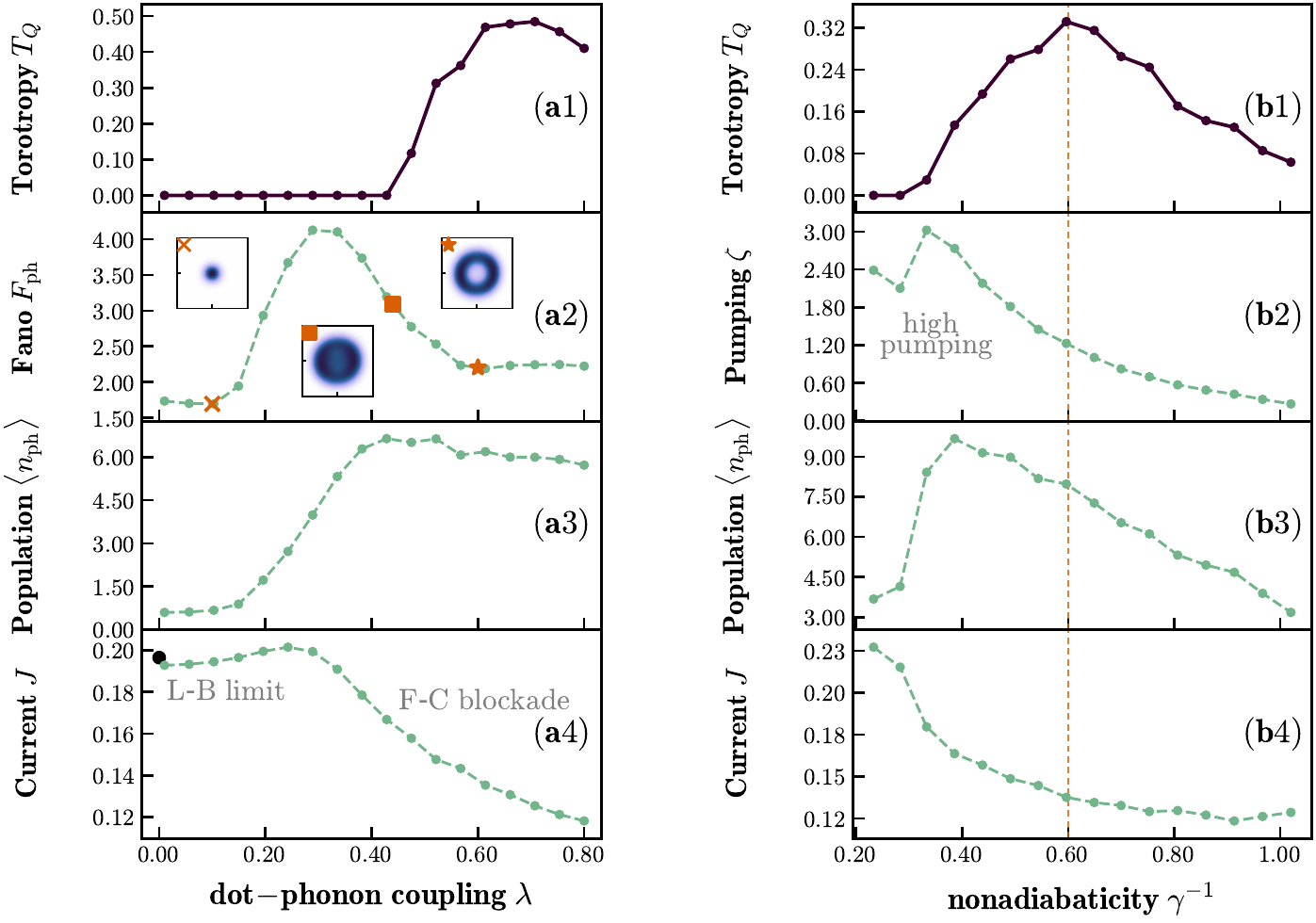} \vspace{.3cm}\newline
    \hspace{1cm}\includegraphics[width=0.75\linewidth]{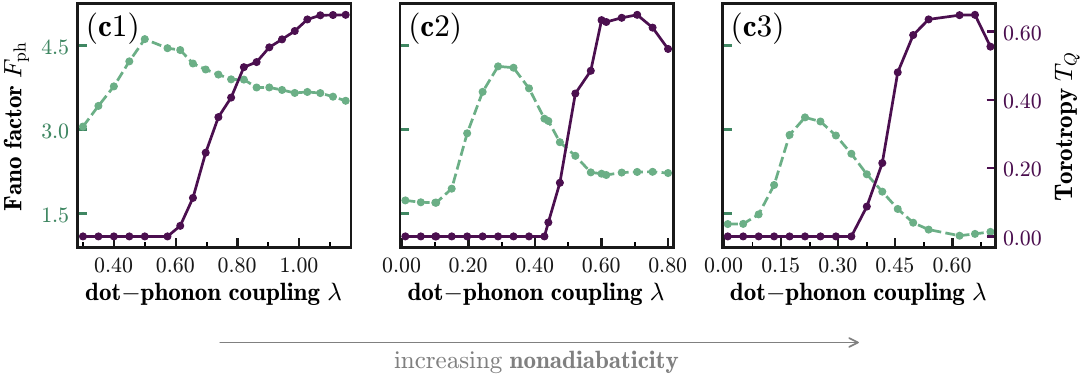}
    \parbox{\textwidth}{\caption{\justifying
Relation between vibrational excitation, fluctuations, and self-oscillation. Torotropy $T_Q$ (a1), phonon-number Fano factor $F_{\rm ph}$ (a2), mean phonon occupation $\langle n_{\rm ph}\rangle$ (a3), and current $J$ (a4) as functions of the dot--phonon coupling $\lambda$ at fixed $\omega_0=1$ and $\Gamma=1$. The Fano factor reaches its maximum before the onset of appreciable torotropy, indicating that strong phonon-number fluctuations develop before the steady state forms a clear ring-like phase-space structure. The current decreases at large $\lambda$ due to Franck--Condon suppression, while the black dot in (a4) marks the noninteracting Landauer benchmark at $\lambda=0$. In (a2) we display the Husimi $Q$-distributions (indicated by orange markers) of the vibrational NESS and the emergence of an annular phase-space structure while sweeping $\lambda$ across the Fano peak. Torotropy $T_Q$ (b1), pumping indicator $\zeta$ (b2), mean phonon occupation $\langle n_{\rm ph}\rangle$ (b3), and current $J$ (b4) as functions of the phonon frequency $\omega_0$ at fixed $\lambda=1$ and $\Gamma=1.5$. The dashed vertical line marks the maximum of $T_Q$. At small $\omega_0$, both $\zeta$ and $\langle n_{\rm ph}\rangle$ remain large while $T_Q$ is negligible, indicating transport-induced vibrational heating without self-oscillations. Finite torotropy develops only in an intermediate frequency window where delayed electronic backaction supports sustained oscillatory motion. The Fano peak near the onset of self-oscillations while increasing $\lambda$, across slow and fast mechanical oscillations: $\omega_0=0.66 \text{ (c1), } 1 \text{ (c2), } 1.5 \text{  (c3)}$, and fixed $\Gamma=1$. The vibrational Hilbert-space cutoff is $M=128$. See Table~\ref{tab::param} for the parameters used in this figure.
    \label{plot::transport_vs_lambda}}}
\end{figure*}

\subsection{Adiabaticity and self-sustained oscillations}
\label{sec:adiabaticity}

Figure~\ref{plot::TQ_lambda_Omega0} shows the torotropy $T_Q$ in the $(\lambda,\omega_0)$ plane for several electronic relaxation rates $\Gamma$. Finite-$T_Q$ regions depend strongly on $\Gamma$. As $\Gamma$ increases, the strongest torotropy shifts toward higher phonon frequencies, while the optimal coupling moves toward smaller $\lambda$. These trends indicate that self-oscillation is controlled by the competition between the electronic relaxation timescale, $\tau_{\rm el}\sim\Gamma^{-1}$, and the mechanical timescale, $\tau_{\rm m}\sim\omega_0^{-1}$.

In the adiabatic regime, $\gamma=\Gamma/\omega_0\gg1$ [Eq.~\eqref{eq:gamma_definition}], the electronic occupation relaxes rapidly compared with the mechanical motion. The electronic force, therefore, follows the oscillator almost instantaneously and produces little delayed backaction. In this limit, transport mainly renormalizes the oscillator frequency and damping, but does not efficiently sustain coherent motion \cite{blanter2004single, a2005quantum, ludwig2008optomechanical}.

Self-oscillation becomes more favorable when the electronic and mechanical timescales become comparable, $\gamma\sim1$. In this regime, the electronic occupation cannot fully follow the oscillator displacement, generating a delayed force that can inject energy into the mechanical motion over a cycle. This mechanism underlies transport-induced self-oscillations in nanoelectromechanical and optomechanical systems \cite{armour2004classical, Fedorets2004PRL, Pistolesi2004PRB, blanter2004single, a2005quantum, ludwig2008optomechanical, Wen2020NatPhys}. The upward shift of the finite-$T_Q$ region with increasing $\Gamma$ is consistent with this picture, since maintaining $\gamma\sim1$ requires larger $\omega_0$ at larger electronic relaxation rates.

The dependence on $\lambda$ reflects the competition between electromechanical feedback and transport suppression. At weak coupling, the leading phonon-assisted tunneling processes scale linearly with $\lambda$, so the corresponding backaction rates scale approximately as $\lambda^2$ \cite{Holstein1959, Koch05, Koch96, mitra2004phonon, avriller2011unified, Schinabeck2020HQMEFCS}. Increasing $\Gamma$ can therefore partly compensate for a smaller $\lambda$ by increasing the number of tunneling events per unit time. This is consistent with the shift of the strongest response toward lower coupling at intermediate $\Gamma$.

At large $\lambda$, however, transport is suppressed by the polaronic dressing and Franck--Condon blockade [see Fig.~\ref{plot::transport_vs_lambda}(a4)], reducing the current available to drive the oscillator. The self-oscillatory regime therefore remains confined to an intermediate coupling window where the weak coupling produces insufficient feedback, while strong coupling suppresses the transport processes responsible for gain.

Figure~\ref{plot::TQ_lambda_Omega0}, therefore, identifies a finite dynamical region where transport backaction and mechanical damping balance to produce sustained oscillatory motion. Importantly, the conditions for mechanical energy injection are not identical to those for self-sustained oscillations. In Sec.~\ref{subsec:heating_vs_limitcycle}, we compare torotropy with vibrational excitation measures and show that large phonon occupation alone does not necessarily imply coherent limit-cycle dynamics.
\begin{figure*}[!tbp]
    \centering
    \includegraphics[width=\linewidth]{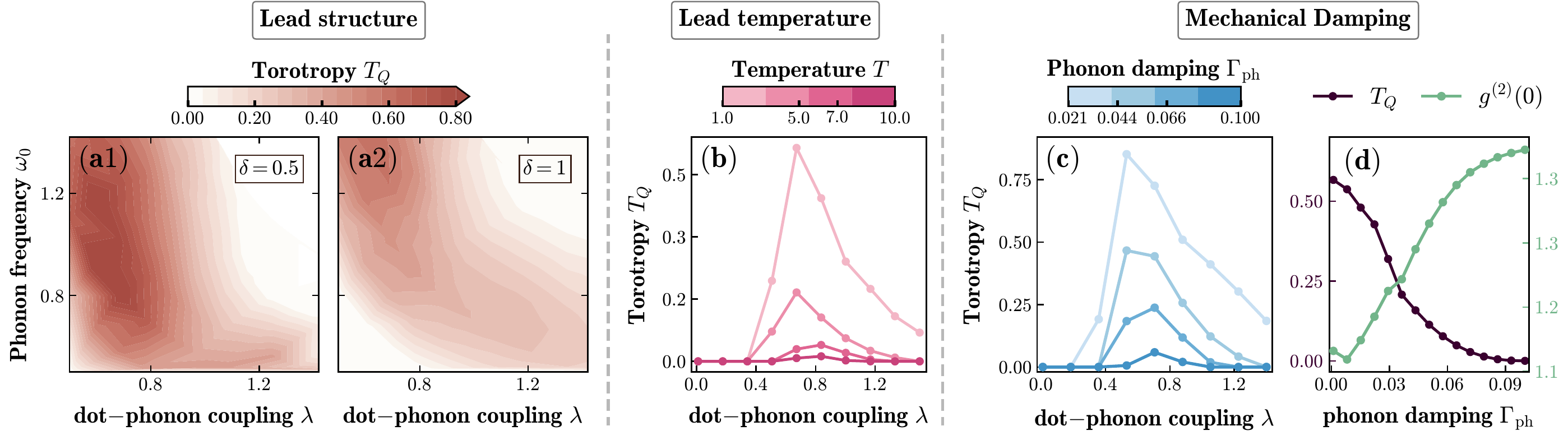}
      \parbox{\textwidth}{\caption{\justifying Effect of lead structure, temperature, and bosonic dissipation on self-oscillations. Torotropy $T_Q$ is measured in the $(\lambda,\omega_0)$ plane for different lead spectral widths $\delta$ (a1, a2). The color scale indicates the magnitude of $T_Q$, highlighting the regions where the steady state develops a limit-cycle. Next, $T_Q$ versus (b) dimensionless dot--phonon coupling $\lambda$ for different lead temperatures $T/\Gamma$ and (c) phonon decay strength $\Gamma_{\rm ph}$ has been explored. $T_Q$ is maximal at intermediate coupling and is reduced as the electronic temperature or phonon decay is increased. In contrast, $g^{(2)}(0)$ primarily reflects phonon-number bunching and increases with temperature and phonon decay (d). The phonon Hilbert-space truncation is $M=128$. See Table~\ref{tab::param} for the parameters used in this figure. 
    \label{plot::environment}}}
\end{figure*}

\subsection{Interplay with vibrational fluctuations}
\label{subsec:heating_vs_limitcycle}

Figure~\ref{plot::transport_vs_lambda}(a) compares torotropy, phonon-number fluctuations, vibrational occupation, and current as functions of the dot--phonon coupling $\lambda$. As $\lambda$ increases, the current $J$ decreases while the phonon occupation $\langle n_{\rm ph}\rangle$ increases and eventually saturates. This behavior is consistent with increasing electromechanical backaction together with the onset of Franck--Condon suppression at strong coupling \cite{Koch05, Koch96,avriller2011unified, Schinabeck2020HQMEFCS}.

The phonon-number Fano factor $F_{\rm ph}$ exhibits a pronounced peak before torotropy becomes appreciable. Transport, therefore, first generates a broad nonequilibrium phonon-number distribution before stabilizing a self-oscillatory NESS. As we increase $\lambda$ further after the Fano peak, $T_Q$ increases until $F_{\rm ph}$ settles towards a lower plateau, and for even stronger electromechanical coupling strengths, $T_Q$ values are suppressed in turn. In Fig.~\ref{plot::transport_vs_lambda}(c), we show that across slow and fast oscillation regimes ($\omega_0=0.66, 1, 1.5$) of the mechanical mode, higher phonon occupation fluctuations can be found just before the onset of self-sustained oscillations while increasing $\lambda$. This behavior is consistent with a threshold scenario in which stochastic tunneling acts as a fluctuating mechanical drive \cite{blanter2004single, Usmani2007PRB}, whereas a finite-$T_Q$ state requires the transport backaction to overcome damping and phase diffusion sufficiently to stabilize the self-sustained oscillations. 

The analysis in Sec.~\ref{sec:adiabaticity} identified the regime where the electronic and the mechanical timescales are favorably matched for self-oscillation. We now compare torotropy with the transport-based pumping indicator proposed in Ref.~\cite{sevitz2025quantum},
\begin{equation}
\zeta :=
\frac{|I_R^{\rm el}|E_{\rm mec}}{\omega_0^2}
=
\frac{|I_R^{\rm el}|\langle \hat n_{\rm ph}\rangle}{\omega_0},
\label{eq:zeta_def}
\end{equation}
where $E_{\rm mec}=\omega_0\langle \hat n_{\rm ph}\rangle$. Figure~\ref{plot::transport_vs_lambda}(b) compares $T_Q$, $\zeta$, $\langle n_{\rm ph}\rangle$, and the current $J$ as functions of $\omega_0$. At small $\omega_0$, both the current and vibrational occupation are large, and $\zeta$ is correspondingly enhanced, while $T_Q$ remains close to zero. Transport, therefore, produces substantial vibrational excitation without forming a clear limit-cycle structure.

As $\omega_0$ increases, $T_Q$ rises sharply and reaches a maximum even though both $J$ and $\zeta$ decrease over much of the same interval. The onset of self-oscillation is therefore not controlled solely by the amount of injected vibrational energy, but also by the phase relation between the transport-induced backaction and the mechanical motion.

At still larger $\omega_0$, $T_Q$, $\zeta$, and $\langle n_{\rm ph}\rangle$ all decrease. The comparison, therefore, distinguishes a low-frequency heating regime from an intermediate window where electronic backaction supports self-sustained oscillations. Throughout this work, we therefore use torotropy as the primary indicator of self-oscillation, while $\zeta$ is treated as a measure of transport-induced vibrational excitation.

\subsection{Effects of lead structure, thermal broadening, and bosonic dissipation} \label{subsec::params_interplay}

Figure~\ref{plot::environment} shows how the finite bandwidth of the fermionic leads affects the formation of self-oscillations. In the Lorentzian reservoirs used here [Eq.~\eqref{eq:spectral_densities}], the width $\delta$ controls the reservoir correlation time, $\tau_\delta\sim\delta^{-1}$, and therefore changes how rapidly the fermionic leads respond to the dot dynamics. For the narrower spectrum, $\delta=0.5$ [Fig.~\ref{plot::environment}(a1)], torotropy is relatively large but concentrated in a restricted region of intermediate coupling and phonon frequency. For the broader spectrum, $\delta=1.0$ [Fig.~\ref{plot::environment}(a2)], self-oscillations tend to occur for larger $\omega_0$ (and smaller $\lambda$), similar to the behavior with increasing the electronic relaxation rate $\Gamma$ [Fig.~\ref{plot::TQ_lambda_Omega0}]. Thus, the reservoir bandwidth acts as an additional dynamical control parameter that reshapes, rather than monotonically strengthens, the transport-induced limit-cycle window.

We next examine how external noise and damping affect the self-oscillatory NESS. Figure~\ref{plot::environment}(c) shows the effect of the weak bosonic bath [Eq.~\eqref{eq:Dph}] coupled to the phonon mode. Increasing the phonon decay rate $\Gamma_{\rm ph}$ monotonically suppresses the torotropy $T_Q$, indicating that the intrinsic mechanical damping competes with the transport-induced phase-space organization. Over the same range, $g^{(2)}(0)$ increases from a value close to unity, showing that the phonon statistics become more bunched as the bosonic bath becomes more strongly coupled. In the plotted range, the state is not fully thermalized, but the trend is consistent with a monotonic trend away from the self-oscillating NESS and towards more bath-dominated fluctuations with $g^{(2)}(0)>1$.

Figure~\ref{plot::environment}(b) shows the effect of the electronic lead temperature on the coupling dependence.  At low temperature, $T_Q$ is strongly nonmonotonic in $\lambda$: it is small at weak coupling, reaches a maximum at intermediate coupling, and decreases again at larger $\lambda$. Increasing $T$ reduces the peak torotropy, consistent with thermal fluctuations washing out the phase-space structure. The behavior of $g^{(2)}(0)$ mainly tracks phonon-number bunching and increases with temperature. Notice that for small dot--phonon couplings $\lambda$, $g^{(2)}(0)$ is close to 2 (the value corresponding to a \textit{thermal} bosonic state).  Together, Fig.~\ref{plot::environment} shows that the NESS in these parameter regimes are sensitive both to mechanical damping and to thermal fluctuations in the electronic reservoirs. Both effects reduce the torotropy, but they leave distinct signatures in $g^{(2)}(0)$.

\subsection{Benchmarks in different parameter regimes} \label{sec::benchmark}

\begin{figure}[!tbp]
    \centering
    \includegraphics[width=\linewidth]{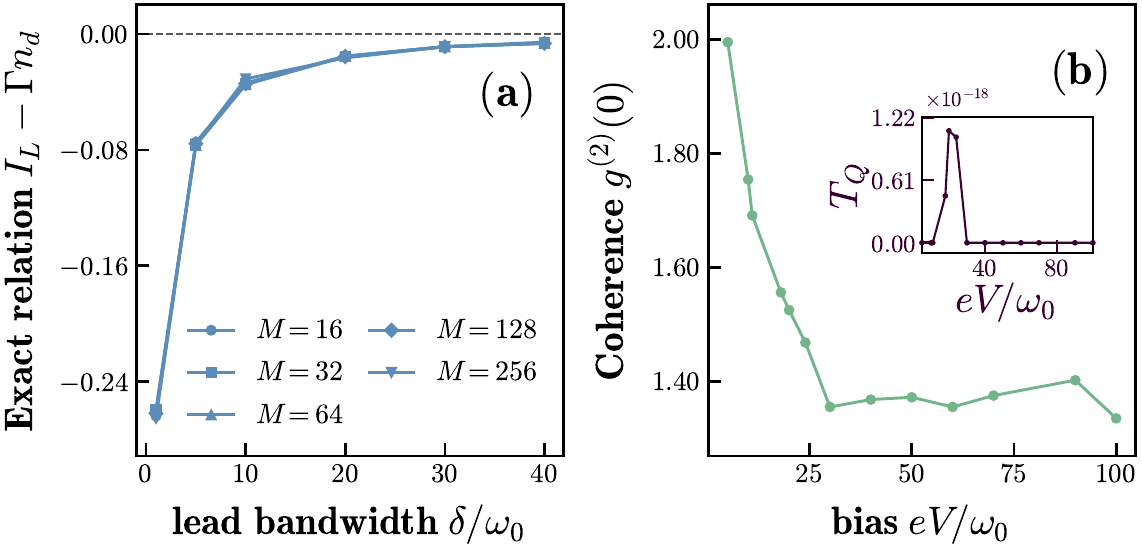}
      \caption{\justifying (a) Accuracy assessment of the simulations with increasing lead bandwidth $\delta$ by demonstrating convergence to the exact current-occupation relation for structureless leads [Eq.~\eqref{eq:exact_curr_occ_rel}], for different vibrational mode occupation cutoffs $M$. (b) Benchmark with quasi-adiabatic regime ($\gamma = 5$) results from Culhane \textit{et al.}~\cite{Culhane2022PRE}. See Table~\ref{tab::param} for the exact parameters used in this figure. \label{plot::oisin_benchmark}}
\end{figure}

We have demonstrated above that limit cycle dynamics in the mechanical mode requires structured lead spectra. We can still assess the accuracy of our method in the wide-band limit using an exact NESS relation proposed in \cite{agarwalla2017anderson}. Assuming symmetrical coupling $\Gamma_\alpha=\Gamma$ in Eq.~\eqref{eq:spectral_densities} and in the limit $\mu_R=-\infty$, the mean current $\langle I_L\rangle$ from the left lead towards the dot (that is, arbitrarily interacting with the mechanical mode) is connected to the mean dot occupation $\langle n_{d}\rangle$ by the exact current-occupation relation
\begin{equation}
    \langle I_{L} \rangle = e\Gamma \langle n_{d}\rangle.
    \label{eq:exact_curr_occ_rel}
\end{equation}
In Fig.~\ref{plot::oisin_benchmark}, we show that our results systematically converge to this exact relation in the appropriate limit of large lead bandwidth $\delta$ for the Lorentzian spectral density [Eq.~\eqref{eq:spectral_densities}]. Moreover, this convergence is obeyed irrespective of the phonon level truncation $M$ employed in the calculations. This confirms that our mesoscopic-lead embedding correctly reproduces the wide-band limit of the fermionic leads, and that the NESS current and occupation are internally consistent. Ref.~\cite{agarwalla2017anderson} showed that using analytically evaluated Franck--Condon factors in a truncated phonon basis \emph{violates} Eq.~\eqref{eq:exact_curr_occ_rel}, because the matrix elements are inconsistent with the finite truncation. Satisfaction of the relation requires that the phonon displacement operator be evaluated self-consistently within the retained Fock space, which is precisely what the binary pseudosite encoding achieves by construction [Eqs.~\eqref{eq:binary_map}--\eqref{eq:binary_increment}].  The agreement across all $M$, therefore, provides direct evidence that our binary encoding correctly implements the Franck--Condon displacement within the truncated phonon space.

In the quasi-adiabatic regime $\gamma = \Gamma / \omega_0 \gg 1$, using independent results from Culhane \textit{et al.}~\cite{Culhane2022PRE}, who employed a semiclassical Fokker--Planck approach. Figure~\ref{plot::oisin_benchmark}(b) shows the zero-delay second-order coherence $g^{(2)}(0)$ as a function of dimensionless bias $eV / \omega_0$ at $\gamma=5$.  Our tensor-network calculations are in quantitative agreement with the Fokker--Planck calculations of Ref.~\cite{Culhane2022PRE} across the full bias range shown, with $g^{(2)}(0)$ decreasing from the thermal value of $2$ at low bias and approaching a plateau near $1.4$ at large bias, consistent with a sub-thermal but non-Poissonian phonon distribution. The inset shows the corresponding torotropy $T_Q$, which remains negligible throughout. This is consistent with our discussion in Sec .~\ref {subsec::phase-space struct}, that is, while high bias can produce ring-like structures in the Wigner function and finite ergotropy in the mechanical mode~\cite{Culhane2022PRE}, the near-zero $T_Q$ confirms that these do not constitute a limit cycle, in line with the distinction drawn in Ref.~\cite{sevitz2026autonomous}.

At $\lambda=0$, the mechanical mode decouples from the dot and the model reduces to a noninteracting resonant level coupled to two fermionic leads. The current is then given by the Landauer expression \cite{Landauer1957,Landauer1970,MeirWingreen1992,HaugJauho2008}
\begin{equation}
    J = \frac{1}{2\pi}\int_{-\infty}^{\infty} d\omega\,
    \mathcal{T}(\omega)\left[f_L(\omega)-f_R(\omega)\right],
    \label{eq:landauer}
\end{equation}
where $\mathcal{T}(\omega)$ is the transmission function and $f_{L,R}$ are the Fermi occupations of the leads [Appendix~\ref{sec:landauer}]. For the parameters of Fig.~\ref{plot::transport_vs_lambda}(a), this yields $ J \simeq 0.1965$, and the small-$\lambda$ current in the figure is consistent with this analytical benchmark.

\subsection{Convergence and computation times} \label{sec::convergence}
The two main numerical controls for the accuracy of the tensor-network representation of our model are the phonon Hilbert space cutoff $M$ and the allowed maximum bond dimension $\chi_{\rm max}$ of the NESS. As discussed earlier in Sec.~\ref{sec::methods}, the cost increases with the number of pseudosites and with the bond dimension required to represent the steady state. But a sufficiently large $M$ is crucial to represent the highly excited phonon mode accurately and must therefore be accompanied by checks in the maximum bond dimension $\chi$ (we have $\chi \leq \chi_{\rm max}$). In our numerical method implementation [see Appendix~\ref{ape:numerical_strategies}], we have adopted several physically motivated truncations in $\chi$ to keep the entanglement, and in turn, the bond dimension growth tractable. Therefore, it is crucial to check the performance and stability of the binary encoding scheme DMRG method that we use for our model. 

\begin{figure}[!tbp]
    \centering
    \includegraphics[width=\linewidth]{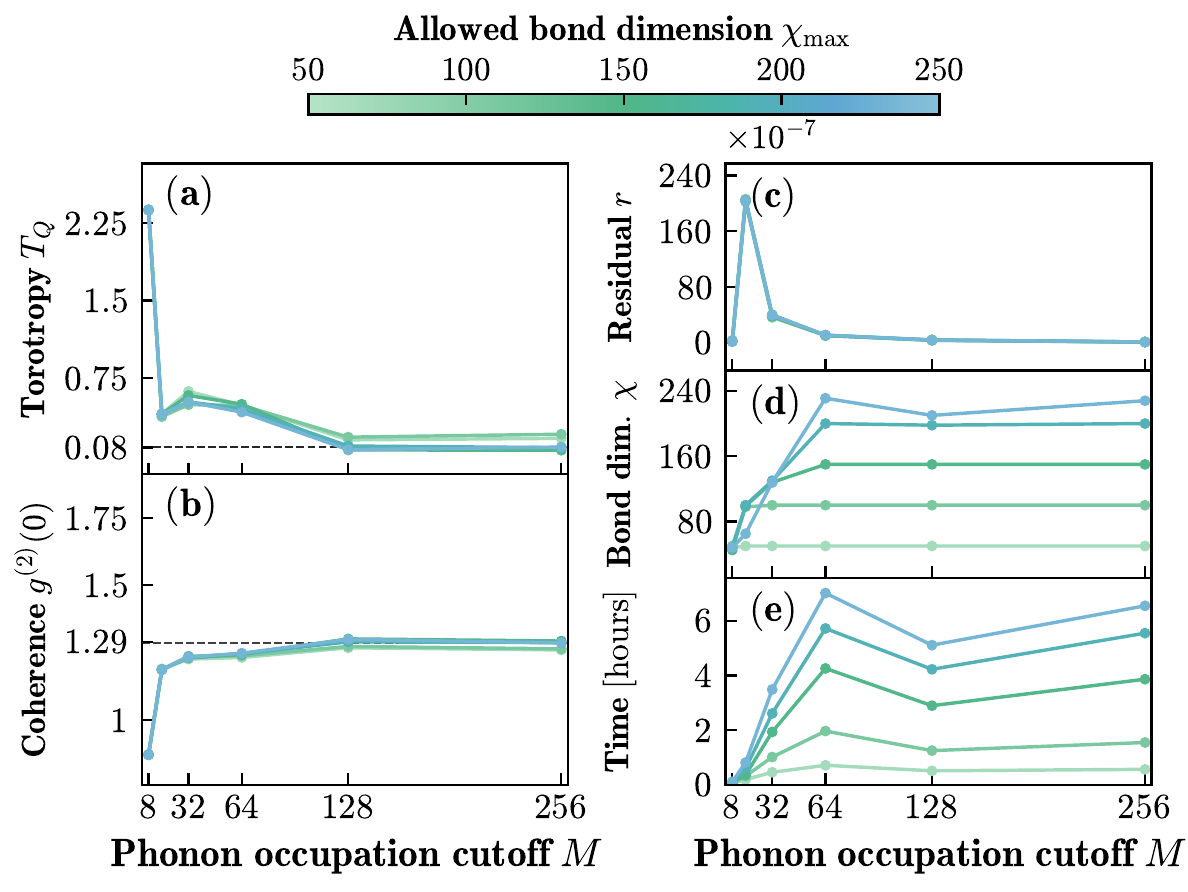}
      \parbox{0.48\textwidth}{\caption{\justifying Convergence with phonon Hilbert-space cutoff $M$ and allowed bond dimension $\chi_{\max}$. (a) Torotropy $T_Q$ and (b) zero-delay second-order coherence $g^{(2)}(0)$ as functions of $M$ for different $\chi_{\max}$. Both observables show strong cutoff artifacts at small $M$ and stabilize for $M\gtrsim128$ in the calculations shown. (c) Liouvillian residual $r$, (d) final maximum bond dimension $\chi$, and (e) wall time required to obtain the NESS. Dashed lines in (a,b) mark the reference values from the largest calculation, $M=256$ and $\chi_{\max}=250$. The color scale denotes the imposed bond dimension cutoff $\chi_{\max}$. Parameters are given in Table~\ref{tab::param}.
\label{plot::convergence_check}}}
\end{figure}

In Fig.~\ref{plot::convergence_check}, we demonstrate the phonon occupation cutoff $M$ and the bond dimension cutoff $\chi_{\text{max}}$-dependence in finding NESS within our tensor-network framework, using relevant observables in our model. We present the numerical convergence results for the torotropy $T_Q$ and zero-delay second-order coherence $g^{(2)}(0)$ of the oscillator NESS, which are expected to exhibit strong and nonmonotonic cutoff-dependence. While $T_Q$ captures the phase-space structure,  $g^{(2)}(0)$ probes phonon occupation fluctuations. Together, they probe complementary aspects of the structure of the oscillator state undergoing self-sustained oscillations. We calculate $T_Q$ and $g^{(2)}(0)$ by varying $M$ while allowing the bond dimension $\chi$ to grow. In Figs.~\ref{plot::convergence_check}(a, b), the smallest cutoff, $M=8$, gives an anomalously large $T_Q$ and a strongly suppressed $g^{(2)}(0)$. This is a truncation artifact since for this cutoff, the highest retained occupation is $n_{\rm max}=M-1=7$, while the mean occupation is already $\langle n_{\rm ph}\rangle\simeq 6.25$ for the particular parameters. This constrains support for the high-occupation tail of the oscillator distribution close to the upper boundary of the truncated Fock space. This cutoff can falsely enhance the annular structure measured by $T_Q$ and suppresses phonon-number fluctuations, leading to an artificially small value of $g^{(2)}(0)$. But as we increase $M$, both of these quantities simultaneously converge to a nonzero $T_Q$ ($=0.08$) value around $M = 128$ with a concurrently converged and sufficiently suppressed Liouvillian residual $r$ [see Fig.~\ref{plot::convergence_check}(c)]. The dashed horizontal lines represent the reference converged values that we obtain at $\chi_{\text{max}} = 250$ and $M=256$. More details regarding the employed convergence checks in our tensor-network calculations can be found in  Appendix~\ref{ape:numerical_strategies}. 

In Fig.~\ref{plot::convergence_check}(d), we show that although larger phonon cutoffs require substantially larger computational resources, the actual maximum bond dimension $\chi$ of the NESS solutions remains below the bond dimension cutoff $\chi_{\text{max}}$ imposed in our calculations. Interestingly, the study shows a convergence in the final bond dimension for $M \geq 128$. We can associate this behavior with the binary encoding scheme because, although $M$ has grown from 128 to 256, for the DMRG treatment in our approach, it only means building up one additional block by introducing one pseudosite with 2 states to the previous block. This keeps the local bond dimension growth moderate in our calculations. We further illustrate the computational complexity of our tensor-network methodology by calculating the overall simulation wall time as a function of occupation cutoff $M$ for different $\chi_{\rm max}$ in Fig.~\ref{plot::convergence_check}(e) and see a similar result as in Fig.~\ref{plot::convergence_check}(d) for the actual maximum bond dimensions $\chi$ for the obtained NESS. We thus see that the overall computational cost generally increases with increasing phonon occupation cutoff $M$, but the runtimes and bond dimensions $\chi$ of the obtained NESS are not strictly monotonic functions of $M$. These observations reflect the fact that the cost of the variational Liouvillian minimization depends not only on the local Hilbert-space dimension, but also on the entanglement structure and conditioning of the vibrational NESS.

\section{Conclusions and outlook}
\label{sec::conclusion}

We have developed a framework for identifying and characterizing autonomous self-sustained oscillations in nanoscale electromechanical transport. We address the technical challenge of describing the simultaneous presence of a highly excited vibrational mode, strong electromechanical coupling, finite-bandwidth of the fermionic reservoirs, and weak mechanical damping. These ingredients are precisely those required for describing transport-induced mechanical self-oscillations in realistic NEMS devices. However, some of these relevant parameter regimes fall outside the scope of standard rate equations, semiclassical Langevin descriptions, and wide-band approximation approaches.

Our main methodological advance is combining three numerically controlled ingredients in a single tensor-network framework. First, we represent the vibrational Hilbert space using a binary pseudosite encoding, enabling large phonon cutoffs. Second, we describe finite-bandwidth fermionic reservoirs via a mesoscopic-lead embedding, yielding an augmented Markovian dynamics. Third, we directly target the NESS by variationally minimizing a positive semidefinite functional of the Liouvillian, avoiding costly real-time propagation. This construction gives direct access to transport observables and to the NESS itself, including its phase-space distribution, phonon-number statistics, and torotropy. More broadly, our numerical strategies offer a route to solving quantum transport problems with one or more of these features, beyond regimes where perturbative, Markovian, or semiclassical approximations are easily justified.

Our calculations verify that the parameter window featuring nonzero torotropy is controlled by two competing requirements. Firstly, the electromechanical coupling must be strong enough to generate strong backaction, but not so strong that polaronic dressing and Franck--Condon suppression block the transport processes that provide gain. Secondly, the location of this self-oscillation window strongly depends on the adiabaticity parameter, that is, the ratio between the electronic relaxation and mechanical oscillation timescales. Large phonon population, strong current, or a large pumping indicator are necessary but not sufficient conditions for establishing self-sustained oscillations in the mechanical mode.

A second physical insight concerns fluctuations near the onset of finite torotropy. With increasing electromechanical coupling $\lambda$, the phonon-number Fano factor can become strongly enhanced just before entering the finite-$T_Q$ window. This indicates a high fluctuation regime in which stochastic tunneling broadens the phonon-number distribution before a clear annular phase-space structure is formed. The effect is more visible in the Fano factor than in the normalized second-order coherence when the mean phonon occupation is already large. The torotropy increases with $\lambda$ until reaching a peak and then decreases for even stronger coupling strengths. Thus, number fluctuations, vibrational occupation, and phase-space structure probe distinct aspects of the mechanical self-oscillatory NESS.

We also find that the reservoirs are active control parameters for generating and sustaining self-sustained oscillations. Finite lead bandwidth reshapes the self-oscillation window through energy-dependent electronic tunneling. Lead temperature and mechanical damping suppress self-oscillations while enhancing phonon-number bunching. Thus thermal noise and bosonic dissipation can destroy self-oscillations even when the oscillator remains excited.

Several benchmarks support the numerical reliability of our tensor-network calculations. In the noninteracting limit, the mean current through the electronic level agrees with the Landauer predictions. In the wide-band limit of the fermionic leads, the calculations converge to an exact current-occupation relation. In the quasi-adiabatic regime, we reproduce semiclassical Fokker--Planck predictions for phonon statistics. Finally, convergence with respect to phonon cutoff, bond dimension, Liouvillian residual, and physical observables confirms the stability of the self-oscillation signatures reported in this work.

The framework introduced here has direct implications for the design and interpretation of experiments on autonomous NEMS devices. It suggests that strong self-sustained oscillations can be found in intermediate electromechanical coupling regimes, with matching of the electronic and mechanical timescales to generate sufficient backaction, and with environmental damping low enough to preserve the annular phase-space structure. It also clarifies why transport measurements alone may be ambiguous in particular regimes (such as the slow vibration regime) where energetic pumping can be large, yet the oscillator does not display self-oscillatory behavior. 

In this work, we have employed our framework to understand the mechanisms behind self-sustained oscillations across a broad parameter landscape in the spinless AH model as a minimal, non-trivial description of NEMS devices. Due to the generality of the framework, the method can be readily extended to more complicated scenarios that are experimentally relevant, such as multilevel quantum dots, Coulomb interactions, multiple vibrational modes, or anharmonic oscillators. These extensions would further clarify how many-body dynamics, reservoir memory, and nonlinear mechanical response modify the balance between transport-induced gain and dissipation. Importantly, by identifying the regimes in which a stable limit cycle survives across different electronic and mechanical timescales, our results also establish controlled operating points for collective phenomena such as  synchronization~\cite{moeckel2014,Wen2020NatPhys,PhysRevLett.112.094102,Weiss_2016}, and makes it possible to explore ways of engineering autonomous phonon sources \cite{Wen2020NatPhys, Urgell2020NatPhys}, mechanical clocks \cite{Culhane2024NJP}, nanoscale engines \cite{sevitz2026autonomous}, or quantum work-storage elements \cite{Culhane2022PRE}. Combining the present tensor-network steady state solver with real-time evolution~\cite{PhysRevLett.105.050404,mandal26} would further uncover any distinct nonequilibrium pathways during the dynamical approach towards such self-oscillatory steady states.

\section{Acknowledgments}
We acknowledge many helpful discussions with Natalia Ares, Stephen R. Clark, Archak Purkayastha, Federico Cerisola, Sofia Sevitz, Antonio Verdú, and Jesus Moreno Meseguer. J.P. acknowledges support from the QuantERA II program (Mf-QDS) and Quantera III program (AQuSeND) that have received funding from the European Union’s Horizon 2020 research and innovation program under Grant Agreement No. 101017733 and from the Agencia Estatal de Investigación, with project codes PCI2022-132915, PCI2024-153474, and SCPP2400C011413XV0  funded by MICIU/AEI/10.13039/501100011033 and by the European Union NextGenerationEU/PRTR and FEDER, UE. M.T.M. is supported by a Royal Society University Research Fellowship. This project is co-funded by the European Union (Quantum Flagship project ASPECTS, Grant Agreement No. 101080167) and UK Research and Innovation (UKRI). Views and opinions expressed are, however, those of the authors only and do not necessarily reflect those of the European Union, Research Executive Agency, or UKRI. Neither the European Union nor UKRI can be held responsible for them.

\section{Data availability}

The open-source code used in our simulations is based on the ITensor library \cite{ITensor}, and can be accessed at the publicly available GitHub repository \cite{github}.

\appendix

\section{Tensor-network implementation}
\label{ape:numerical_strategies}

In this section, we present details regarding the implementation of the numerical techniques for representing the vibrational NESS [Eq.~\eqref{eq:ness_condition}] accurately as a tensor network state. 
The main ideas described in this section are as follows. First, we discuss the importance of geometry in ordering the Liouville-space matrix product state (MPS) sites. Second, we discuss the particular strategies, such as homotopy and adaptive DMRG controls for the variational solution of the NESS [Eq.~\eqref{eq:variational_energy}]. Third, we explain how the convergence to the NESS is controlled not only by the Liouvillian residual [Eq.~\eqref{eq:residual}], but also by satisfying other physical constraints, such as stability in transport balance and vibrational excitation statistics in the obtained NESS.

\subsection{Geometry of the MPS}
\label{subsec:geometry_aware_ordering}

The numerical efficiency of an MPS representation strongly depends on the particular ordering of its degrees of freedom \cite{Perez-Garcia, Schollwoeck2005RMP, Schollwoeck2011, Wolf2014PRB,
HeMillis2017PRB}. This is especially important for impurity and transport problems, where correlations are generated near the interacting region and then rapidly spread into bath or auxiliary modes, thereby making the dynamics much harder to describe. A useful ordering should therefore keep the dominant correlation channels short-ranged, even if this ordering does \textit{not} coincide with the physical geometry of the sites in the Hamiltonian \cite{KohnSantoro2021PRB, KohnSantoroQuench2021, casagrande2021analysis}.

We apply this principle to both the binary phonon pseudosites and the fermionic auxiliary lead modes. The truncated phonon Hilbert space is encoded in $N=\log_2 M$ hard-core-boson pseudosites. In this basis [Eqs.~\eqref{eq:bdag_factorized}-\eqref{eq:binary_increment}], the ladder operators $\hat b$ and $\hat b^\dagger$ are nonlocal over the pseudosites. We therefore group the phonon pseudosites into a compact physical-tilde block and place the most significant (highest-excitation) pseudosites closest to the dot. This does not render $\hat b$ local, but confines its binary-string structure to a contiguous MPO region and keeps the electron-phonon coupling next to the interacting dot.

In the Liouville-space MPS, each physical degree of freedom sits next to its tilde copy. For the single-auxiliary-mode-per-lead setup used here, the ordering is chosen to match the superfermion structure of the Liouvillian: each auxiliary lead dissipator couples a fermionic mode only to its tilde partner. Placing these partners adjacent makes the dissipative lead terms local within each physical-tilde dimer \cite{Dzhioev2011, Brenes2020, brenes2023particle}.

The dot dimer is placed between the phonon pseudosite block and the auxiliary lead dimers, so that dot–phonon and dot–lead couplings appear as short MPO strings. The only remaining long-range terms are the physically required nonlocal binary phonon ladder operators. This mirrors efficient MPS orderings for Anderson impurity models \cite{Wolf2014PRB, HeMillis2017PRB, rams2020breaking, KohnSantoro2021PRB, KohnSantoroQuench2021}, where separating or interleaving bath sites according to dominant processes can strongly reduce entanglement growth.  

For more structured reservoir spectral densities than the Lorentzian one in [Eq.~\eqref{eq:spectral_densities}], which require additional auxiliary lead modes, the same ordering principle applies: place modes most relevant to the active transport window closest to the dot. Increasing the number of leads enhances the resolution of reservoir memory, structured spectral features, and sharp Fermi-edge physics, but also lengthens the chain, making careful MPS geometry even more important.

\subsection{Initialization and warm start}
\label{subsec:initialization_warm_start}
The initial MPS provides a stable starting point for the variational search. We use a short low-bond-dimension warmup~\cite{Mascarenhas2015, CuiCiracBanuls2015PRL,  casagrande2021analysis} to reduce the initial variational energy before entering the full homotopy and adaptive DMRG procedure described in the next Sec.~\ref{subsec:homotopy_adaptive_dmrg}.

The Liouville-space MPS contains physical and auxiliary copies of all degrees of freedom. Normalization and expectation values are evaluated with the left vacuum (defined in Eq.~\ref{eq:left_vac_f})
\begin{equation}
    \langle\!\langle \mathds{1}|\rho\rangle\!\rangle = \operatorname{Tr}[\rho],
    \label{eq:trace_functional}
\end{equation}
represented as a product of local maximally entangled bras over
physical-auxiliary pairs.  During the optimization, the MPS is repeatedly
trace-normalized,
\begin{equation}
    |\rho\rangle\!\rangle \leftarrow \frac{|\rho\rangle\!\rangle}{\langle\!\langle \mathds{1}|\rho\rangle\!\rangle}.
    \label{eq:normalization}
\end{equation}
This prevents numerical drift in the trace and makes expectation values comparable across sweeps and homotopy stages, as described below.

\subsection{Homotopy continuation and adaptive DMRG sweeps}
\label{subsec:homotopy_adaptive_dmrg}

To improve the stability of the variational search using DMRG, we use a homotopy strategy. The optimization is initialized at a nearby parameter point that is easier to converge and is then
continued toward the target parameters. For example, the phonon-bath coupling is ramped geometrically, with an analogous interpolation for $T_{\rm ph}$.  We use geometric interpolation because the relevant relaxation scales can vary strongly across the parameter range. Each converged stage is used as the initial state for the next one. This continuation procedure reduces sensitivity to the initial state and improves convergence in regions with broad phonon distributions or slow relaxation.

At each homotopy stage, Eq.~\eqref{eq:variational_energy} is minimized by two-site DMRG sweeps~\cite{White1992DMRG,White1993DMRG,Schollwoeck2005RMP,Schollwoeck2011}. The local effective equations are solved with Krylov methods. We adapt the Krylov dimension when the local solver stagnates, and increase the MPS bond dimension when truncation errors remain significant. The normalized residual $r$ [Eq.~\eqref{eq:residual}] is monitored together with the maximum truncation error per sweep. After reaching the target parameters, we perform additional polishing sweeps. In this final stage, the bond dimension is fixed or increased only slowly, and the auxiliary density-matrix perturbation \cite{White2005SingleSiteDMRG, Hubig2015SubspaceExpansion} used to enrich the local DMRG basis during previous steps is gradually decreased to zero.  

\breakablealgcaption{Initialization of the vectorized MPS}{alg:init_mps}
{\small
\begin{algorithmic}[1]
\Require Model parameters $\boldsymbol{\theta}_0$; phonon cutoff $N_{\max}=2^{n_{\rm ph}}$; discretized fermionic and bosonic reservoirs.
\Ensure Normalized warm-start state $|\rho_0\rangle\!\rangle$.

\State Encode the phonon Fock space using Eq.~\eqref{eq:binary_map}.
\State Build the doubled Liouville-space MPS chain containing the dot, phonon pseudosites, reservoirs, and auxiliary copies.
\State Construct the left vacuum $\langle\!\langle\mathds{1}|$ [Eq.~\eqref{eq:trace_functional}] as described in Sec.~\ref{subsec::superf}.
\State Prepare a factorized thermal-like initial state $|\rho_0\rangle\!\rangle$ for the dot, phonon mode, and reservoirs.
\State Normalize $|\rho_0\rangle\!\rangle$ using Eq.~\eqref{eq:normalization}.
\State Build $\mathcal{L}(\boldsymbol{\theta})$ and $\mathcal{A}$ from Eq.~\eqref{eq:variational_energy}.
\State Run a short low-bond-dimension DMRG warmup minimizing Eq.~\eqref{eq:variational_energy}.
\State Evaluate residual Eq.~\eqref{eq:residual}, current balance Eq.~\eqref{eq:current_balance}, and phonon occupation.
\If{phonon cutoff or current balance is inadequate}
    \State Increase $N_{\max}$ or use easier parameters $\boldsymbol{\theta}_{0}$ (initial homotopy point), then repeat initialization.
\EndIf
\State \Return $|\rho_0\rangle\!\rangle$ and initial solver parameters.

\end{algorithmic}}
\vspace{-.2cm}
\par\noindent\hrulefill

\begin{figure}[!tbp]
\centering
\includegraphics[width=\columnwidth, trim=0.cm 0.0cm 0.0cm 0.cm]{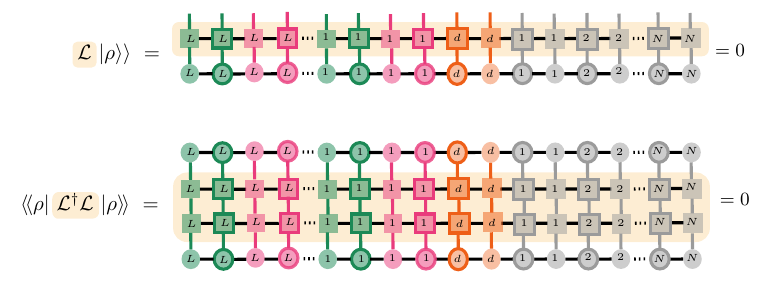}\vspace{0.3cm}
\parbox{\linewidth}{\caption{\justifying  Variational steady-state condition depicted in tensor network form. The NESS satisfies $\mathcal{L}|\rho\rangle\!\rangle=0$ and is obtained by variationally minimizing the positive semidefinite form $\langle\!\langle\rho|\mathcal{L}^\dagger\mathcal{L}|\rho\rangle\!\rangle$. \label{fig::LdagL}}}
\end{figure}
\subsection{Convergence diagnostics and phonon cutoff checks}
\label{subsec:convergence_cutoff_checks}

Since the Liouvillian is represented as a compressed MPO and the MPS has a finite bond dimension, convergence is also checked through physical consistency conditions. A primary transport diagnostic is the current balance,
\begin{equation}
    I_L \simeq J,
    \label{eq:current_balance}
\end{equation}
where $I_L$ is the particle current injected by the Markovian relaxation of the left mesoscopic lead mode, while $J$ is the coherent tunneling current across the left lead--mode--dot interface. Since these denote independently evaluated current observables, agreement between them provides a steady-state consistency check.  For the phonon mode, we monitor $\langle n_{\rm ph}\rangle$, $\operatorname{Var}(n_{\rm ph})$, $F_{\rm ph}$ [Eq.~\eqref{eq:fano}], and $g^{(2)}(0)$ [Eq.~\eqref{eq:g2}]. The final state is accepted only when the residual, current balance, variational energy, and phonon observables are stable under additional sweeps and moderate increases of the numerical control parameters.

The phonon cutoff is checked explicitly. If the phonon distribution approaches the upper edge of the retained Fock space, the calculation is repeated with a larger $M$ [Eq.~\eqref{eq:boson_basis}]. This avoids an artificial suppression of the mechanical amplitude and prevents cutoff-induced shifts in the apparent onset of oscillatory behavior. Conversely, when the distribution is strongly concentrated in the lowest phonon levels, increasing $M$ does not improve the physical accuracy but enlarges the variational search space. This situation occurs, for example, when electromechanical pumping is weak because of a small $\lambda$, a large $\omega_0$, or a narrow lead bandwidth $\delta$. For small $\omega_0$, however, the vibrational spectrum becomes dense, and the steady-state distribution can extend over many nearby Fock states so that a higher cutoff is required.

To study low-bias transport, sharp Fermi-edge effects, or more complicated spectral structures, additional auxiliary modes can be introduced. Placing more modes near the chemical potentials $\mu_\alpha$ and near relevant vibronic transition energies improves the resolution of the Fermi functions and of $J_\alpha(\omega)$ in the active transport window. The price is computational. Each extra fermionic auxiliary mode adds a physical-tilde pair in Liouville space, lengthening the MPS chain and increasing the number of dot--lead couplings in the Liouvillian MPO. At fixed bond dimension, the cost therefore grows at least linearly with the number of auxiliary modes, while the required bond dimension may also increase if additional modes become strongly correlated with the dot--phonon subsystem. In practice, this gives a controlled accuracy-cost tradeoff: sharper spectral and Fermi-edge resolution can be obtained systematically, but must be accompanied by convergence checks in $L_\alpha$, phonon cutoff, and MPS bond dimension.

Once convergence is reached, the reduced phonon state is obtained by tracing out
the dot and reservoir degrees of freedom,
\begin{equation}
    \rho_{\rm ph} = \operatorname{Tr}_{\rm dot,res}[\rho_{\rm ss}]. \label{eq:rho_ph}
\end{equation}

We then compute the Husimi-$Q$ function as in Eq.~\eqref{eq:Husimi}. Torotropy $T_Q$ is extracted from radial cuts of $Q(\alpha)$ around the phase-space center $\langle b\rangle$.  In this work, $T_Q$ is used as a phase-space diagnostic of annular, limit-cycle-like structure.  This distinction is important because a large phonon occupation alone signals vibrational energy pumping, but does not by itself establish coherent self-oscillation. The complete numerical workflow is summarized in Algorithms \ref{alg:init_mps} and \ref{alg:dmrg_ness}.

\breakablealgcaption{Variational steady-state solver}{alg:dmrg_ness}
{\small
\begin{algorithmic}[1]
\Require Warm-start state $|\rho_0\rangle\!\rangle$ from Algorithm~\ref{alg:init_mps}; target parameters $\boldsymbol{\theta}_{\rm target}$; homotopy path $\{\boldsymbol{\theta}_k\}$; initial bond and Krylov dimensions.
\Ensure Steady state $|\rho_{\rm ss}\rangle\!\rangle$, residual $r$, and steady-state observables.

\State Set $|\rho\rangle\!\rangle\leftarrow|\rho_0\rangle\!\rangle$.
\ForAll{homotopy stages $\boldsymbol{\theta}_k$}
    \State Build $\mathcal{L}(\boldsymbol{\theta}_k)$ and $\mathcal{A}_k=\mathcal{L}^{\dagger}(\boldsymbol{\theta}_k)\mathcal{L}(\boldsymbol{\theta}_k)$ as in Eq.~\eqref{eq:variational_energy}.
    \State Match MPO and MPS site indices.
    \Repeat
        \State Run DMRG sweeps minimizing Eq.~\eqref{eq:variational_energy}.
        \State Increase Krylov dimension if local convergence stalls.
        \State Normalize $|\rho\rangle\!\rangle$ using Eq.~\eqref{eq:normalization}.
        \State Evaluate current balance Eq.~\eqref{eq:current_balance} and residual Eq.~\eqref{eq:residual}.
    \Until{current balance and residual are stable at stage $k$}
\EndFor

\State Enter adaptive polishing at the target parameters.
\Repeat
    \State Run single-sweep DMRG refinements.
    \State Monitor truncation error, residual Eq.~\eqref{eq:residual}, current balance Eq.~\eqref{eq:current_balance}, and phonon observables.
    \State Increase bond dimension if the truncation error remains large.
    \State Reduce auxiliary density-matrix perturbation once bond dimension is stable.
    \State Check convergence of $g^{(2)}(0)$ from Eq.~\eqref{eq:g2}, $F_{\rm ph}$ from Eq.~\eqref{eq:fano}, and the variational energy Eq.~\eqref{eq:variational_energy}.
\Until{residual, current balance, energy, and phonon statistics have converged}

\State Normalize the final state using Eq.~\eqref{eq:normalization}.
\State Compute $\rho_{\rm ph}$ from Eq.~\eqref{eq:rho_ph}, $Q(\alpha)$ from Eq.~\eqref{eq:Husimi}, and torotropy $T_Q$.
\State Set $|\rho_{\rm ss}\rangle\!\rangle\leftarrow|\rho\rangle\!\rangle$.
\State \Return $|\rho_{\rm ss}\rangle\!\rangle$, $r$, and
       $\bigl(\langle n_{\rm ph}\rangle,g^{(2)}(0),F_{\rm ph},I_L,J,T_Q\bigr)$.

\end{algorithmic}}
\vspace{-.2cm}
\par\noindent\hrulefill

\section{Noninteracting current benchmark}\label{sec:landauer}

At $\lambda=0$ the phonon mode decouples from the dot, and the model reduces to a noninteracting spinless resonant level coupled to two structured fermionic reservoirs.  The current can then be computed from the Landauer formula
\begin{equation}
    J = \frac{1}{2\pi}\int_{-\infty}^{\infty} d\omega\,\mathcal{T}(\omega)\left[ f_L(\omega)-f_R(\omega)\right],
    \label{eq:landauer_benchmark}
\end{equation}
with
\begin{equation}
    \mathcal{T}(\omega) = J_L(\omega)J_R(\omega) \left|G_d^R(\omega)\right|^2.
    \label{eq:transmission_benchmark}
\end{equation}
For the Lorentzian hybridization functions of Eq.~\eqref{eq:spectral_densities}, the corresponding retarded self-energy is
\begin{equation}
    \Sigma_\alpha^R(\omega) =  \frac{\kappa_\alpha^2}{\omega-\omega_\alpha+i\delta_\alpha},
    \qquad  \kappa_\alpha^2=\frac{\Gamma_\alpha\delta_\alpha}{2},
    \label{eq:lorentzian_self_energy}
\end{equation}
so that
\begin{equation}
    -2\,{\rm Im}\,\Sigma_\alpha^R(\omega) =  J_\alpha(\omega).
\end{equation}

Using the parameters of Fig.~\ref{plot::transport_vs_lambda}, $\epsilon=0$, $\mu_L =-\mu_R =5$, $\Gamma=\delta=1$, $\omega_L=-\omega_R=1$, $T=1$, one has
\begin{equation}
    J_L(\omega)=\frac{1}{(\omega-1)^2+1}, \qquad J_R(\omega)=\frac{1}{(\omega+1)^2+1}.
\end{equation}
The total self-energy is
\begin{equation}
    \Sigma_L^R(\omega)+\Sigma_R^R(\omega) = \frac{1}{2}\frac{1}{\omega-1+i} + \frac{1}{2}\frac{1}{\omega+1+i} = \frac{\omega+i}{\omega^2+2i\omega-2},
\end{equation}
and therefore
\begin{equation}
    G_d^R(\omega) = \frac{1}{\omega-\epsilon-\Sigma_L^R(\omega)-\Sigma_R^R(\omega)} =    \frac{\omega^2+2i\omega-2}{\omega^3+2i\omega^2-3\omega-i}.
\end{equation}
This gives the transmission function
\begin{equation}
    \mathcal{T}(\omega) = \frac{1}{\omega^6-2\omega^4+5\omega^2+1}.
    \label{eq:benchmark_transmission_closed}
\end{equation}
Keeping the finite Fermi functions at $T=1$ and $\mu_L=-\mu_R=5$ gives
\begin{equation}
    J = \frac{1}{2\pi}\int d\omega\, \mathcal{T}(\omega) \left[ f_L(\omega)-f_R(\omega)\right] \simeq 0.1965.
    \label{eq:benchmarkeq_T}
\end{equation}
The small-$\lambda$ current in Fig.~\ref{plot::transport_vs_lambda} is consistent with these noninteracting benchmarks.

\nocite{apsrev42Control}
\bibliographystyle{apsrev4-2}
\bibliography{references}

@article{elenewski2017master,
    author = {Elenewski, Justin E. and Gruss, Daniel and Zwolak, Michael},
    title = {Communication: Master equations for electron transport: The limits of the Markovian limit},
    journal = {J. Chem. Phys.},
    volume = {147},
    number = {15},
    pages = {151101},
    year = {2017},
    month = {10},
    issn = {0021-9606},
    doi = {10.1063/1.5000747},
    url = {https://doi.org/10.1063/1.5000747},
}

@article{subotnik2009nonequilibrium,
    author = {Subotnik, Joseph E. and Hansen, Thorsten and Ratner, Mark A. and Nitzan, Abraham},
    title = {Nonequilibrium steady state transport via the reduced density matrix operator},
    journal = {J. Chem. Phys.},
    volume = {130},
    number = {14},
    pages = {144105},
    year = {2009},
    month = {04},
    issn = {0021-9606},
    doi = {10.1063/1.3109898},
    url = {https://doi.org/10.1063/1.3109898},
}

@article{ajisaka2012nonequilibrium,
  title = {Nonequilibrium particle and energy currents in quantum chains connected to mesoscopic Fermi reservoirs},
  author = {Ajisaka, Shigeru and Barra, Felipe and Mej\'{\i}a-Monasterio, Carlos and Prosen, Toma\v{z}},
  journal = {Phys. Rev. B},
  volume = {86},
  issue = {12},
  pages = {125111},
  numpages = {5},
  year = {2012},
  month = {Sep},
  publisher = {American Physical Society},
  doi = {10.1103/PhysRevB.86.125111},
  url = {https://link.aps.org/doi/10.1103/PhysRevB.86.125111}
}

@misc{github,
    title = {{ASPECTS} code repository},
howpublished = {\url{https://github.com/aspects-quantum/Auto26.git}},
    year = {2026}
}

@article{guzman2023useful,
doi = {10.1088/1361-6633/ad8803},
url = {https://doi.org/10.1088/1361-6633/ad8803},
year = {2024},
month = {nov},
publisher = {IOP Publishing},
volume = {87},
number = {12},
pages = {122001},
author = {Antonio Marín Guzmán, José and Erker, Paul and Gasparinetti, Simone and Huber, Marcus and Yunger Halpern, Nicole},
title = {Key issues review: useful autonomous quantum machines},
journal = {Reports on Progress in Physics},

}

@article{allahverdyan2004,
doi = {10.1209/epl/i2004-10101-2},
url = {https://doi.org/10.1209/epl/i2004-10101-2},
year = {2004},
month = {aug},
publisher = {},
volume = {67},
number = {4},
pages = {565},
author = {A. E. Allahverdyan and R. Balian and Th. M. Nieuwenhuizen},
title = {Maximal work extraction from finite quantum systems},
journal = {Europhysics Letters}
}

@article{PhysRevLett.105.050404,
  author = {Prior, Javier and Chin, Alex W. and Huelga, Susana F. and Plenio, Martin B.},
  title = {Efficient Simulation of Strong System-Environment Interactions},
  journal = {Phys. Rev. Lett.},
  volume = {105},
  issue = {5},
  pages = {050404},
  year = {2010},
  doi = {10.1103/PhysRevLett.105.050404},
  url = {https://doi.org/10.1103/PhysRevLett.105.050404}
}

@article{mahadeviya26,
  author = {Mahadeviya, Khalak and Moreira, Saulo V and Mandal, Sheikh Parvez and Pandit, Mahasweta and Prior, Javier and Mitchison, Mark},
  title = {Current fluctuations in nonequilibrium open quantum systems beyond weak coupling: a reaction coordinate approach},
  journal = {New J. Phys.},
  volume = {28},
  pages = {044511},
  year = {2026},
  doi = {10.1088/1367-2630/ae528c},
  url = {https://doi.org/10.1088/1367-2630/ae528c}
}

@article{mandal26,
  author = {Mandal, Sheikh Parvez and Pandit, Mahasweta and Mahadeviya, Khalak and Mitchison, Mark T. and Prior, Javier},
  title = {Heat operator approach to quantum stochastic thermodynamics in the strong-coupling regime},
  journal = {Phys. Rev. Res.},
  volume = {8},
  issue = {1},
  pages = {013321},
  year = {2026},
  doi = {10.1103/8plx-nfvq},
  url = {https://doi.org/10.1103/8plx-nfvq}
}

@article{PhysRevB.74.195305,
  author = {Gustavsson, S. and Leturcq, R. and Simovi\ifmmode \check{c}\else \v{c}\fi{}, B. and Schleser, R. and Studerus, P. and Ihn, T. and Ensslin, K. and Driscoll, D. C. and Gossard, A. C.},
  title = {Counting statistics and super-Poissonian noise in a quantum dot: Time-resolved measurements of electron transport},
  journal = {Phys. Rev. B},
  volume = {74},
  issue = {19},
  pages = {195305},
  year = {2006},
  doi = {10.1103/PhysRevB.74.195305},
  url = {https://doi.org/10.1103/PhysRevB.74.195305}
}

@article{Schinabeck2018HQMEVibBath,
  author = {Schinabeck, C. and H{\"a}rtle, R. and Thoss, M.},
  title = {Hierarchical quantum master equation approach to electronic-vibrational coupling in nonequilibrium transport through nanosystems: Reservoir formulation and application to vibrational instabilities},
  journal = {Phys. Rev. B},
  volume = {97},
  pages = {235429},
  year = {2018},
  doi = {10.1103/PhysRevB.97.235429},
  url = {https://doi.org/10.1103/PhysRevB.97.235429}
}

@article{Koenig2012,
  author = {Koenig, D. R. and Weig, E. M.},
  title = {Voltage-sustained self-oscillation of a nano-mechanical electron shuttle},
  journal = {Appl. Phys. Lett.},
  volume = {101},
  pages = {213111},
  year = {2012},
  doi = {10.1063/1.4767359},
  url = {https://doi.org/10.1063/1.4767359},
}

@article{Jonsson2008,
  title={Self-Organization of Irregular Nanoelectromechanical Vibrations<? format?> in Multimode Shuttle Structures},
  author={Jonsson, L Magnus and Santandrea, Fabio and Gorelik, Leonid Y and Shekhter, Robert I and Jonson, Mats},
  journal={Physical review letters},
  volume={100},
  number={18},
  pages={186802},
  year={2008},
  publisher={APS},
  doi = {10.1103/PhysRevLett.100.186802},
  url = {https://doi.org/10.1103/PhysRevLett.100.186802}
}

@article{Song2014,
  title={Self-sustained oscillations in nanoelectromechanical systems induced by Kondo resonance},
  author={Song, Taegeun and Kiselev, Mikhail N and Kikoin, Konstantin and Shekhter, Robert I and Gorelik, Leonid Y},
  journal={New Journal of Physics},
  volume={16},
  number={3},
  pages={033043},
  year={2014},
  publisher={IOP Publishing},
  doi = {10.1088/1367-2630/16/3/033043},
  url = {https://doi.org/10.1088/1367-2630/16/3/033043}
}

@article{Skorobagatko2013,
  author = {Skorobagatko, G. A. and Krive, I. V. and Shekhter, R. I.},
  title = {Polaronic effects in electron shuttling},
  journal = {Low Temp. Phys.},
  volume = {35},
  pages = {949--956},
  year = {2009},
  doi = {10.1063/1.3276063},
  url = {https://doi.org/10.1063/1.3276063},
}

@article{Aresbistability24,
  author = {Tabanera-Bravo, Jorge and Vigneau, Florian and Monsel, Juliette and Aggarwal, Kushagra and Bresque, L\'ea and Fedele, Federico and Cerisola, Federico and Briggs, G. A. D. and Anders, Janet and Auff\`eves, Alexia and Parrondo, Juan M. R. and Ares, Natalia},
  title = {Stability of long-sustained oscillations induced by electron tunneling},
  journal = {Phys. Rev. Res.},
  volume = {6},
  issue = {1},
  pages = {013291},
  year = {2024},
  doi = {10.1103/PhysRevResearch.6.013291},
  url = {https://doi.org/10.1103/PhysRevResearch.6.013291}
}

@article{ChenReichman2016NCA,
  title = {Anderson--Holstein model in two flavors of the noncrossing approximation},
  author = {Chen, Hsing-Ta and Cohen, Guy and Millis, Andrew J. and Reichman, David R.},
  journal = {Physical Review B},
  volume = {93},
  pages = {174309},
  year = {2016},
  doi = {10.1103/PhysRevB.93.174309}
}

@article{agarwalla2017anderson,
    author = {Agarwalla, Bijay Kumar and Segal, Dvira},
    title = {The Anderson impurity model out-of-equilibrium: Assessing the accuracy of simulation techniques with an exact current-occupation relation},
    journal = {The Journal of Chemical Physics},
    volume = {147},
    number = {5},
    pages = {054104},
    year = {2017},
    month = {08},
    issn = {0021-9606},
    doi = {10.1063/1.4996562},
    url = {https://doi.org/10.1063/1.4996562}
}

@article{HartleThoss2009PRL,
  title = {Vibrational Nonequilibrium Effects in the Conductance of Single Molecules with Multiple Electronic States},
  author = {H{\"a}rtle, R. and Benesch, C. and Thoss, M.},
  journal = {Physical Review Letters},
  volume = {102},
  pages = {146801},
  year = {2009},
  doi = {10.1103/PhysRevLett.102.146801}
}

@article{FlindtNovotny2005NoiseSpectrum,
  title = {Current noise spectrum of a quantum shuttle},
  author = {Flindt, Christian and Novotn{\'y}, Tom{\'a}{\v{s}} and Jauho, Antti-Pekka},
  journal = {Physica E: Low-dimensional Systems and Nanostructures},
  volume = {29},
  number = {3--4},
  pages = {411--418},
  year = {2005},
  doi = {10.1016/j.physe.2005.05.040}
}

@article{BennettClerk2010Scattering,
  title = {Scattering Approach to Backaction in Coherent Nanoelectromechanical Systems},
  author = {Bennett, Steven D. and Maassen, Jesse and Clerk, Aashish A.},
  journal = {Physical Review Letters},
  volume = {105},
  pages = {217206},
  year = {2010},
  doi = {10.1103/PhysRevLett.105.217206}
}

@article{ITensor,
    title = {The {ITensor} {Software} {Library} for {Tensor} {Network} {Calculations}},
    issn = {2949-804X},
    url = {https://www.scipost.org/SciPostPhysCodeb.4},
    doi = {10.21468/SciPostPhysCodeb.4},
    language = {english},
    urldate = {2025-04-11},
    journal = {SciPost Physics Codebases},
    author = {Fishman, Matthew and White, Steven and Stoudenmire, Edwin Miles},
    volume = {4},
    year = {2022},
    pages = {004},
}

@article{moeckel2014,
  title={Synchronizing a single-electron shuttle to an external drive},
  author={Moeckel, Michael J and Southworth, Darren R and Weig, Eva M and Marquardt, Florian},
  journal={New Journal of Physics},
  volume={16},
  number={4},
  pages={043009},
  year={2014},
  publisher={IOP Publishing},
  url={https://iopscience.iop.org/article/10.1088/1367-2630/16/4/043009}
}

@article{haken1964nonlinear,
  author = {Haken, Hermann},
  title = {A nonlinear theory of laser noise and coherence. I},
  journal = {Z. Phys.},
  volume = {181},
  number = {1},
  pages = {96--124},
  year = {1964},
  doi = {10.1007/BF01383921},
  url = {https://doi.org/10.1007/BF01383921}
}

@article{van1926lxxxviii,
  author = {Van der Pol, Balth},
  title = {LXXXVIII. On “relaxation-oscillations”},
  journal = {Philos. Mag.},
  volume = {2},
  number = {11},
  pages = {978--992},
  year = {1926},
  doi = {10.1080/14786442608564127},
  url = {https://doi.org/10.1080/14786442608564127}
}

@article{ginoux2012van,
  author = {Ginoux, Jean-Marc and Letellier, Christophe},
  title = {Van der Pol and the history of relaxation oscillations: Toward the emergence of a concept},
  journal = {Chaos},
  volume = {22},
  number = {2},
  pages = {023120},
  year = {2012},
  doi = {10.1063/1.3670008},
  url = {https://doi.org/10.1063/1.3670008}
}

@article{lau2016redox,
  author = {Lau, Chit Siong and Sadeghi, Hatef and Rogers, Gregory and Sangtarash, Sara and Dallas, Panagiotis and Porfyrakis, Kyriakos and Warner, Jamie and Lambert, Colin J and Briggs, G Andrew D and Mol, Jan A},
  title = {Redox-dependent Franck--Condon blockade and avalanche transport in a graphene--fullerene single-molecule transistor},
  journal = {Nano Lett.},
  volume = {16},
  number = {1},
  pages = {170--176},
  year = {2016},
  doi = {10.1021/acs.nanolett.5b03434},
  url = {https://doi.org/10.1021/acs.nanolett.5b03434}
}

@article{Koch05,
  author = {Koch, Jens and von Oppen, Felix},
  title = {Franck-Condon Blockade and Giant Fano Factors in Transport through Single Molecules},
  journal = {Phys. Rev. Lett.},
  volume = {94},
  issue = {20},
  pages = {206804},
  year = {2005},
  doi = {10.1103/PhysRevLett.94.206804},
  url = {https://doi.org/10.1103/PhysRevLett.94.206804}
}

@article{Koch96,
  author = {Koch, Jens and Raikh, M. E. and von Oppen, Felix},
  title = {Pair Tunneling through Single Molecules},
  journal = {Phys. Rev. Lett.},
  volume = {96},
  issue = {5},
  pages = {056803},
  year = {2006},
  doi = {10.1103/PhysRevLett.96.056803},
  url = {https://doi.org/10.1103/PhysRevLett.96.056803}
}

@book{schleich2015quantum,
  author = {Schleich, Wolfgang P},
  title = {Quantum optics in phase space},
  publisher = {John Wiley \& Sons},
  year = {2001},
  doi = {10.1002/3527602976},
  url = {https://doi.org/10.1002/3527602976}
}

@article{Blencowe2005NJP,
  author = {Blencowe, M. P. and Imbers, J. and Armour, A. D.},
  title = {Dynamics of a nanomechanical resonator coupled to a superconducting single-electron transistor},
  journal = {New J. Phys.},
  volume = {7},
  pages = {236},
  year = {2005},
  doi = {10.1088/1367-2630/7/1/236},
  url = {https://doi.org/10.1088/1367-2630/7/1/236}
}

@article{Fedorets2004PRL,
  author = {Fedorets, D. and Gorelik, L. Y. and Shekhter, R. I. and Jonson, M.},
  title = {Quantum Shuttle Phenomena in a Nanoelectromechanical Single-Electron Transistor},
  journal = {Phys. Rev. Lett.},
  volume = {92},
  number = {16},
  pages = {166801},
  year = {2004},
  doi = {10.1103/PhysRevLett.92.166801},
  url = {https://doi.org/10.1103/PhysRevLett.92.166801}
}

@article{Pistolesi2004PRB,
  author = {Pistolesi, F.},
  title = {Full counting statistics of a charge shuttle},
  journal = {Phys. Rev. B},
  volume = {69},
  number = {24},
  pages = {245409},
  year = {2004},
  doi = {10.1103/PhysRevB.69.245409},
  url = {https://doi.org/10.1103/PhysRevB.69.245409}
}

@article{PootVanDerZant2012PhysRep,
  author = {Poot, M. and van der Zant, H. S. J.},
  title = {Mechanical systems in the quantum regime},
  journal = {Phys. Rep.},
  volume = {511},
  number = {5},
  pages = {273--335},
  year = {2012},
  doi = {10.1016/j.physrep.2011.12.004},
  url = {https://doi.org/10.1016/j.physrep.2011.12.004}
}

@article{Schwarz2016PRB,
  author = {Schwarz, Frauke and Goldstein, Moshe and Dorda, Antonius and Arrigoni, Enrico and Weichselbaum, Andreas and von Delft, Jan},
  title = {Lindblad-driven discretized leads for nonequilibrium steady-state transport in quantum impurity models: Recovering the continuum limit},
  journal = {Phys. Rev. B},
  volume = {94},
  number = {15},
  pages = {155142},
  year = {2016},
  doi = {10.1103/PhysRevB.94.155142},
  url = {https://doi.org/10.1103/PhysRevB.94.155142}
}

@article{Arrigoni2013PRL,
  author = {Arrigoni, Enrico and Knap, Michael and von der Linden, Wolfgang},
  title = {An Auxiliary Quantum Master Equation Approach for Nonequilibrium Steady-State Properties of Strongly Correlated Impurity Models},
  journal = {Phys. Rev. Lett.},
  volume = {110},
  number = {8},
  pages = {086403},
  year = {2013},
  doi = {10.1103/PhysRevLett.110.086403},
  url = {https://doi.org/10.1103/PhysRevLett.110.086403}
}

@article{Dorda2014PRB,
  author = {Dorda, Antonius and Nuss, Martin and von der Linden, Wolfgang and Arrigoni, Enrico},
  title = {Auxiliary master equation approach to nonequilibrium correlated impurities},
  journal = {Phys. Rev. B},
  volume = {89},
  number = {16},
  pages = {165105},
  year = {2014},
  doi = {10.1103/PhysRevB.89.165105},
  url = {https://doi.org/10.1103/PhysRevB.89.165105}
}

@article{JeckelmannWhite1998PRB,
  author = {Jeckelmann, Eric and White, Steven R.},
  title = {Density-matrix renormalization-group study of the polaron problem in the Holstein model},
  journal = {Phys. Rev. B},
  volume = {57},
  number = {11},
  pages = {6376--6386},
  year = {1998},
  doi = {10.1103/PhysRevB.57.6376},
  url = {https://doi.org/10.1103/PhysRevB.57.6376}
}

@article{Verstraete2004PRL,
  author = {Verstraete, Frank and Garc{\'i}a-Ripoll, Juan Jos{\'e} and Cirac, J. Ignacio},
  title = {Matrix Product Density Operators: Simulation of Finite-Temperature and Dissipative Systems},
  journal = {Phys. Rev. Lett.},
  volume = {93},
  number = {20},
  pages = {207204},
  year = {2004},
  doi = {10.1103/PhysRevLett.93.207204},
  url = {https://doi.org/10.1103/PhysRevLett.93.207204}
}

@article{avriller2011unified,
  author = {Avriller, R},
  title = {Unified description of charge transfer mechanisms and vibronic dynamics in nanoscale junctions},
  journal = {Journal of Physics: Condensed Matter},
  volume = {23},
  number = {10},
  pages = {105301},
  year = {2011},
  doi = {10.1088/0953-8984/23/10/105301},
  url = {https://doi.org/10.1088/0953-8984/23/10/105301}
}

@article{ZwolakVidal2004PRL,
  author = {Zwolak, Michael and Vidal, Guifr{\'e}},
  title = {Mixed-State Dynamics in One-Dimensional Quantum Lattice Systems: A Time-Dependent Superoperator Renormalization Algorithm},
  journal = {Phys. Rev. Lett.},
  volume = {93},
  number = {20},
  pages = {207205},
  year = {2004},
  doi = {10.1103/PhysRevLett.93.207205},
  url = {https://doi.org/10.1103/PhysRevLett.93.207205}
}

@article{Landauer1957,
  author = {Landauer, Rolf},
  title = {Spatial Variation of Currents and Fields Due to Localized Scatterers in Metallic Conduction},
  journal = {IBM J. Res. Dev.},
  volume = {1},
  number = {3},
  pages = {223--231},
  year = {1957},
  doi = {10.1147/rd.13.0223},
  url = {https://doi.org/10.1147/rd.13.0223}
}

@article{Landauer1970,
author = {Rolf Landauer},
title = {Electrical resistance of disordered one-dimensional lattices},
journal = {The Philosophical Magazine: A Journal of Theoretical Experimental and Applied Physics},
volume = {21},
number = {172},
pages = {863--867},
year = {1970},
publisher = {Taylor \& Francis},
doi = {10.1080/14786437008238472},
URL = {https://doi.org/10.1080/14786437008238472},
}

@article{MeirWingreen1992,
  author = {Meir, Yigal and Wingreen, Ned S.},
  title = {Landauer Formula for the Current through an Interacting Electron Region},
  journal = {Phys. Rev. Lett.},
  volume = {68},
  number = {16},
  pages = {2512--2515},
  year = {1992},
  doi = {10.1103/PhysRevLett.68.2512},
  url = {https://doi.org/10.1103/PhysRevLett.68.2512}
}

@book{HaugJauho2008,
  author = {Haug, Hartmut and Jauho, Antti-Pekka},
  title = {Quantum Kinetics in Transport and Optics of Semiconductors},
  edition = {2},
  series = {Springer Series in Solid-State Sciences},
  volume = {123},
  publisher = {Springer},
  address = {Berlin, Heidelberg},
  year = {2008},
  doi = {10.1007/978-3-540-73564-9},
  url = {https://doi.org/10.1007/978-3-540-73564-9}
}

@article{Prosen2008NJP,
  author = {Prosen, Toma{\v{z}}},
  title = {Third Quantization: A General Method to Solve Master Equations for Quadratic Open Fermi Systems},
  journal = {New J. Phys.},
  volume = {10},
  number = {4},
  pages = {043026},
  year = {2008},
  doi = {10.1088/1367-2630/10/4/043026},
  url = {https://doi.org/10.1088/1367-2630/10/4/043026}
}

@article{rams2020breaking,
  author = {Rams, Marek M and Zwolak, Michael},
  title = {Breaking the entanglement barrier: Tensor network simulation of quantum transport},
  journal = {Phys. Rev. Lett.},
  volume = {124},
  number = {13},
  pages = {137701},
  year = {2020},
  doi = {10.1103/PhysRevLett.124.137701},
  url = {https://doi.org/10.1103/PhysRevLett.124.137701}
}

@article{CuiCiracBanuls2015PRL,
  author = {Cui, Jian and Cirac, J. Ignacio and Ba{\~n}uls, Mari Carmen},
  title = {Variational Matrix Product Operators for the Steady State of Dissipative Quantum Systems},
  journal = {Phys. Rev. Lett.},
  volume = {114},
  number = {22},
  pages = {220601},
  year = {2015},
  doi = {10.1103/PhysRevLett.114.220601},
  url = {https://doi.org/10.1103/PhysRevLett.114.220601}
}

@article{Wen2020NatPhys,
  author = {Wen, Yutian and Ares, N. and Schupp, F. J. and Pei, T. and Briggs, G. A. D. and Laird, E. A.},
  title = {A coherent nanomechanical oscillator driven by single-electron tunnelling},
  journal = {Nat. Phys.},
  volume = {16},
  number = {1},
  pages = {75--82},
  year = {2020},
  doi = {10.1038/s41567-019-0683-5},
  url = {https://doi.org/10.1038/s41567-019-0683-5}
}

@article{Urgell2020NatPhys,
  author = {Urgell, C. and Yang, W. and de Bonis, S. L. and Samanta, C. and Esplandiu, M. J. and Dong, Q. and Jin, Y. and Bachtold, A.},
  title = {Cooling and self-oscillation in a nanotube electromechanical resonator},
  journal = {Nat. Phys.},
  volume = {16},
  number = {1},
  pages = {32--37},
  year = {2020},
  doi = {10.1038/s41567-019-0682-6},
  url = {https://doi.org/10.1038/s41567-019-0682-6}
}

@article{Schollwoeck2005RMP,
  author = {Schollw{\"o}ck, Ulrich},
  title = {The density-matrix renormalization group},
  journal = {Rev. Mod. Phys.},
  volume = {77},
  number = {1},
  pages = {259--315},
  year = {2005},
  doi = {10.1103/RevModPhys.77.259},
  url = {https://doi.org/10.1103/RevModPhys.77.259}
}

@article{WegewijsNowack2005NJP,
  author = {Wegewijs, M. R. and Nowack, K. C.},
  title = {Nuclear wavefunction interference in single-molecule electron transport},
  journal = {New J. Phys.},
  volume = {7},
  pages = {239},
  year = {2005},
  doi = {10.1088/1367-2630/7/1/239},
  url = {https://doi.org/10.1088/1367-2630/7/1/239}
}

@book{KamenevBook2011,
  author = {Kamenev, Alex},
  title = {Field Theory of Non-Equilibrium Systems},
  publisher = {Cambridge University Press},
  address = {Cambridge},
  year = {2011},
  doi = {10.1017/CBO9781139003667},
  url = {https://doi.org/10.1017/CBO9781139003667}
}

@article{Galperin2007JPCM,
  author = {Galperin, M. and Ratner, M. A. and Nitzan, A.},
  title = {Molecular transport junctions: vibrational effects},
  journal = {J. Phys.: Condens. Matter},
  volume = {19},
  pages = {103201},
  year = {2007},
  doi = {10.1088/0953-8984/19/10/103201},
  url = {https://doi.org/10.1088/0953-8984/19/10/103201}
}

@article{RyndykHartungCuniberti2006PRB,
  author = {Ryndyk, D. A. and Hartung, M. and Cuniberti, G.},
  title = {Nonequilibrium molecular vibrons: An approach based on the nonequilibrium Green function technique},
  journal = {Phys. Rev. B},
  volume = {73},
  pages = {045420},
  year = {2006},
  doi = {10.1103/PhysRevB.73.045420},
  url = {https://doi.org/10.1103/PhysRevB.73.045420}
}

@article{Anderson1961,
  author = {Anderson, P. W.},
  title = {Localized Magnetic States in Metals},
  journal = {Phys. Rev.},
  volume = {124},
  number = {1},
  pages = {41--53},
  year = {1961},
  doi = {10.1103/PhysRev.124.41},
  url = {https://doi.org/10.1103/PhysRev.124.41}
}

@article{Holstein1959,
  author = {Holstein, T.},
  title = {Studies of Polaron Motion: Part I. The Molecular-Crystal Model},
  journal = {Ann. Phys.},
  volume = {8},
  number = {3},
  pages = {325--342},
  year = {1959},
  doi = {10.1016/0003-4916(59)90002-8},
  url = {https://doi.org/10.1016/0003-4916(59)90002-8}
}

@article{HewsonMeyer2002,
  title={Numerical renormalization group study of the Anderson-Holstein impurity model},
  author={Hewson, AC and Meyer, D},
  journal={Journal of Physics: Condensed Matter},
  volume={14},
  number={3},
  pages={427--445},
  year={2002},
  doi = {10.1088/0953-8984/14/3/312},
  url ={https://iopscience.iop.org/article/10.1088/0953-8984/14/3/312}
}

@article{Novotny2003PRL,
  author = {Novotn{\'y}, Tom{\'a}{\v{s}} and Donarini, Andrea and Jauho, Antti-Pekka},
  title = {Quantum Shuttle in Phase Space},
  journal = {Phys. Rev. Lett.},
  volume = {90},
  pages = {256801},
  year = {2003},
  doi = {10.1103/PhysRevLett.90.256801},
  url = {https://doi.org/10.1103/PhysRevLett.90.256801}
}

@article{Utami2006PRB,
  title = {Quantum noise in the electromechanical shuttle: Quantum master equation treatment},
  author = {Utami, D. Wahyu and Goan, Hsi-Sheng and Holmes, C. A. and Milburn, G. J.},
  journal = {Phys. Rev. B},
  volume = {74},
  issue = {1},
  pages = {014303},
  numpages = {20},
  year = {2006},
  month = {Jul},
  publisher = {American Physical Society},
  doi = {10.1103/PhysRevB.74.014303},
  url = {https://link.aps.org/doi/10.1103/PhysRevB.74.014303}
}

@article{alford2025subtleties,
  title = {Subtleties in the pseudomodes formalism},
  author = {Alford, Wynter and Bettmann, Laetitia P. and Landi, Gabriel T.},
  journal = {Phys. Rev. B},
  volume = {113},
  issue = {20},
  pages = {205143},
  numpages = {21},
  year = {2026},
  month = {May},
  publisher = {American Physical Society},
  doi = {10.1103/79dp-vz29},
  url = {https://link.aps.org/doi/10.1103/79dp-vz29}
}

@article{brenes2023particle,
  author = {Brenes, Marlon and Guarnieri, Giacomo and Purkayastha, Archak and Eisert, Jens and Segal, Dvira and Landi, Gabriel},
  title = {Particle current statistics in driven mesoscale conductors},
  journal = {Phys. Rev. B},
  volume = {108},
  number = {8},
  pages = {L081119},
  year = {2023},
  doi = {10.1103/PhysRevB.108.L081119},
  url = {https://doi.org/10.1103/PhysRevB.108.L081119}
}

@article{Wolf2014PRB,
  author = {Wolf, F. Alexander and McCulloch, Ian P. and Schollw{\"o}ck, Ulrich},
  title = {Solving nonequilibrium dynamical mean-field theory using matrix product states},
  journal = {Phys. Rev. B},
  volume = {90},
  pages = {235131},
  year = {2014},
  doi = {10.1103/PhysRevB.90.235131},
  url = {https://doi.org/10.1103/PhysRevB.90.235131}
}

@article{KohnSantoro2021PRB,
  author = {Kohn, Lucas and Santoro, Giuseppe E.},
  title = {Efficient mapping for Anderson impurity problems with matrix product states},
  journal = {Phys. Rev. B},
  volume = {104},
  pages = {014303},
  year = {2021},
  doi = {10.1103/PhysRevB.104.014303},
  url = {https://doi.org/10.1103/PhysRevB.104.014303}
}

@article{KohnSantoroQuench2021,
  title={Quench dynamics of the Anderson impurity model at finite temperature using matrix product states: entanglement and bath dynamics},
  author={Kohn, Lucas and Santoro, Giuseppe E},
  journal={Journal of Statistical Mechanics: Theory and Experiment},
  volume={2022},
  number={6},
  pages={063102},
  year={2022},
  publisher={IOP Publishing and SISSA},
  doi = {10.1088/1742-5468/ac729b},
  url = {https://doi.org/10.1088/1742-5468/ac729b}
}

@article{wingreen1989inelastic,
  author = {Wingreen, Ned S and Jacobsen, Karsten W and Wilkins, John W},
  title = {Inelastic scattering in resonant tunneling},
  journal = {Phys. Rev. B},
  volume = {40},
  number = {17},
  pages = {11834--11850},
  year = {1989},
  doi = {10.1103/PhysRevB.40.11834},
  url = {https://doi.org/10.1103/PhysRevB.40.11834}
}

@book{Mahan2000,
  title={Many-particle physics},
  author={Mahan, Gerald D},
  year={2013},
  publisher={Springer Science \& Business Media},
  url={https://link.springer.com/book/10.1007/978-1-4613-1469-1#bibliographic-information}
}

@article{Perez-Garcia,
  author       = {Perez-Garcia, D and Verstraete, Frank and Wolf, MM and Cirac, JI},
  issn         = {1533-7146},
  journal      = {Quantum Information \& Computation},
  number       = {5-6},
  pages        = {401--430},
  title        = {Matrix product state representations},
  volume       = {7},
  year         = {2007},
  doi          = {https://doi.org/10.26421/QIC7.5-6-1}
}

@article{LangFirsov1963,
  title={Kinetic theory of semiconductors with low mobility},
  author={Lang, IG and Firsov, Yu A},
  journal={Sov. Phys. JETP},
  volume={16},
  number={5},
  pages={1301},
  year={1963},
  url={http://refhub.elsevier.com/S0022-2313(26)00068-2/sref26}
}

@article{Bode2012Review,
  author = {Bode, Niels and Kusminskiy, Silvia Viola and Egger, Reinhold and von Oppen, Felix},
  title = {Current-induced forces in mesoscopic systems: A scattering-matrix approach},
  journal = {Beilstein J. Nanotechnol.},
  volume = {3},
  pages = {144--162},
  year = {2012},
  doi = {10.3762/bjnano.3.15},
  url = {https://doi.org/10.3762/bjnano.3.15}
}

@article{HeMillis2017PRB,
  author = {He, Zhuoran and Millis, Andrew J.},
  title = {Entanglement entropy and computational complexity of the Anderson impurity model out of equilibrium: Quench dynamics},
  journal = {Phys. Rev. B},
  volume = {96},
  pages = {085107},
  year = {2017},
  doi = {10.1103/PhysRevB.96.085107},
  url = {https://doi.org/10.1103/PhysRevB.96.085107}
}

@article{casagrande2021analysis,
  author = {Casagrande, Heitor P and Poletti, Dario and Landi, Gabriel T},
  title = {Analysis of a density matrix renormalization group approach for transport in open quantum systems},
  journal = {Comput. Phys. Commun.},
  volume = {267},
  pages = {108060},
  year = {2021},
  doi = {10.1016/j.cpc.2021.108060},
  url = {https://doi.org/10.1016/j.cpc.2021.108060}
}

@article{Flindt2010PRB,
  author = {Flindt, Christian and Novotn{\'y}, Tom{\'a}{\v{s}} and Braggio, Alessandro and Jauho, Antti-Pekka},
  title = {Counting statistics of non-Markovian quantum stochastic processes},
  journal = {Phys. Rev. B},
  volume = {82},
  pages = {155407},
  year = {2010},
  doi = {10.1103/PhysRevB.82.155407},
  url = {https://doi.org/10.1103/PhysRevB.82.155407}
}

@article{Piovano2011PRB,
  author = {Piovano, G. and Cavaliere, F. and Paladino, E. and Sassetti, M.},
  title = {Coherent properties of nanoelectromechanical systems},
  journal = {Phys. Rev. B},
  volume = {83},
  number = {24},
  pages = {245311},
  year = {2011},
  doi = {10.1103/PhysRevB.83.245311},
  url = {https://doi.org/10.1103/PhysRevB.83.245311}
}

@article{White2005SingleSiteDMRG,
  title = {Density matrix renormalization group algorithms with a single center site},
  author = {White, Steven R.},
  journal = {Phys. Rev. B},
  volume = {72},
  issue = {18},
  pages = {180403(R)},
  numpages = {4},
  year = {2005},
  month = {Nov},
  publisher = {American Physical Society},
  doi = {10.1103/PhysRevB.72.180403},
  url = {https://link.aps.org/doi/10.1103/PhysRevB.72.180403}
}

@article{Hubig2015SubspaceExpansion,
  author = {Hubig, Claudius and McCulloch, Ian P. and Schollw{\"o}ck, Ulrich and Wolf, F. Alexander},
  title = {Strictly single-site DMRG algorithm with subspace expansion},
  journal = {Phys. Rev. B},
  volume = {91},
  pages = {155115},
  year = {2015},
  doi = {10.1103/PhysRevB.91.155115},
  url = {https://doi.org/10.1103/PhysRevB.91.155115}
}

@article{White1992DMRG,
  author = {White, Steven R.},
  title = {Density Matrix Formulation for Quantum Renormalization Groups},
  journal = {Phys. Rev. Lett.},
  volume = {69},
  pages = {2863--2866},
  year = {1992},
  doi = {10.1103/PhysRevLett.69.2863},
  url = {https://doi.org/10.1103/PhysRevLett.69.2863}
}

@article{White1993DMRG,
  author = {White, Steven R.},
  title = {Density-Matrix Algorithms for Quantum Renormalization Groups},
  journal = {Phys. Rev. B},
  volume = {48},
  pages = {10345--10356},
  year = {1993},
  doi = {10.1103/PhysRevB.48.10345},
  url = {https://doi.org/10.1103/PhysRevB.48.10345}
}

@article{Thomas2013PRB,
  title = {Electron waiting times in non-Markovian quantum transport},
  author = {Thomas, Konrad H. and Flindt, Christian},
  journal = {Phys. Rev. B},
  volume = {87},
  issue = {12},
  pages = {121405(R)},
  numpages = {4},
  year = {2013},
  month = {Mar},
  publisher = {American Physical Society},
  doi = {10.1103/PhysRevB.87.121405},
  url = {https://link.aps.org/doi/10.1103/PhysRevB.87.121405}
}

@article{Schinabeck2016PRB,
  title = {Hierarchical quantum master equation approach to electronic-vibrational coupling in nonequilibrium transport through nanosystems},
  author = {Schinabeck, C. and Erpenbeck, A. and H\"artle, R. and Thoss, M.},
  journal = {Phys. Rev. B},
  volume = {94},
  issue = {20},
  pages = {201407(R)},
  numpages = {6},
  year = {2016},
  month = {Nov},
  publisher = {American Physical Society},
  doi = {10.1103/PhysRevB.94.201407},
  url = {https://link.aps.org/doi/10.1103/PhysRevB.94.201407}
}

@article{mitra2004phonon,
  author = {Mitra, Aditi and Aleiner, Igor and Millis, A. J.},
  title = {Phonon effects in molecular transistors: Quantal and classical treatment},
  journal = {Phys. Rev. B},
  volume = {69},
  number = {24},
  pages = {245302},
  year = {2004},
  doi = {10.1103/PhysRevB.69.245302},
  url = {https://doi.org/10.1103/PhysRevB.69.245302}
}

@article{de2020manipulation,
  author = {De, Bitan and Muralidharan, Bhaskaran},
  title = {Manipulation of non-linear heat currents in the dissipative anderson--holstein model},
  journal = {J. Phys.: Condens. Matter},
  volume = {32},
  number = {3},
  pages = {035305},
  year = {2020},
  doi = {10.1088/1361-648X/ab3f82},
  url = {https://doi.org/10.1088/1361-648X/ab3f82}
}

@article{dalton2001theory,
  author = {Dalton, BJ and Barnett, Stephen M and Garraway, Barry M},
  title = {Theory of pseudomodes in quantum optical processes},
  journal = {Phys. Rev. A},
  volume = {64},
  number = {5},
  pages = {053813},
  year = {2001},
  doi = {10.1103/PhysRevA.64.053813},
  url = {https://doi.org/10.1103/PhysRevA.64.053813}
}

@article{sevitz2025quantum,
  author = {Sevitz, Sofia and Cerisola, Federico and Anders, Janet},
  title = {Quantum master equation for nanoelectromechanical systems beyond the wide-band limit},
  year = {2025},
  journal = {arXiv preprint},
  doi = {10.48550/arXiv.2506.20593},
  url = {https://doi.org/10.48550/arXiv.2506.20593},
}

@article{ribeiro2015non,
  title = {Non-Markovian effects in electronic and spin transport},
  author = {Ribeiro, Pedro and Vieira, Vitor R.},
  journal = {Phys. Rev. B},
  volume = {92},
  issue = {10},
  pages = {100302(R)},
  numpages = {5},
  year = {2015},
  month = {Sep},
  publisher = {American Physical Society},
  doi = {10.1103/PhysRevB.92.100302},
  url = {https://link.aps.org/doi/10.1103/PhysRevB.92.100302}
}

@article{luo2013non,
    author = {Luo, JunYan and Jiao, HuJun and Xiong, BiTao and He, Xiao-Ling and Wang, Changrong},
    title = {Non-Markovian dynamics and noise characteristics in continuous measurement of a solid-state charge qubit},
    journal = {Journal of Applied Physics},
    volume = {114},
    number = {17},
    pages = {173703},
    year = {2013},
    month = {11},
    issn = {0021-8979},
    doi = {10.1063/1.4828870},
    url = {https://doi.org/10.1063/1.4828870}
}

@article{Gorelik1998PRL,
  author = {Gorelik, L. Y. and Isacsson, A. and Voinova, M. V. and Kasemo, B. and Shekhter, R. I. and Jonson, M.},
  title = {Shuttle Mechanism for Charge Transfer in Coulomb Blockade Nanostructures},
  journal = {Phys. Rev. Lett.},
  volume = {80},
  pages = {4526--4529},
  year = {1998},
  doi = {10.1103/PhysRevLett.80.4526},
  url = {https://doi.org/10.1103/PhysRevLett.80.4526}
}

@article{Steele2009Science,
  author = {Steele, G. A. and H{\"u}ttel, A. K. and Witkamp, B. and Poot, M. and Meerwaldt, H. B. and Kouwenhoven, L. P. and van der Zant, H. S. J.},
  title = {Strong Coupling Between Single-Electron Tunneling and Nanomechanical Motion},
  journal = {Science},
  volume = {325},
  pages = {1103--1107},
  year = {2009},
  doi = {10.1126/science.1176076},
  url = {https://doi.org/10.1126/science.1176076}
}

@article{Jenkins2013PhysRep,
  author = {Jenkins, Alejandro},
  title = {Self-oscillation},
  journal = {Phys. Rep.},
  volume = {525},
  number = {2},
  pages = {167--222},
  year = {2013},
  doi = {10.1016/j.physrep.2012.10.007},
  url = {https://doi.org/10.1016/j.physrep.2012.10.007}
}

@article{liu2024fine,
  title={Fine structure of current noise spectra in nanoelectromechanical resonators},
  author={Liu, Dong E and Levchenko, Alex},
  journal={Low Temperature Physics},
  volume={50},
  number={12},
  pages={1162--1167},
  year={2024},
  publisher={AIP Publishing},
  url={https://doi.org/10.1063/10.0034370}
}

@article{BenArosh2021PRResearch,
  author = {Ben Arosh, Lior and Cross, M. C. and Lifshitz, Ron},
  title = {Quantum limit cycles and the Rayleigh and the van der Pol oscillators},
  journal = {Phys. Rev. Res.},
  volume = {3},
  pages = {013130},
  year = {2021},
  doi = {10.1103/PhysRevResearch.3.013130},
  url = {https://doi.org/10.1103/PhysRevResearch.3.013130}
}

@article{EpsteinShowalter1996JPhysChem,
  author = {Epstein, Irving R. and Showalter, Kenneth},
  title = {Nonlinear Chemical Dynamics: Oscillations, Patterns, and Chaos},
  journal = {J. Phys. Chem.},
  volume = {100},
  number = {31},
  pages = {13132--13147},
  year = {1996},
  doi = {10.1021/jp953547m},
  url = {https://doi.org/10.1021/jp953547m}
}

@article{BellPedersen2005NatRevGenet,
  author = {Bell-Pedersen, Deborah and Cassone, Vincent M. and Earnest, David J. and Golden, Susan S. and Hardin, Paul E. and Thomas, Terry L. and Zoran, Mark J.},
  title = {Circadian rhythms from multiple oscillators: lessons from diverse organisms},
  journal = {Nat. Rev. Genet.},
  volume = {6},
  pages = {544--556},
  year = {2005},
  doi = {10.1038/nrg1633},
  url = {https://doi.org/10.1038/nrg1633}
}

@article{Patke2020NatRevMolCellBiol,
  author = {Patke, Alina and Young, Michael W. and Axelrod, Sofia},
  title = {Molecular mechanisms and physiological importance of circadian rhythms},
  journal = {Nat. Rev. Mol. Cell Biol.},
  volume = {21},
  pages = {67--84},
  year = {2020},
  doi = {10.1038/s41580-019-0179-2},
  url = {https://doi.org/10.1038/s41580-019-0179-2}
}

@article{Dzhioev2011,
  author = {Dzhioev, Alan A and Kosov, Daniel S},
  title = {Super-fermion representation of quantum kinetic equations for the electron transport problem},
  journal = {J. Chem. Phys.},
  volume = {134},
  number = {4},
  pages = {044121},
  year = {2011},
  doi = {10.1063/1.3548065},
  url = {https://doi.org/10.1063/1.3548065}
}

@article{Leturcq2009NatPhys,
  author = {Leturcq, Renaud and Stampfer, Christoph and Inderbitzin, Kevin and Durrer, Lukas and Hierold, Christofer and Mariani, Eros and Schultz, Maximilian G. and von Oppen, Felix and Ensslin, Klaus},
  title = {Franck--Condon blockade in suspended carbon nanotube quantum dots},
  journal = {Nat. Phys.},
  volume = {5},
  number = {5},
  pages = {327--331},
  year = {2009},
  doi = {10.1038/nphys1234},
  url = {https://doi.org/10.1038/nphys1234}
}

@article{ludwig2008optomechanical,
  author = {Ludwig, Max and Kubala, Bj{\"o}rn and Marquardt, Florian},
  title = {The optomechanical instability in the quantum regime},
  journal = {New J. Phys.},
  volume = {10},
  number = {9},
  pages = {095013},
  year = {2008},
  doi = {10.1088/1367-2630/10/9/095013},
  url = {https://doi.org/10.1088/1367-2630/10/9/095013}
}

@article{Park2000Nature,
  author = {Park, H. and Park, J. and Lim, A. K. L. and Anderson, E. H. and Alivisatos, A. P. and McEuen, P. L.},
  title = {Nanomechanical oscillations in a single-{C}$_{60}$ transistor},
  journal = {Nature},
  volume = {407},
  number = {6800},
  pages = {57--60},
  year = {2000},
  doi = {10.1038/35024031},
  url = {https://doi.org/10.1038/35024031}
}

@article{Lassagne2009Science,
  author = {Lassagne, B. and Tarakanov, Y. and Kinaret, J. and Garcia-Sanchez, D. and Bachtold, A.},
  title = {Coupling mechanics to charge transport in carbon nanotube mechanical resonators},
  journal = {Science},
  volume = {325},
  number = {5944},
  pages = {1107--1110},
  year = {2009},
  doi = {10.1126/science.1174290},
  url = {https://doi.org/10.1126/science.1174290}
}

@article{Siddiqui2007,
  author = {Siddiqui, L. and Ghosh, A. W. and Datta, S.},
  title = {Phonon runaway in carbon nanotube quantum dots},
  journal = {Phys. Rev. B},
  volume = {76},
  pages = {085433},
  year = {2007},
  doi = {10.1103/PhysRevB.76.085433},
  url = {https://doi.org/10.1103/PhysRevB.76.085433}
}

@article{Kolkowitz2012Science,
  author = {Kolkowitz, S. and Bleszynski Jayich, A. C. and Unterreithmeier, Q. P. and Bennett, S. D. and Rabl, P. and Harris, J. G. E. and Lukin, M. D.},
  title = {Coherent sensing of a mechanical resonator with a single-spin qubit},
  journal = {Science},
  volume = {335},
  number = {6076},
  pages = {1603--1606},
  year = {2012},
  doi = {10.1126/science.1216821},
  url = {https://doi.org/10.1126/science.1216821}
}

@article{Teufel2011Nature,
  author = {Teufel, J. D. and Donner, T. and Li, D. and Harlow, J. W. and Allman, M. S. and Cicak, K. and Sirois, A. J. and Whittaker, J. D. and Lehnert, K. W. and Simmonds, R. W.},
  title = {Sideband cooling of micromechanical motion to the quantum ground state},
  journal = {Nature},
  volume = {475},
  number = {7356},
  pages = {359--363},
  year = {2011},
  doi = {10.1038/nature10261},
  url = {https://doi.org/10.1038/nature10261}
}

@article{Aspelmeyer2014RMP,
  author = {Aspelmeyer, Markus and Kippenberg, Tobias J. and Marquardt, Florian},
  title = {Cavity optomechanics},
  journal = {Rev. Mod. Phys.},
  volume = {86},
  number = {4},
  pages = {1391--1452},
  year = {2014},
  doi = {10.1103/RevModPhys.86.1391},
  url = {https://doi.org/10.1103/RevModPhys.86.1391}
}

@article{Stolpp2021CPC,
  author = {Stolpp, Jan and K{\"o}hler, Thomas and Manmana, Salvatore R and Jeckelmann, Eric and Heidrich-Meisner, Fabian and Paeckel, Sebastian},
  title = {Comparative study of state-of-the-art matrix-product-state methods for lattice models with large local {Hilbert} spaces without {U(1)} symmetry},
  journal = {Computer Physics Communications},
  volume={269},
  pages={108106},
  year={2021},
  publisher={Elsevier},
  doi = {10.1016/j.cpc.2021.108106},
  url = {https://doi.org/10.1016/j.cpc.2021.108106}
}

@article{Zhao2023PRR,
  author = {Zhao, Pei-Yuan and Ding, Ke and Yang, Shuo},
  title = {Chebyshev pseudosite matrix product state approach for the spectral functions of electron-phonon coupling systems},
  journal = {Phys. Rev. Res.},
  volume = {5},
  pages = {023026},
  year = {2023},
  doi = {10.1103/PhysRevResearch.5.023026},
  url = {https://doi.org/10.1103/PhysRevResearch.5.023026}
}

@article{blanter2004single,
  author = {Blanter, Ya. M. and Usmani, O. and Nazarov, Yu. V.},
  title = {Single-Electron Tunneling with Strong Mechanical Feedback},
  journal = {Phys. Rev. Lett.},
  volume = {93},
  number = {13},
  pages = {136802},
  year = {2004},
  doi = {10.1103/PhysRevLett.93.136802},
  url = {https://doi.org/10.1103/PhysRevLett.93.136802}
}

@article{Usmani2007PRB,
  author = {Usmani, O. and Blanter, Ya. M. and Nazarov, Yu. V.},
  title = {Strong feedback and current noise in nanoelectromechanical systems},
  journal = {Phys. Rev. B},
  volume = {75},
  pages = {195312},
  year = {2007},
  doi = {10.1103/PhysRevB.75.195312},
  url = {https://doi.org/10.1103/PhysRevB.75.195312}
}

@article{Willick2020PRR,
  author = {Willick, Kyle and Baugh, Jonathan},
  title = {Self-driven oscillation in Coulomb blockaded suspended carbon nanotubes},
  journal = {Phys. Rev. Res.},
  volume = {2},
  pages = {033040},
  year = {2020},
  doi = {10.1103/PhysRevResearch.2.033040},
  url = {https://doi.org/10.1103/PhysRevResearch.2.033040}
}

@article{Erker2017PRX,
  author = {Erker, P. and Mitchison, M. T. and Silva, R. and Woods, M. P. and Brunner, N. and Huber, M.},
  title = {Autonomous Quantum Clocks: Does Thermodynamics Limit Our Ability to Measure Time?},
  journal = {Phys. Rev. X},
  volume = {7},
  issue = {3},
  pages = {031022},
  year = {2017},
  doi = {10.1103/PhysRevX.7.031022},
  url = {https://doi.org/10.1103/PhysRevX.7.031022}
}

@article{Wachtler2019NJP,
  author = {W{\"a}chtler, C. W. and Strasberg, P. and Klapp, S. H. L. and Schaller, G. and Jarzynski, C.},
  title = {Stochastic thermodynamics of self-oscillations: the electron shuttle},
  journal = {New J. Phys.},
  volume = {21},
  pages = {073009},
  year = {2019},
  doi = {10.1088/1367-2630/ab2727},
  url = {https://doi.org/10.1088/1367-2630/ab2727}
}

@article{Culhane2024NJP,
doi = {10.1088/1367-2630/ad202b},
url = {https://doi.org/10.1088/1367-2630/ad202b},
year = {2024},
month = {feb},
publisher = {IOP Publishing},
volume = {26},
number = {2},
pages = {023047},
author = {Culhane, Oisín and Kewming, Michael J and Silva, Alessandro and Goold, John and Mitchison, Mark T},
title = {Powering an autonomous clock with quantum electromechanics},
journal = {New Journal of Physics},

}

@article{Culhane2022PRE,
  author = {Culhane, Ois{\'\i}n and Mitchison, Mark T. and Goold, John},
  title = {Extractable work in quantum electromechanics},
  journal = {Phys. Rev. E},
  volume = {106},
  number = {3},
  pages = {L032104},
  year = {2022},
  doi = {10.1103/PhysRevE.106.L032104},
  url = {https://doi.org/10.1103/PhysRevE.106.L032104}
}

@article{ArrangoizArriola2019Nature,
  author = {Arrangoiz-Arriola, Patricio and Wollack, E. Alex and Wang, Zhaoyou and Pechal, Marek and Jiang, Wentao and McKenna, Timothy P. and Witmer, Jeremy D. and Van Laer, Rapha{\"e}l and Safavi-Naeini, Amir H.},
  title = {Resolving the energy levels of a nanomechanical oscillator},
  journal = {Nature},
  volume = {571},
  pages = {537--540},
  year = {2019},
  doi = {10.1038/s41586-019-1386-x},
  url = {https://doi.org/10.1038/s41586-019-1386-x}
}

@article{OConnell2010Nature,
  author = {O{'}Connell, A. D. and Hofheinz, M. and Ansmann, M. and Bialczak, R. C. and Lenander, M. and Lucero, E. and Neeley, M. and Sank, D. and Wang, H. and Weides, M. and Wenner, J. and Martinis, J. M. and Cleland, A. N.},
  title = {Quantum ground state and single-phonon control of a mechanical resonator},
  journal = {Nature},
  volume = {464},
  pages = {697--703},
  year = {2010},
  doi = {10.1038/nature08967},
  url = {https://doi.org/10.1038/nature08967}
}

@article{Rossi2018Nature,
  author = {Rossi, Massimiliano and Mason, David and Chen, Junxin and Tsaturyan, Yeghishe and Schliesser, Albert},
  title = {Measurement-based quantum control of mechanical motion},
  journal = {Nature},
  volume = {563},
  pages = {53--58},
  year = {2018},
  doi = {10.1038/s41586-018-0643-8},
  url = {https://doi.org/10.1038/s41586-018-0643-8}
}

@article{a2005quantum,
  author = {A Clerk, Aashish and Bennett, Steven},
  title = {Quantum nanoelectromechanics with electrons, quasi-particles and Cooper pairs: effective bath descriptions and strong feedback effects},
  journal = {New J. Phys.},
  volume = {7},
  number = {1},
  pages = {238--238},
  year = {2005},
  doi = {10.1088/1367-2630/7/1/238},
  url = {https://doi.org/10.1088/1367-2630/7/1/238}
}

@article{armour2004classical,
  author = {Armour, AD and Blencowe, MP and Zhang, Yong},
  title = {Classical dynamics of a nanomechanical resonator coupled to a single-electron transistor},
  journal = {Phys. Rev. B},
  volume = {69},
  number = {12},
  pages = {125313},
  year = {2004},
  doi = {10.1103/PhysRevB.69.125313},
  url = {https://doi.org/10.1103/PhysRevB.69.125313}
}

@article{Karwat2018PRA,
  author = {Karwat, Pawe{\l} and Reiter, Doris E. and Kuhn, Tilmann and Hess, Ortwin},
  title = {Coherent phonon lasing in a thermal quantum nanomachine},
  journal = {Phys. Rev. A},
  volume = {98},
  pages = {053855},
  year = {2018},
  doi = {10.1103/PhysRevA.98.053855},
  url = {https://doi.org/10.1103/PhysRevA.98.053855}
}

@article{Thoss2011,
  author = {H\"artle, R. and Thoss, M.},
  title = {Resonant electron transport in single-molecule junctions: Vibrational excitation, rectification, negative differential resistance, and local cooling},
  journal = {Phys. Rev. B},
  volume = {83},
  issue = {11},
  pages = {115414},
  year = {2011},
  doi = {10.1103/PhysRevB.83.115414},
  url = {https://doi.org/10.1103/PhysRevB.83.115414}
}

@article{Albrecht2013,
  author = {Albrecht, K. F. and Martin-Rodero, A. and Monreal, R. C. and M\"uhlbacher, L. and Levy Yeyati, A.},
  title = {Long transient dynamics in the Anderson-Holstein model out of equilibrium},
  journal = {Phys. Rev. B},
  volume = {87},
  issue = {8},
  pages = {085127},
  year = {2013},
  doi = {10.1103/PhysRevB.87.085127},
  url = {https://doi.org/10.1103/PhysRevB.87.085127}
}

@article{Schmidt2008,
  author = {Schmidt, T. L. and Werner, P. and M\"uhlbacher, L. and Komnik, A.},
  title = {Transient dynamics of the Anderson impurity model out of equilibrium},
  journal = {Phys. Rev. B},
  volume = {78},
  issue = {23},
  pages = {235110},
  year = {2008},
  doi = {10.1103/PhysRevB.78.235110},
  url = {https://doi.org/10.1103/PhysRevB.78.235110}
}

@article{sevitz2026autonomous,
  title={Autonomous conversion of particle-exchange to quantum self-oscillations},
  author={Sevitz, Sofia and Cerisola, Federico and Hovhannisyan, Karen V and Anders, Janet},
  journal={Quantum Science and Technology},
  volume={11},
  number={1},
  pages={015059},
  year={2026},
  publisher={IOP Publishing},
  url={https://iopscience.iop.org/article/10.1088/2058-9565/ae2e39}
}

@article{hubener2007vibrational,
  author = {H{\"u}bener, Hannes and Brandes, Tobias},
  title = {Vibrational coherences in single electron tunneling through nanoscale oscillators},
  journal = {Phys. Rev. Lett.},
  volume = {99},
  number = {24},
  pages = {247206},
  year = {2007},
  doi = {10.1103/PhysRevLett.99.247206},
  url = {https://doi.org/10.1103/PhysRevLett.99.247206}
}

@article{Schollwoeck2011,
  author = {Schollw{\"o}ck, Ulrich},
  title = {The density-matrix renormalization group in the age of matrix product states},
  journal = {Ann. Phys.},
  volume = {326},
  pages = {96--192},
  year = {2011},
  doi = {10.1016/j.aop.2010.09.012},
  url = {https://doi.org/10.1016/j.aop.2010.09.012}
}

@article{Mascarenhas2015,
  author = {Mascarenhas, Eduardo and Flayac, Hugo and Savona, Vincenzo},
  title = {Matrix-product-operator approach to the nonequilibrium steady state of driven-dissipative quantum arrays},
  journal = {Phys. Rev. A},
  volume = {92},
  pages = {022116},
  year = {2015},
  doi = {10.1103/PhysRevA.92.022116},
  url = {https://doi.org/10.1103/PhysRevA.92.022116}
}

@article{Brenes2020,
  author = {Brenes, Marlon and Mendoza-Arenas, Juan Jos{\'e} and Purkayastha, Archak and Mitchison, Mark T. and Clark, Stephen R. and Goold, John},
  title = {Tensor-Network Method to Simulate Strongly Interacting Quantum Thermal Machines},
  journal = {Phys. Rev. X},
  volume = {10},
  pages = {031040},
  year = {2020},
  doi = {10.1103/PhysRevX.10.031040},
  url = {https://doi.org/10.1103/PhysRevX.10.031040}
}

@article{Tamascelli2018,
  author = {Tamascelli, Dario and Smirne, Andrea and Huelga, Susana F. and Plenio, Martin B.},
  title = {Nonperturbative Treatment of non-Markovian Dynamics of Open Quantum Systems},
  journal = {Phys. Rev. Lett.},
  volume = {120},
  pages = {030402},
  year = {2018},
  doi = {10.1103/PhysRevLett.120.030402},
  url = {https://doi.org/10.1103/PhysRevLett.120.030402}
}

@article{Schinabeck2020HQMEFCS,
  author = {Schinabeck, C. and Thoss, M.},
  title = {Hierarchical quantum master equation approach to current fluctuations in nonequilibrium charge transport through nanosystems},
  journal = {Phys. Rev. B},
  volume = {101},
  pages = {075422},
  year = {2020},
  doi = {10.1103/PhysRevB.101.075422},
  url = {https://doi.org/10.1103/PhysRevB.101.075422}
}

@article{Imamoglu1994,
  author = {Imamo{\=g}lu, A.},
  title = {Stochastic wave-function approach to non-Markovian systems},
  journal = {Phys. Rev. A},
  volume = {50},
  pages = {3650--3653},
  year = {1994},
  doi = {10.1103/PhysRevA.50.3650},
  url = {https://doi.org/10.1103/PhysRevA.50.3650}
}

@article{Vigneau2022PRR,
  author = {Vigneau, Florian and Monsel, Juliette and Tabanera, Jorge and Aggarwal, Kushagra and Bresque, L{\'e}a and Fedele, Federico and Cerisola, Federico and Briggs, G. A. D. and Anders, Janet and Parrondo, Juan M. R. and Auff{\`e}ves, Alexia and Ares, Natalia},
  title = {Ultrastrong coupling between electron tunneling and mechanical motion},
  journal = {Phys. Rev. Res.},
  volume = {4},
  number = {4},
  pages = {043168},
  year = {2022},
  doi = {10.1103/PhysRevResearch.4.043168},
  url = {https://doi.org/10.1103/PhysRevResearch.4.043168}
}

@article{2026roadmap,
  author = {Campbell, Steve and d’Amico, Irene and Ciampini, Mario A and Anders, Janet and Ares, Natalia and Artini, Simone and Auff{\`e}ves, Alexia and Bassman Oftelie, Lindsay and Bettmann, Laetitia P and Bonan{\c{c}}a, Marcus VS and others},
  title = {Roadmap on quantum thermodynamics},
  journal = {Quantum Sci. Technol.},
  volume = {11},
  number = {1},
  pages = {012501},
  year = {2026},
  doi = {10.1088/2058-9565/ae1e27},
  url = {https://doi.org/10.1088/2058-9565/ae1e27},
}

@article{deVegaAlonso2017RMP,
  author = {de Vega, In{\'e}s and Alonso, Daniel},
  title = {Dynamics of non-Markovian open quantum systems},
  journal = {Rev. Mod. Phys.},
  volume = {89},
  number = {1},
  pages = {015001},
  year = {2017},
  doi = {10.1103/RevModPhys.89.015001},
  url = {https://doi.org/10.1103/RevModPhys.89.015001}
}

@article{lacerda2024entropy,
  author = {Lacerda, Artur M and Kewming, Michael J and Brenes, Marlon and Jackson, Conor and Clark, Stephen R and Mitchison, Mark T and Goold, John},
  title = {Entropy production in the mesoscopic-leads formulation of quantum thermodynamics},
  journal = {Phys. Rev. E},
  volume = {110},
  number = {1},
  pages = {014125},
  year = {2024},
  doi = {10.1103/PhysRevE.110.014125},
  url = {https://doi.org/10.1103/PhysRevE.110.014125}
}

@article{Strathearn2018NatComm,
  author = {Strathearn, Aidan and Kirton, Peter and Kilda, Dainius and Keeling, Jonathan and Lovett, Brendon W.},
  title = {Efficient non-Markovian quantum dynamics using time-evolving matrix product operators},
  journal = {Nat. Commun.},
  volume = {9},
  pages = {3322},
  year = {2018},
  doi = {10.1038/s41467-018-05617-3},
  url = {https://doi.org/10.1038/s41467-018-05617-3}
}

@article{Weimer2021RMP,
  author = {Weimer, Hendrik and Kshetrimayum, Augustine and Or{\'u}s, Rom{\'a}n},
  title = {Simulation methods for open quantum many-body systems},
  journal = {Rev. Mod. Phys.},
  volume = {93},
  number = {1},
  pages = {015008},
  year = {2021},
  doi = {10.1103/RevModPhys.93.015008},
  url = {https://doi.org/10.1103/RevModPhys.93.015008}
}

@article{jovchev2013,
  author = {Jovchev, Andre and Anders, Frithjof B.},
  title = {Influence of vibrational modes on quantum transport through a nanodevice},
  journal = {Phys. Rev. B},
  volume = {87},
  issue = {19},
  pages = {195112},
  year = {2013},
  doi = {10.1103/PhysRevB.87.195112},
  url = {https://doi.org/10.1103/PhysRevB.87.195112}
}

@article{sapmaz2006tunneling,
  author = {Sapmaz, S and Jarillo-Herrero, P and Blanter, Ya M and Dekker, C and Van Der Zant, HSJ},
  title = {Tunneling in suspended carbon nanotubes assisted by longitudinal phonons},
  journal = {Phys. Rev. Lett.},
  volume = {96},
  number = {2},
  pages = {026801},
  year = {2006},
  doi = {10.1103/PhysRevLett.96.026801},
  url = {https://doi.org/10.1103/PhysRevLett.96.026801}
}

@CONTROL{apsrev42Control,
  title = "0"
}

@article{PhysRevLett.112.094102,
  title = {Quantum Synchronization of a Driven Self-Sustained Oscillator},
  author = {Walter, Stefan and Nunnenkamp, Andreas and Bruder, Christoph},
  journal = {Phys. Rev. Lett.},
  volume = {112},
  issue = {9},
  pages = {094102},
  numpages = {5},
  year = {2014},
  month = {Mar},
  publisher = {American Physical Society},
  doi = {10.1103/PhysRevLett.112.094102},
  url = {https://link.aps.org/doi/10.1103/PhysRevLett.112.094102}
}

@article{Weiss_2016,
doi = {10.1088/1367-2630/18/1/013043},
url = {https://doi.org/10.1088/1367-2630/18/1/013043},
year = {2016},
month = {jan},
publisher = {IOP Publishing},
volume = {18},
number = {1},
pages = {013043},
author = {Weiss, Talitha and Kronwald, Andreas and Marquardt, Florian},
title = {Noise-induced transitions in optomechanical synchronization},
journal = {New Journal of Physics},
}

\end{document}